\pdfoutput=1
\documentclass[11pt]{article}

\usepackage[a4paper,margin=2.5cm]{geometry}
\usepackage[T1]{fontenc}
\usepackage{lmodern}
\rmfamily
\DeclareFontShape{T1}{lmr}{bx}{sc}{<->ssub*cmr/bx/sc}{}
\usepackage{microtype}
\usepackage{amsmath,amssymb,amsthm}
\usepackage{mathtools}
\usepackage{booktabs}
\usepackage{xspace}
\usepackage{graphicx}
\usepackage[hyphens]{url}
\usepackage[hidelinks]{hyperref}

\def\..{\,\mathpunct{\ldotp\ldotp}} %

\newcommand{\F}{\mathbf F}
\newtheorem{problem}{Problem}
\newtheorem{proposition}{Proposition}

\newcommand{\Prob}{\operatorname{Prob}}

\renewcommand{\epsilon}{\varepsilon}
\newcommand{\VFunc}{{\scshape{VFunc}}\xspace}
\newcommand{\VFilter}{{\scshape{VFilter}}\xspace}
\newcommand{\CompVFunc}{{\scshape{CompVFunc}}\xspace}
\newcommand{\PHunc}{{\scshape{PHunc}}\xspace}
\newcommand{\LcpMmphf}{{\scshape{LcpMmphf}}\xspace}
\newcommand{\LcpTwoMmphf}{{\scshape{Lcp2Mmphf}}\xspace}

\title{$\epsilon$--Cost Sharding: Scaling Hypergraph--Based
Data Structures to Trillions of Keys}

\author{Sebastiano Vigna\\Dipartimento di Informatica, Universit\`a degli Studi di Milano}
\date{}

\begin{document}
\bibliographystyle{plain}
\maketitle
\begin{abstract}
	We describe a simple yet very scalable implementation of static functions
	(\VFunc) and of static filters (\VFilter) based on hypergraphs. We introduce
	the idea of \emph{$\epsilon$-cost sharding}, which makes it possible to build
	structures with trillions of keys while increasing the memory locality of
	hypergraph-based constructions. Unlike the commonly
	used HEM sharding method~\cite{BPZPPHNOS}, $\epsilon$-cost sharding does not
	require storing additional information, and does not introduce
	dependencies in the computation chain; its only costs are a few
	arithmetic instructions and a relative increase $\epsilon$ in space
	usage. We apply $\epsilon$-cost sharding to the classical MWHC
	construction~\cite{MWHFPHM}, but we obtain the best results with
	Dietzfelbinger and Walzer's \emph{fuse graphs}~\cite{DiWDPRUH}. For
	large key sets, the resulting structures have an overhead of $10.5$\% with
	respect to the information-theoretical lower bound, and a query time within a
	few nanoseconds of that of the unsharded version, which is the fastest
	hypergraph-based construction within the same space bounds. Besides comparing our
	structures with an unsharded version, we contrast their tradeoffs with those
	of \emph{bumped ribbon} constructions~\cite{DDHSWRFSSRAM}, a space-saving
	alternative to hypergraph-based static functions and filters, which have
	almost optimal space usage, but slower sequential construction (although
	their construction can be parallelized very efficiently) and, in general,
	slower queries.
	We also introduce \PHunc, a static function that stores its
	values in an array indexed by the PHastR minimal perfect hash
	function~\cite{VigP}: in all our experiments, the structure with the fastest
	queries among those that scale to large key sets in comparable space is one
	of \VFunc, \VFilter, and \PHunc. We build a
	trillion-key $2$-bit filter offline on a single machine in $78$\,ns/key,
	using little more memory than the filter itself. Finally, we show that
	$\epsilon$-cost sharding and fuse graphs carry over with no effort to data
	structures that reduce to static functions, such as monotone minimal perfect
	hash functions and compressed static functions, with the fastest queries
	available.
\end{abstract}

\section{Introduction}\label{sec:intro}

\emph{Static functions} are data structures designed to store arbitrary
functions from finite sets to integers; that is, given a set of $n$ keys $K$,
and for each $x\in K$ a value $v_x$, with $0\leq v_x<
	2^b$, a static function will retrieve $v_x$ given $x\in K$ in constant time.
Closely related are \emph{minimal perfect hash functions (MPHFs)}, where only
the set $K$ is given, and the data structure yields an injective numbering of
$K$ into the first $n$ natural numbers.

While these tasks can be easily implemented using hash tables,
static functions and MPHFs are allowed to return \emph{any} value if the
queried key is not in the original set $K$; this relaxation makes it possible to
break the information-theoretical lower bound of storing the set $K$. Indeed,
constructions for static functions achieve just $O(nb)$ bits of space and MPHFs
$O(n)$ bits of space, regardless of the size of the keys. This property makes
static functions and MPHFs powerful tools for handling, for instance, large sets
of strings, and they are important building blocks of space-efficient data
structures.

Historically, the first construction of static functions was proposed
by Majewski, Wormald, Havas, and Czech~\cite{MWHFPHM}; it is based on the idea
of solving a random linear system greedily: the solution is guaranteed to exist
for sufficiently sparse systems because of \emph{peelability} properties of an
associated random hypergraph~\cite{MolCRYBF}. Their technique was later refined
(and sometimes rediscovered~\cite{CKRBF}) to reduce the space usage~\cite{GOVFSCCSMPHF,DiWDPRUH}.

The main problem with hypergraph peeling is that it does not scale
well due to its poor cache behavior. Visiting a hypergraph is a linear
operation, but the memory access pattern is very irregular, and the data
structure is not amenable to parallelization.

A different approach is given by variants of the \emph{ribbon}
construction~\cite{DDHSWRFSSRAM}, which again uses a random linear system, but of
a specific form that, albeit not peelable as a hypergraph, can be solved easily
by Gaussian elimination; this makes it possible to build static functions and
filters whose space is within $1$\% of the optimum. The construction is
also highly parallelizable. The downside, as we will see, is that queries and
sequential construction are slow.

As noted by Dietzfelbinger and Pagh~\cite{DiPSDSRAM}, every construction for a
static function provides a construction for a static \emph{filter}, which is a
folklore name for an approximate dictionary with false positives. It is
sufficient to store a static function mapping every key to a $b$-bit hash, and
at query time check that the output of the function is equal to the hash of the
key. Thus, a construction for static functions is a construction for static
filters (but not vice versa).

In this paper, we propose \emph{$\epsilon$-cost sharding}, a new way to build
hypergraph-based static functions (in particular, those based on fuse
graphs~\cite{DiWDPRUH}) for very large key sets. Unlike the commonly used HEM
sharding method~\cite{BPZPPHNOS}, $\epsilon$-cost sharding stores no offsets or
seeds for the shards, and its only costs are a few arithmetic instructions and a
relative increase $\epsilon$ in space. The static functions and filters we
obtain use about $10.5$\% more space than the information-theoretical lower
bound for large key sets, so they are not competitive with ribbon-based
constructions in terms of space, but in general they provide faster queries (much
faster on positive queries, which are the case of static functions) and faster sequential
construction, which for most realistic large-scale applications are more
relevant than a ${\approx}10$\% size gain. Moreover, for large key sets
their construction can be parallelized, and becomes very fast, although bumped
ribbons parallelize even better. Our structures can be easily built offline, and
we report timings for the offline construction of large filters (up to a
trillion keys).

We also compare our structures with \PHunc, a static function that stores its
values in an array indexed by the PHastR minimal perfect hash
function~\cite{VigP}. A query to \PHunc must wait for the result of the minimal
perfect hash function before accessing the array, whereas a query to our
structures issues independent memory accesses; nonetheless, \PHunc turns out to
be a strong alternative, as its space overhead is about two bits per key, it
builds faster in parallel, and its queries are faster when the compiler cannot
vectorize the query loop (under highly parallel workloads on large key sets,
they are within a few percent of those of \VFunc even when the loop is
vectorized). Which of the two
approaches is preferable depends on the application
(Section~\ref{sec:results}).

The same machinery applies, with no additional effort, to every data structure
that can be reduced to static functions, and we show this on two classical
problems. Monotone minimal perfect hash functions based on longest common
prefixes~\cite{BBPMMPH,BBPTPMMPH2}, once built on our static functions, have a
query time up to an order of magnitude smaller than that of the learned
construction LeMonHash~\cite{FLSVLMMPH}, whose query time grows quickly with the
size of the key set (Section~\ref{sec:mmphf}). Compressed
static functions~\cite{GOVFSCCSMPHF}, in which the codeword of each value is
stored using a run of translated edges, answer queries up to five times faster
than the original HEM-sharded implementation in the same space, and up to
thirteen times faster than a recent construction based on bumped
ribbons~\cite{HLVWLSFDS} (Section~\ref{sec:csf}); in this case, the correlation
between translated edges raises the peelability threshold, and we describe an
empirical law, fitted on simulations, that takes it into account.

All our implementations are distributed as free software as part of the Rust
crate \texttt{sux},\footnote{\url{https://crates.io/crates/sux}} except for
the variants used only for comparison, which are part of our benchmark framework
(Section~\ref{sec:protocol}).

\section{Notation and Tools}
\label{sec:notation}

We use von Neumann's definition and notation for natural numbers,
identifying $n$ with the set $\{\,0,1,\ldots,n-1\,\}$, so $2=\{\,0,1\,\}$ and
$2^b$ is the set of $b$-bit numbers. We use $\ln$ for the natural logarithm,
$\lg$ for the binary logarithm, and $\log$ in Big-Oh notation.

A \emph{$k$-hypergraph} (or \emph{$k$-uniform hypergraph}) on a vertex set $V$
is a subset $E$ of $\binom Vk$, the set
of subsets of $V$ of cardinality $k$. An element of $E$ is called a \emph{hyperedge};
since we will only be discussing hypergraphs, we will call it simply an
\emph{edge}. The \emph{degree} of a vertex is the number of edges that contain
it. An \emph{induced subgraph} is given by a subset of vertices and all edges
incident on those vertices (and no other vertices). The $d$-core of a hypergraph
is its maximal induced subgraph in which every vertex has degree at least $d$.

A hypergraph is \emph{peelable} if it is possible to sort its edges in a list so
that for each edge there is a vertex that does not appear in the following
elements of the list (such an order can be found easily in linear time using a
visit). If we fix an Abelian group $\mathbf A$ and assign to each edge $e$ of a
$k$-hypergraph a value $v_e\in\mathbf A$, we can associate with the hypergraph a
system of equations in which variables are vertices and edges represent
equations: more precisely, an edge $e\subseteq V$ represents the equation
$\sum_{v\in e}v=v_e$. This correspondence induces a bijection between
systems with $|E|$ equations, each containing $k$ variables, and $k$-hypergraphs
with values assigned to edges. We will switch frequently between the two views.

\section{Problems and Previous Work}

Let us first state our problems more formally.

\begin{problem}[Static functions]
\label{prob:func}
We are given a set of keys $K$, $|K|=n$, out of a universe $U$, and
a function $f:K\to 2^b$ associating a $b$-bit value with each key. Our problem is to
store information so that given $x \in U$ we can compute in constant time a
value that will be $f(x)$ if $x\in K$.
\end{problem}

The key point here is that we give no condition when $x\not\in K$ (except that
the computation must still be performed in constant time), so the
information-theoretical lower bound for storing $K$ as a subset of $U$ does not
apply. However, storing the set of values requires in general $nb$ bits.

In some literature, the alternative term \emph{retrieval problem} is used
to denote the same problem~\cite{DiPSDSRAM}.

\begin{problem}[Static approximate dictionaries, or static filters]
\label{prob:filter}
We are given a set of keys $K$, $|K|=n$, out of a universe $U$, and a $\rho$
such that $0<\rho<1$. Our problem is to store information so that given $x \in
	U$ we can compute in constant time whether $x\in K$ with a false-positive
error rate of $\rho$. The error is one-sided: every $x\in K$ is reported as
present, and every $x\not\in K$ is reported as present with probability at most
$\rho$. There are no false negatives.
\end{problem}

This problem is also known as the \emph{approximate membership
problem}~\cite{DiPSDSRAM}.

In general, storing a function with a $b$-bit output requires at least $nb$
bits, whereas storing a filter requires at least $n\lg(1/\rho)$
bits~\cite{CFGEAMT}. In both cases we will use the term \emph{overhead} to
denote the increase in space with respect to the optimum.

\subsection{MWHC} In their seminal paper~\cite{MWHFPHM}, Majewski, Wormald, Havas, and Czech (MWHC
hereinafter) introduced the first compact construction for static functions
using the connection between systems of equations and hypergraphs. To store a function
$f:K\to 2^b$ with $b$-bit values, they generate a random system with $n=|K|$
equations in $\lceil cn\rceil$ variables, for some \emph{expansion factor} $c\geq1$, of the form
\begin{equation}
	\label{eq:MWHC}
	w_{h_0(x)}\oplus w_{h_1(x)}\oplus \cdots
	\oplus w_{h_{k-1}(x)}= f(x) \quad x \in S.
\end{equation}
Here $h_i:U\to \lceil cn\rceil$ are $k$ fully random hash functions, and the $w_i$'s are
variables in some Abelian group with operation $\oplus$. Due to bounds on the
peelability of random hypergraphs, when $k\geq 3$ and the expansion factor $c$
(i.e., essentially the ratio between the number of variables
and the number of equations) is above a certain threshold $\gamma_k$,
the system can almost always be triangulated in linear time by peeling the
associated hypergraph, and thus solved; in other words, we have both a
probabilistic guarantee that the system is solvable and that the solution can
be found in linear time.

The data structure is then a solution of the system (i.e., the values of the
variables $w_i$): storing the solution clearly makes it possible to compute
$f(x)$ in constant time. The space usage is approximately $\gamma_k b$ bits per
key. The constant $\gamma_k$ attains its minimum at $k=3$ ($\gamma_3\approx
	1.2218$), providing static functions with $23$\% space overhead~\cite{MolCRYBF}, as
in practice the number of vertices must be about $1$\% above the threshold.

The original paper~\cite{MWHFPHM} used arithmetic modulo some integer $m$ for
$\oplus$. All modern constructions use $b$-bit vectors and XOR instead, as in
practical applications there is no advantage in using a modulus that is not a
power of $2$, and modulo operations are very slow.

As we noted in the introduction, Dietzfelbinger and Pagh~\cite{DiPSDSRAM} argue
that every construction for a static function provides a construction for a
static filter by mapping each key to a $b$-bit hash, providing a false-positive
error rate of $2^{-b}$. This observation led to implementations of static
filters based on the MWHC construction, such as those in the Sux4J library of
succinct data structures~\cite{Sux4J31}. A few years later, Graf and Lemire
named the same structures “Xor filters”~\cite{GrLXF}; their paper, however, does
not compare with previous MWHC-based implementations. Moreover, “Xor filter”,
albeit catchy, is a bit of a misnomer, as essentially all constructions of this
kind are based on the XOR operation. We will thus use the MWHC acronym to refer
to the construction and keep the “static” label to avoid confusion.

\subsection{HEM}\label{sec:HEM} Botelho, Pagh, and Ziviani~\cite{BPZPPHNOS}
introduced a practical external-memory algorithm called Heuristic External
Memory (HEM) to construct static functions (and MPHFs) for sets that are too
large to store their hypergraph in memory. The idea is to map all keys
to random-looking \emph{signatures} of $\Theta(\log n)$ bits that guarantee uniqueness
with high probability (typically, $128$-bit good-quality signatures are
sufficient for all applications). At that point, signatures are sorted and
divided into shards using the $h$ highest bits, resulting in $2^h$ shards;
alternatively, one can generalize and map a larger number of high bits to a
range that is not a power of two using fixed-point
arithmetic~\cite{GOVFSCCSMPHF}, which makes it possible to size the shards more
accurately. One then computes a separate structure for the keys in each shard
and stores all such structures. To do so, one has to keep track of
\emph{offsets}, that is, the starting position of the representation of each
structure. Moreover, for each shard one keeps track of a \emph{local seed},
which is necessary to obviate a possible failure of the hypergraph
construction in each shard. While devised for MPHFs based on the MWHC technique,
HEM can be made to work with any construction technique based on hypergraphs,
and it has long been the standard approach to the construction of large-scale
static functions and filters: besides making it possible to build structures
larger than the available memory, sharding increases the locality of the
construction and enables parallel computation.

Partitioning is by now standard practice in large-scale minimal perfect
hashing, too: PTHash is built in parallel and in external memory by
partitioning the key set~\cite{PiTPEMCMPHFP}, and PHOBIC~\cite{HLPSWPHOBIC}
analyzes explicitly the tradeoff between partition size and space.

\subsection{GOV}
A first improvement over MWHC is due to Genuzio, Ottaviano, and
Vigna~\cite{GOVFSCCSMPHF}, who proposed generating
random systems in the same fashion as MWHC, but at the threshold of solvability
of linear systems on $\mathbf F_2$, which is lower than that of peelability,
and moreover decreases towards one as the number of variables per equation
increases~\cite{PiSSTkX}. Gaussian elimination, however, is cubic, so the
authors shard the keys heavily using a variant of HEM and solve each shard
using \emph{lazy Gaussian elimination}, a heuristic technique that makes
it possible to quickly solve systems with tens of thousands of variables.
The overhead in this case is about $10$\% with $3$ variables per equation,
or $4$\% using $4$ variables.

\subsection{Fuse Graphs}

Dietzfelbinger and Walzer~\cite{DiWDPRUH} revolutionized the field by
introducing a class of random graphs, \emph{fuse graphs}, that are almost always
peelable if sufficiently sparse, as in the general case, but have peelability
thresholds close to the solvability threshold of linear systems on $\mathbf F_2$
for sufficiently large key sets. This approach instantly made it possible to
build static functions by peeling with an overhead similar to that of GOV, but
with a much faster construction time. Fuse graphs are currently the state of the art among
hypergraph-based constructions, although they have some limitations that we will
discuss later.

\subsection{Bumped Ribbons}

Finally, Dietzfelbinger, Dillinger, Hübschle-Schneider, Sanders, and Walzer
introduced \emph{bumped ribbons}~\cite{DDHSWRFSSRAM}, building on the
\emph{ribbon} filter of Dillinger and Walzer~\cite{DiWRFPSBX}: a class of
systems of equations of particular form that can be solved in practical linear
time, and in which a good locality makes it possible to use a large number of
variables, thus bypassing the limits of GOV and lowering the space overhead
below $1$\%. This is a goal that cannot be reached by hypergraph peeling, but,
as we will see, using a large number of variables has a high cost in
terms of construction and query time. We will use the acronym proposed by the
authors, \emph{BuRR} (for \emph{Bumped Ribbon Retrieval}), to refer to the
construction.

\subsection{Other Constructions}
There are many other constructions of filters, some of which even support
updates: for example, blocked Bloom
filters~\cite{PSSCHSEBF}, cuckoo filters~\cite{FAKMCFPBTB}, Morton
filters~\cite{MorFCSCF}, quotient filters~\cite{BFJDTCHF}, and so on. A very large
number of these variants has been
painstakingly tested in the BuRR paper~\cite{DDHSWRFSSRAM}, with a very detailed
analysis of the tradeoffs within each construction, so we will not repeat the
same analysis here. Their conclusion is that it is difficult to beat the speed of
hypergraph-based constructions (in particular, the MWHC construction, therein
called “Xor filter”) due to the simplicity of their operation, and to the
fact that they issue independent memory accesses.

Minimal perfect hash functions follow a parallel line of development, which
starts with Hash, Displace, and Compress~\cite{BBDHDC} and continues with
PTHash~\cite{PiTPTH}, PtrHash~\cite{RagPTR}, and PHast+~\cite{BeSPHast}. In our
experiments we use PHastR~\cite{VigP}, a minimal perfect hash function based,
like PHast+, on bucket placement, which is available in \texttt{sux}, to build
\PHunc, a static function that stores its values in an array indexed by PHastR.

Our goal is to make hypergraph-based structures scalable to arbitrary sizes on
commodity hardware, that is, on a single machine whose memory need not be much
larger than the structure being built.

\section{\texorpdfstring{$\epsilon$}{ε}--Cost Sharding}
\label{sec:eps}

We already discussed the HEM approach to the construction of static functions
for large sets of keys. HEM slows down queries significantly for two main
reasons:
\begin{itemize}
	\item shards have approximately, but not exactly, the same size, so it is in
	      general necessary to keep track of their sizes---more commonly, of the
	      prefix sums of their sizes;
	\item if the construction of the data structure is probabilistic, the
	      construction of each shard may fail, which requires storing a
	      \emph{local seed} per shard, used to restart the construction in case
	      of failure.
\end{itemize}

The consequence is that it is necessary to store these two pieces of data
(typically, conflated in a single $64$-bit word, as a few bits of seeding are
sufficient), and at query time retrieving them might be costly in terms of cache
misses; more importantly, they introduce a memory dependency in the chain of
operations that are necessary to compute the output. For example, in the GOV
construction~\cite{GOVFSCCSMPHF} the necessity of creating shards small enough to
be solved yields a fairly large offset/seed array, which has a significant
impact on query time.\footnote{BuRR does not have this problem
	because, as its authors observe~\cite{DDHSWRFSSRAM}, the difference between
	shards can be conflated in the bumping process.}

We thus propose \emph{$\epsilon$-cost sharding}. Our observation is that
sharding is essentially a \emph{balls-and-bins problem} when we treat the shard
assignment as random. Using a Chernoff bound, we will now show how, given an
error $\epsilon$ and a probability $\theta$, one can compute easily and
accurately a maximum number of bins such that, with probability at least
$1-\theta$, the ratio between the maximum and the average number of balls in a
bin is bounded by $1+\epsilon$; the tight result of Raab and
Steger~\cite{RaLBIB} on the maximum number of balls in a bin shows that this
number is essentially the largest possible. At that point, since (as is usually
the case, and in particular in our use cases) it is possible to build a
structure on a shard using slightly more space than strictly necessary, we can
size all structures on the largest shard, knowing that with probability at least
$1-\theta$ we will not increase the space usage by more than a factor of
$1+\epsilon$, and dispense completely with the offset array and the associated
memory dependency.\footnote{We remark that a similar observation has been made by Ragnar
	Groot Koerkamp in~\cite[Section~3.1]{RagPTR}, when discussing partitioning
	techniques for the construction of minimal perfect hash functions, albeit
	without obtaining~\eqref{eq:ballsbins}.}

As for the local seed, we can again act on the shard size so that, with good
probability, all shards can be built using the same seed. This requirement does
not conflict with the previous one, as larger shards mean an even smaller space
overhead. A failure simply means that the whole construction has to be
restarted.

Depending on the type of construction at hand, one might have analytical bounds
on the probability of success. Otherwise, one can resort to
simulation to estimate the shard size that yields a good probability of
success. In any case, if no other bound on the shard size applies, one can in
general limit the number of shards to a few thousand, as after that
point the additional memory used at construction time will be negligible with
respect to the size of the data structure.

\begin{proposition}
	\label{prop:ballsbins}
	Let $M$ be the maximum number of balls in a bin when throwing $n$ balls
	independently and uniformly at random into $S$ bins, and let $\epsilon>0$
	and $0<\theta<1$. If
	\begin{equation}
		\label{eq:chernoff}
		S\ln\frac S\theta \leq \frac{n\epsilon^2}{2+\epsilon},
	\end{equation}
	then $M<(1+\epsilon)n/S$ with probability greater than $1-\theta$.
\end{proposition}

\begin{proof}
	The number $X$ of balls in a given bin has a binomial distribution with mean
	$\mu=n/S$. By the multiplicative Chernoff bound~\cite[Theorem~4.4]{MiUPC},
	for every $\delta>0$
	\[
		\Prob\bigl[X\geq(1+\delta)\mu\bigr] <
		\biggl(\frac{e^\delta}{(1+\delta)^{1+\delta}}\biggr)^\mu =
		e^{-\mu((1+\delta)\ln(1+\delta)-\delta)} \leq
		e^{-\frac{\delta^2\mu}{2+\delta}},
	\]
	where the last inequality follows from
	$\ln(1+\delta)\geq2\delta/(2+\delta)$, which holds for $\delta\geq0$ because
	the two sides are equal for $\delta=0$ and the derivative of their
	difference is $\delta^2/\bigl((1+\delta)(2+\delta)^2\bigr)\geq0$. By the
	union bound over the $S$ bins,
	\[
		\Prob\Bigl[M\geq(1+\epsilon)\frac nS\Bigr] <
		S e^{-\frac{\epsilon^2n}{(2+\epsilon)S}},
	\]
	and the right-hand side is at most $\theta$ if and only if
	$\ln(S/\theta)\leq\epsilon^2n/\bigl((2+\epsilon)S\bigr)$, that is, if and
	only if~\eqref{eq:chernoff} holds.
\end{proof}

Setting $y=S/\theta$, \eqref{eq:chernoff} becomes $y\ln y\leq
	n\epsilon^2/\bigl((2+\epsilon)\theta\bigr)$, that is, $\ln y\leq
	W\bigl(n\epsilon^2/\bigl((2+\epsilon)\theta\bigr)\bigr)$, where $W$ is
the Lambert $W$ function. Recalling that $e^{W(x)}=x/W(x)$, we obtain
\begin{equation}
	\label{eq:ballsbins}
	S\leq\frac{n\epsilon^2}{(2+\epsilon)\,W\Bigl(\frac{n\epsilon^2}{(2+\epsilon)\theta}\Bigr)}.
\end{equation}
This bound limits the number of shards we can use while keeping, with
probability at least $1-\theta$, the ratio between the maximum and the mean
number of balls in a bin below $1+\epsilon$.

The bound is essentially tight: from the results of Raab and
Steger~\cite{RaLBIB} it follows that in our range of parameters, with high
probability, $M\geq n/S+(1-O(\epsilon))\sqrt{2(n/S)\ln S}$, so the ratio between
the maximum and the mean is at most $1+\epsilon$ with non-vanishing probability
only if $S\ln S\leq(1+O(\epsilon))n\epsilon^2/2$. With respect to this limit,
\eqref{eq:chernoff} pays only the additive term $\ln(1/\theta)$ beside $\ln S$.

If we are interested only in power-of-two sharding, we use $h$ high bits when
defining shards, where $h$ is the largest integer such that $S=2^h$
satisfies~\eqref{eq:chernoff}, that is,
\begin{equation}
	\label{eq:logballsbins}
	2^h\bigl(h\ln2+\ln(1/\theta)\bigr)\leq\frac{n\epsilon^2}{2+\epsilon}
\end{equation}
($h=0$ if no positive integer satisfies the inequality, as a single bin has no
excess). Since the left-hand side is increasing in $h$, a linear search computes
$h$ exactly; by the approximation $W(x)\approx\ln x-\ln\ln x$, $h=\lg n-O(\lg\lg
	n)$ for fixed $\epsilon$ and $\theta$.

Our implementation uses $\theta=10\%$, a larger budget than the one we will use
for duplicate edges, because exceeding the target overhead does not require
restarting the construction, but just costs slightly more space; moreover, the
probability of exceeding the target decreases very quickly with the size of the
excess, as the exponent of the bound grows quadratically in the excess. After
sharding, our implementation compares the size $M$ of the largest shard with the
mean $n/S$, and restarts the construction with a different seed if
\[
	M > \frac nS + \max\biggl( 5\epsilon\frac nS,\ 2.5\sqrt{\frac nS\ln S}\biggr),
\]
that is, if the realized overhead exceeds five times the target, and the excess
of $M$ over the mean exceeds $2.5/\sqrt2\approx1.77$ times its typical value
$\sqrt{2(n/S)\ln S}$. When the number of shards satisfies~\eqref{eq:chernoff}
and $\epsilon\leq1$, the first condition has probability less than $\theta^{10}$
(that is, $10^{-10}$ in our implementation). Indeed, by the argument of
Proposition~\ref{prop:ballsbins} with $\delta=5\epsilon$, the probability that
$M\geq(1+5\epsilon)n/S$ is less than $Se^{-25\epsilon^2n/((2+5\epsilon)S)}$, and
\[
	\frac{25\epsilon^2n}{(2+5\epsilon)S}\geq\frac{10\epsilon^2n}{(2+\epsilon)S}\geq10\ln\frac S\theta,
\]
where the first inequality holds because $25(2+\epsilon)\geq10(2+5\epsilon)$
when $\epsilon\leq1$, and the second one follows from~\eqref{eq:chernoff};
hence, the probability is less than $S(\theta/S)^{10}\leq\theta^{10}$. Thus, the
check is just a safety net: the realized space overhead is checked at
construction time, rather than assumed.

Note that, as $W$ grows very slowly, \eqref{eq:ballsbins} says essentially that
to get a space cost of $\epsilon$ each shard should contain a small multiple of
$2/\epsilon^2$ keys: for example, with $\epsilon=10^{-3}$ and $\theta=10\%$,
from a few million to a few tens of millions of keys.

We now apply this idea to two hypergraph-based constructions.

\subsection{MWHC}
\label{sec:mwhc}

We first apply $\epsilon$-cost sharding to the MWHC construction~\cite{MWHFPHM},
using $3$-hypergraphs. The asymptotic behavior of random $3$-hypergraphs kicks in
very quickly, and graphs are peelable with overwhelming probability even for
relatively small sizes.

While the MWHC construction has been mostly superseded by fuse
graphs~\cite{DiWDPRUH}, due to their significantly lower space overhead, it is an
interesting term of comparison because, as we will see, fuse graphs pose some
obstacles to $\epsilon$-cost sharding that random $3$-hypergraphs do not. In
this case we set $\epsilon=0.01$, which, as we will see, leads to very fine
sharding and to a very fast parallel construction.

Random $3$-hypergraphs with more than about $10^5$ edges almost never exhibit a
nontrivial $2$-core, so when we cannot peel such a hypergraph the reason is
almost always a duplicate edge. This phenomenon reflects the difference
between the model in which the peelability thresholds are proven (all
hypergraphs without duplicate edges with a certain ratio of edges to vertices
are equiprobable) and the model obtained by the random process generating the
hypergraphs in the data structures we use (edges are generated at random with
replacement, possibly yielding hypergraphs with duplicate edges).

We can however compute analytically the maximum number of shards for which
the probability of a duplicate edge is below a certain threshold, as such
probabilities are well known. If we consider a sequence of random uniform values
$x_0$,~$x_1$,~$x_2$, $\dots$\, out of $N$, and we look for
the first repeated value, that
is, for the least $m$ such that $x_m=x_i$ for some $i<m$, the probability
mass function of $m$ for a uniform random source with $N$ possible outputs is
\[
	\Prob_N(m) = \frac{m}N\prod_{i=0}^{m-1} \biggl( 1 - \frac iN\biggr).
\]
The cumulative distribution function is even simpler:
\[
	F_N(m) =\sum_{i=0}^m \Prob_N(i) = 1 - \prod_{i=0}^{m} \biggl( 1 - \frac
	iN\biggr).
\]
Both expressions can be bounded easily as follows:
\begin{equation}
	\label{eq:approx}
	\prod_{i=0}^{m-1} \biggl( 1 - \frac iN\biggr)=
	e^{\sum_{i=0}^{m-1}\ln(1-i/N)}\leq e^{\sum_{i=0}^{m-1}-i/N}=
	e^{-m(m-1)/2N}.
\end{equation}
In fact, when $m^3\ll N^2$ the bound is a good approximation, so $e^{-m^2/2N}$
approximates the probability that there is no duplicate value after generating
uniformly $m$ values out of $N$.

We will use~\eqref{eq:approx} in the following sections to size shards: we will
impose that the right-hand side is at least $1-\eta$, and conclude that the
probability of a duplicate is at most $\eta$. Since~\eqref{eq:approx} is an
\emph{upper} bound, this step is not conservative, and it needs the matching
lower bound. Using $\ln(1-x)\geq -x-x^2$, which holds for $0\leq x\leq 1/2$, we
obtain for $m\leq N/2$
\begin{equation}
	\label{eq:approxlow}
	\prod_{i=0}^{m-1} \biggl( 1 - \frac iN\biggr)\geq
	e^{-m(m-1)/2N}\,e^{-m^3/3N^2}.
\end{equation}
The two bounds thus differ by a factor of at most $e^{m^3/3N^2}$. Moreover, if
$m^2\leq 3N/2$ we have $m^3/3N^2\leq m/2N$, so the right-hand side
of~\eqref{eq:approxlow} is at least $e^{-m^2/2N}$: replacing $m(m-1)$ with $m^2$
in~\eqref{eq:approx} yields a \emph{lower} bound, and imposing $e^{-m^2/2N}\geq
	1-\eta$ guarantees that the probability of a duplicate is at most $\eta$.
The condition $m^2\leq 3N/2$ is equivalent to $e^{-m^2/2N}\geq e^{-3/4}$, that
is, it holds as long as the estimated probability of a duplicate is at most
$1-e^{-3/4}\approx 53$\%. In our setting $m$ is the number of keys in a shard
and $N$ the number of possible edges, and both conditions hold in every
configuration we use. When we use $e^{-m(m-1)/2N}$ instead
(Section~\ref{sec:small}), we overestimate the probability that there is no
duplicate (and underestimate that of a duplicate) by at most the factor above: $m^3/N^2$ is at most $3.2\cdot10^{-2}$ (for two keys) and less than
$3\cdot10^{-6}$ from $10^5$ keys on, so the error is less than $1.1$\% and
quickly becomes negligible.

In our case, if $c$ is the expansion factor, $k$ the arity of the hypergraph,
and $m$ the number of keys of a shard, we have $(c m/k)^k$ possible
edges.\footnote{We use the $k$-partite construction~\cite{BWZCRRPH}.}
The probability that there are no duplicates for $k=3$ is thus
\[\exp\biggl(-\frac{m^2}{2(cm/3)^3}\biggr) = \exp\biggl(-\frac{1}{2(c/3)^3 m}\biggr).\]
When sharding, the $n$ keys will be divided into $S$ shards, each of which will
have a size very close to $n/S$. Since we need to bound the probability at the
same time for all shards, we want to impose that for some $\eta < 1$, where
$\eta$ is a budget on the probability of having to retry the construction
because of a duplicate edge (a parameter distinct from the space parameter
$\epsilon$; our implementation uses by default $\eta=5\%$, see
Section~\ref{sec:subdup}),
\[
	\exp\left(\frac{-1}{2(c/3)^3( n/S)}\right)^{S}
	\geq 1 - \eta.
\]
Taking logarithms,
\[
	S \cdot \frac{-1}{2(c/3)^3( n/S)} \geq \ln (1-\eta),
\]
that is,
\begin{equation}
	\label{eq:dupmwhc}
	S\leq \sqrt{-2n(c/3)^3 \ln (1-\eta)},
\end{equation}
which, if we assume $S=2^h$, becomes
\[
	h \leq \frac 12\left(\lg n + 1 + 3\lg (c/3) + \lg(-\ln(1-\eta))\right)
	\approx \frac 12\left(\lg n + 1 + 3\lg (c /3) + \lg\eta\right).
\]
For example, even with a strict $\eta=10^{-3}$ we can divide $10^{12}$ keys
into $8192$ shards with $c=1.23$. Our implementation enforces this
constraint, combined with~\eqref{eq:logballsbins}, when
sharding.

\subsection{Fuse Graphs}
\label{sec:fuse}

Let us first quickly recall the definition of a fuse graph of arity $k$~\cite{DiWDPRUH}: given a
set of $n$ keys, we select an expansion factor $c$ and a number of segments
$\ell$. Then, we divide $\lceil cn\rceil$ vertices into $\ell + k - 1$ segments of the same size
(this might require some rounding); finally, we choose a random edge by
choosing a random segment $j\in [0\..\ell)$ and a random vertex in each of the
segments $j$, $j+1$, \ldots, $j+k-1$. We will focus on the case $k=3$.

The rounding has a subtle effect on peelability. The first vertex of an edge
lies in the first $\ell$ segments, and the last $k-1$ segments are only
partially loaded, so peelability depends essentially on the density $\ell
	s/n$ of the first $\ell$ segments, where $s$ is the segment size.
Implementations round $\lceil cn\rceil$ up to a multiple of $s$ and then
subtract $k-1$ segments, so for $k=3$ this density lies in
$[c-2s/n\..c-s/n)$. The difference is immaterial for large graphs, but on
graphs with a few dozen segments it can bring the density well below the
threshold. Our
implementation thus increases $\ell$, when necessary, to the smallest value
whose density reaches an empirical threshold law, fitted on simulations, at
which about $70$\% of the attempts succeed. The law depends on $\ell$ (the
partially loaded final segments make short graphs easier to peel) and on $s$
(the load of smaller segments fluctuates more). The adjustment is necessary
only up to $10^5$ keys, where it increases space by less than $0.5$\% on
average.

For large $n$ and suitable $c$ and $\ell$, fuse graphs are almost always
peelable. The first problem is that the asymptotic
behavior kicks in fairly slowly---one needs a few million vertices. Moreover, while
empirically large graphs with, say, $k=3$ are peelable with $c\approx 1.105$,
the exact values for $\ell$ are not known analytically. Walzer's analysis of
\emph{spatial coupling}~\cite{WalPCOT} explains why fuse graphs peel so close to
the orientability threshold, but it does not provide the finite-size values of
$c$ and $\ell$ that an implementation needs. The second
problem is that for small $n$ the expansion factor becomes significantly
larger, reaching that of general random uniform hypergraphs.

Graf and Lemire~\cite{GrLBFF} propose a highly optimized implementation of fuse
graphs for filters that we use as the basis of our own implementation. In
particular, they provide empirical estimates for $c$ and $\ell$ when $k=3$ or
$k=4$ that we partially use. We depart from their estimates in two cases:
\begin{itemize}
	\item For small sizes (up to $2\cdot 10^7$ keys) we use up to $16$ shards
	      solved by augmented peeling, which we describe in
	      Section~\ref{sec:small}; the expansion factor and the segment size are
	      chosen using a law for the threshold of solvability.
	\item When sharding large key sets, we use $c=1.105$ and compute the base-2
	      logarithm of the segment size using the estimate
	      \begin{equation}
		      \label{eq:nln}
		      0.41 \cdot \ln n \ln \ln n - 3
	      \end{equation}
	      which is $O(\log n \log \log n)$, rather than the estimate from~\cite{GrLBFF},
	      \begin{equation}
		      \label{eq:ln}
		      \log_{3.33} n + 2.25,
	      \end{equation}
	      which is $O(\log n)$ (for shards of at most $2\cdot 10^7$ keys we use
	      instead $c=1.11$ and~\eqref{eq:ln}). Larger segments reduce the
	      locality of the construction, and thus slow down sequential
	      construction, but, as we will discuss now, make much finer sharding
	      possible when signatures are not subsigned.
\end{itemize}

Even if we lack precise probability estimates, with
$c=1.105$ fuse graphs almost never exhibit a nontrivial $2$-core for large $n$
(we never create peelable shards smaller than $10$ million keys). Simulating the
sparsest graphs that the rounding of $\ell$ produces at each segment size, and
composing the results over the actual sizes of the shards, the probability
that a construction attempt fails because of a nontrivial $2$-core is at most
$0.3$\% (for segments of $2^{17}$ vertices) and negligible for larger
segments, well within the retry budget $\eta=5$\%. When we cannot complete
a peeling, once again the reason is a duplicate edge. Duplicate edges are much
more likely than in the previous case because the choice of vertices is limited
to the ones in certain segments, and the problem is only exacerbated by
$\epsilon$-cost sharding, as we are building more structures with the same seed,
using smaller segments.

Using the same framework as in the previous section, if $c$ is the expansion
factor, $k$ the arity of the hypergraph, $m$ the number of keys of a shard, and
$s_c$ the function (parameterized by the expansion factor) returning the size of
a segment given the number of keys, we have $(cm/{s_c(m)} - k +
	1)s_c(m)^k\approx cm\cdot s_c( m)^{k-1}$ possible edges. In our case the
probability that there are no duplicates is thus approximately
\[\exp\biggl(-\frac{m^2}{2cm\cdot s_c(m)^2}\biggr) =
	\exp\biggl(-\frac{m}{2c\cdot s_c(m)^2}\biggr).\] As in the previous section, when
sharding, the $n$ keys will be divided into $S$ shards, each of which will have a
size very close to $n/S$. Since we need to bound the probability at the same
time for all shards, we require that
\[
	\exp\left(\frac{-n/S}{2c\cdot s_c\bigl(n / S\bigr)^2}\right)^{S}
	\geq 1 - \eta.
\]
Taking logarithms,
\[
	S \cdot \frac{-n/S}{2c\cdot s_c\bigl(n / S\bigr)^2} \geq \ln (1-\eta),
\]
that is,
\[
	s_c\bigl(n/S\bigr)\geq \sqrt{-\frac{n}{2c\ln (1-\eta)}}.
\]
We take a base-2 logarithm on both sides and replace $\lg\bigl(s_c(n/S)\bigr)$
with~\eqref{eq:nln}, obtaining
\[
	0.41 \ln\Bigl(\frac n{S}\Bigr)\ln\ln\Bigl(\frac n{S}\Bigr) - 3
	\geq \frac12\lg \frac{-n}{2c\ln (1-\eta)},
\]
and then
\[
	\ln\ln\Bigl(\frac nS\Bigr) \geq W
	\biggl( \frac1{0.82} \Bigl(\lg\frac{-n}{2c\ln (1-\eta)} + 6\Bigr)\biggr)
\]
where $W$ is the Lambert function. Solving for $S$, and recalling that $e^{W(x)} =
	x/ W(x)$, we finally obtain
\begin{equation}
	\label{eq:dupfuse}
	S \leq
	n\cdot \exp \left(-\frac{ \frac1{0.82} \Bigl(\lg\frac{-n}{2c\ln (1-\eta)} + 6\Bigr)}{W \left( \frac1{0.82} \Bigl(\lg\frac{-n}{2c\ln (1-\eta)} + 6\Bigr)\right)}\right),
\end{equation}
which, if we assume $S=2^h$, becomes
\begin{equation}
	\label{eq:logdupfuse}
	h\leq \lg n -
	\frac{ \frac1{0.82} \Bigl(\lg\frac{-n}{2c\ln (1-\eta)} + 6\Bigr)}{\ln 2 \cdot W \left( \frac1{0.82} \Bigl(\lg\frac{-n}{2c\ln (1-\eta)} + 6\Bigr)\right)}.
\end{equation}
For example, we can divide $10^{12}$ keys into $256$ shards with
$\eta=10^{-3}$; with the budget $\eta=5\%$ used by our implementation
the closed form gives $2048$ shards, and the exact evaluation
performed by the implementation, which accounts for the rounding of
segment sizes, gives $1024$. This bound, however, turns out to be very
pessimistic: as we will see in Section~\ref{sec:subdup}, the
interaction between subsigning and the generation of edges makes
duplicate edges impossible at this scale, and we will be able to use
up to $8192$ shards.

The right-hand side of~\eqref{eq:logdupfuse} is of the form $\lg n - O( \lg n /
	\lg \lg n)$, whereas the largest $h$ satisfying~\eqref{eq:logballsbins} is
of the form $\lg n - O(\lg \lg n)$, so, in the long run, the former will be the
dominating bound---at least when edges are generated using full signatures: as
we will see in Section~\ref{sec:subdup}, subsigning changes the picture
radically. Note that in this case Lambert's function must be computed
accurately, rather than using approximations such as $W(x)\approx \ln x - \ln
	\ln x$. In practice, with $\epsilon=10^{-3}$ the
bound~\eqref{eq:logballsbins} never binds for fuse graphs: up to $2\cdot10^7$
keys the number of shards is given by a fixed rule (Section~\ref{sec:small}),
and beyond that
point it is determined by the minimum shard size of $10^7$ keys and by duplicate
edges, which in the injective regime of Section~\ref{sec:subdup} amounts to
keeping shards above ${\approx}10^8$ keys; the space cost of sharding is thus
below $\epsilon$.

We remark that with full signatures using the estimate~\eqref{eq:ln} instead
of~\eqref{eq:nln} would lead to little sharding, even for trillions of keys.

\subsection{Peel by Signature, Peel by Index}
\label{sec:peelsig}

An important issue when peeling hypergraphs, albeit often overlooked in the
literature, is memory usage. Since one of the
purposes of sharding is to parallelize peelings, memory usage becomes even more
important.

Let us briefly recall how the peeling visit works: one first chooses a
supporting data structure $Q$, which can be a queue or a stack; we will also need
an edge stack $P$ where we accumulate peeled edges.

Then, we (almost) follow the standard visit algorithm: we iterate over the
vertices, and when we find a vertex of degree one we enqueue it in $Q$. We then
dequeue vertices from $Q$ until it is empty, and for each dequeued vertex of
degree one we remove the only incident edge, pushing it on $P$, updating the
incidence lists of the other vertices in the edge (if the vertex has degree
zero, we discard it). Any of these vertices that now has degree one
is enqueued in $Q$; this maintains the invariant that each vertex in $Q$
has degree at most one. Once $Q$ is empty, we continue to iterate over the
vertices.

At the end of the visit, the peeling is successful if $P$ contains all
edges. At that point, edges can be popped from $P$, and for each edge we can
assign a value to the vertex from which the edge was peeled. Thus, this vertex
must also be stored in $P$, but it is sufficient to store its index in the edge,
which we call the \emph{side} of the vertex.\footnote{Technically, edges are sets, but in practice they
	are represented as a sequence of vertices, so sides are well defined.}

There are many variants that can be faster on some architectures: for
example, one can preload $Q$ with all vertices of degree one. Note that, as in
the standard case, the stack-based visit is not a depth-first visit: a
depth-first visit would also require pushing on the stack the state of the
iteration on the other vertices. In any case, however, each vertex is
enqueued in $Q$ at most once.

Whereas the random hypergraphs of the MWHC construction have very bad locality
(edges are, indeed, almost entirely random), fuse graphs have edges whose
vertices are relatively near, as they lie in consecutive segments: in fact, the
smaller the segments, the higher the locality.

The choice of visit for the peeling process can make a large difference:
stack-based visits are known, in general, to have a higher locality than
queue-based visits, and this difference is extremely pronounced in the case of
fuse graphs, as noted by Graf and Lemire~\cite{GrLBFF}: not only is the peeling
visit much faster using a stack, but the subsequent assignment of values also
becomes faster, because in $P$ nearby edges often belong to the same segment.

All current constructions for hypergraph-based structures start by hashing the
keys to unique signatures (usually, to $64$ bits for small sizes, $128$ for large
sizes) and then work on the signatures. This approach avoids costly recalculations
each time we have to turn a key into an edge, because we can simply generate the
edge from the signature. It also combines well with $\epsilon$-cost sharding, because
we can reuse the sharding signatures, taking care not to use the sharding
bits, as they will be the same for all signatures in the same shard.

If we need to represent a static function, along with each signature we
also have to keep track of the associated value; in the case of filters, we can
compute hashes of the desired size directly from the signature~\cite{Sux4J31}. One
first has to pass through all the signatures to generate the edges and store them in
incidence lists, one for each vertex, and again in this phase fuse graphs help, as
count sorting the signatures by the segment of their associated edge increases the
locality of the construction~\cite{GrLBFF}.

A significant reduction of the space used for incidence lists can be obtained
using \emph{Djamal's XOR trick}~\cite{BelXOR,BBOCOPRH}: rather than storing an edge list per
vertex, it is sufficient to XOR together the edges in the list, because we
follow an edge during the visit only if the vertex we discover is of degree one,
and thus the XOR contains just the edge we are interested in.

There are at this point two main strategies:
\begin{itemize}
	\item Peel-by-signature: incidence lists are stored as XOR of signatures (and
	      possibly associated values). This is, for example, the strategy used
	      in~\cite{GrLBFF}.
	\item Peel-by-index: incidence lists are stored as XOR of indices into the list
	      of signatures (and possibly values).
\end{itemize}
In the peel-by-signature case, if the goal is to peel the graph completely,
it is possible to drop the signature/value pairs and just keep the graph
representation. Thus the memory usage is reduced to a signature/value pair per
vertex. In the peel-by-index case, one has to keep the signature/value
pairs to retrieve an edge given its index, so the memory usage is a
signature/value pair per key plus an index per vertex, but after a failed peeling
it is possible to generate and solve a system of equations taking care of the
$2$-core, that is, the unpeeled edges. Some small additional space is needed to
store degrees and sides, but it is common to both representations.

The other important difference is locality. In the peel-by-signature case, during
the visit we have direct access to signatures, from which we can generate
edges associated with keys, whereas in the peel-by-index case we have a
costly memory indirection from an edge index to a signature.

A further choice concerns the representation of the stack of peeled edges. One
possibility is, once again, to store signature/value pairs, which provide
high locality in the assignment phase~\cite{GrLBFF}. The memory cost of the visit is then an
index per vertex for the visit stack, plus a signature/value pair for each edge.
We call this strategy the \emph{high-memory visit}, as it more than doubles
the memory usage with respect to the graph representation.

As noted in~\cite{GOVFSCCSMPHF}, however, if one takes care of leaving in the
incidence list of a vertex the edge that was peeled from that vertex (it is
sufficient to virtually zero the degree for the peeling algorithm to work), $P$ can
accumulate vertex indices, rather than edges, as the edges (in whichever form)
can be retrieved from incidence lists. At that point, for queue-based visits $P$
and $Q$ can be stored one after another in the same vector, because for each
vertex dequeued from $Q$ there is at most a vertex pushed on $P$.

We note that this consideration can be extended to stack-based visits, by using
a single vector to store both $Q$ and $P$: one stack grows upwards, the other
downwards. The visit and assignment phases can then be performed using an index
per vertex, at the cost of a slightly slower assignment phase: we call this
strategy the \emph{low-memory visit}. In the case of static functions using
$128$-bit signatures, the low-memory visit uses about a sixth of the memory of
the high-memory visit.

Depending on the hardware and on the size of the key set, sometimes the
low-memory visit is much faster than the high-memory visit, in spite of the
lesser locality. While we prefer the high-memory visit for low parallelism, when
significant parallelism is possible we switch to the low-memory visit, which
avoids out-of-memory errors due to excessive allocation, and in general
significantly reduces the memory necessary for the construction, improving the
overall performance. When we use augmented peeling (Section~\ref{sec:small}), we
use the high-memory visit, and we peel by signature, storing in the graph also
the index of each edge, which is needed to perform removals.

\subsection{Small Shards and Augmented Peeling}
\label{sec:small}

Fuse graphs approach their asymptotic behavior slowly: below a few dozen million
keys the expansion factor necessary for peelability increases, and since we
never create peelable shards smaller than $10^7$ keys, construction below $2\cdot
	10^7$ keys would be sequential. For small key sets we thus solve the shards by
\emph{augmented peeling}, a technique proposed by Stefan Walzer, Stefan Hermann,
and Markus Wünstel,\footnote{Personal communication, July 2026. Augmented
	peeling is the subject of the master's thesis of Markus
	Wünstel~\cite{WunAPAISESSCRDS}, advised by Stefan Walzer and Stefan
	Hermann.} which combines the speed of peeling with the threshold of
solvability of linear systems on $\F_2$.

In their proposal, the segments of the fuse graph are glued into a cycle, so
that no segment is lightly loaded, and the graph is peeled; whenever the visit
gets stuck, that is, there are edges left but no vertex of degree one, an edge
is \emph{removed}, and the visit resumes. The removed edge is the leftmost
remaining one, that is, the one with the smallest starting position: in this
way, removals rip a hole in the graph, and peeling proceeds from there as from
the lightly loaded end of a fuse graph.

At the end of the visit, every vertex is
of one of three kinds: a \emph{pivot}, that is, a vertex from which an edge has
been peeled; an \emph{irrelevant} vertex, which belongs to no edge; or a
\emph{lost} vertex, which belongs to some edge but is not a pivot. Since there
is a pivot for each peeled edge, if $m'$ of the $m$ edges have been removed and
$N$ vertices are relevant, there are exactly $m'+N-m$ lost vertices. The values
of the lost vertices can be chosen freely, and they determine, by back
substitution in reverse peeling order, the values of the pivots; each removed
edge thus becomes a linear equation in the values of the lost vertices, and the
system associated with the graph is solvable if and only if this small system is
solvable. The proposal uses as unknowns the last $1.2m'$ lost vertices, sets the
remaining ones to zero, and solves the resulting dense system by Gaussian
elimination.

Augmented peeling is a transposed form of the \emph{inactivation decoding} of
Raptor codes~\cite{ShLRC}, which, like the lazy Gaussian elimination of
GOV~\cite{GOVFSCCSMPHF}, was inspired by the structured Gaussian elimination of
LaMacchia and Odlyzko~\cite{LaOSLSLSFF}. Decoding an erasure code amounts to solving a linear
system by repeatedly using an equation with a single unknown; when no such
equation is left, an unknown is \emph{inactivated}, that is, set aside as a
symbolic value, and the inactivated unknowns are finally computed by Gaussian
elimination on a small dense system. If we transpose the system of a shard, so
that vertices become equations and edges become unknowns, peeling a vertex of
degree one becomes solving an equation with a single unknown, removing an edge
becomes inactivating an unknown, and the dense system of augmented peeling
becomes the transpose of that of inactivation decoding. Lazy Gaussian
elimination applies instead the same scheme to the original
system.\footnote{A previous design of \VFunc{} used lazy Gaussian elimination
instead of augmented peeling. Up to the limit on the number of unknowns
described in Section~\ref{sec:small}, both succeed on exactly the same systems, but
since every equation of the original system initially contains three unknowns,
lazy Gaussian elimination must set aside some unknowns before it can solve any
equation, so it turns out to be significantly slower.}

We use an approximation of the same removal rule, which in our case is
particularly simple: since shards are sorted by the segment of the first vertex
of their edges, the first remaining edge of a shard has its first vertex in the
leftmost segment still containing edges, so a pointer to it suffices. Our
strategy differs, however, in two respects. First, we do not glue the segments
into a cycle: a cyclic graph has no lightly loaded end, so it cannot be peeled
without removals, and on the large shards of the previous sections augmented
peeling would be too slow. Gluing could be made a parameter of the format, but
then every query would have to reduce vertex indices modulo the length of the
cycle, even on large shards, which cannot be glued. To keep queries unchanged,
and the format of our structures independent of the solving strategy, we adapt
augmented peeling to the same fuse graphs we use for large shards; as a
consequence, the lightly loaded final segments still cost some space (about
$1.2$ segments, as we will see).

Second, in our adapted layout a fixed number of unknowns is not sufficient, so
we choose the unknowns adaptively. To this end, we note that not all lost
vertices are useful:

\begin{proposition}
	\label{prop:lost}
	The values associated with the vertices of the removed edges, expressed as
	functions of the values of the lost vertices, depend only on the vertices
	lost after the first removal.
\end{proposition}

\begin{proof}
	Let $e_0$,~$e_1$,~$\dots\,$ be the peeled edges in peeling order, $v_i$ the
	pivot of $e_i$, and $t$ the number of edges peeled before the first removal.
	We show first that the value of $v_i$, for $i\geq t$, depends only on
	vertices lost after the first removal. When $e_i$ is peeled the first
	removal has already happened, and the other vertices of $e_i$ are either
	pivots of edges peeled later, or vertices that will be lost; in the latter
	case, since they belong to $e_i$, they become lost after the first removal.
	As the value of $v_i$ is the XOR of the value associated with $e_i$ and of the
	values of the other vertices of $e_i$, the claim follows by induction in
	reverse peeling order.

	Consider now a removed edge $r$ and one of its vertices $v$. If $v$ is the
	pivot of $e_i$, then $i\geq t$: when an edge $e_i$ with $i<t$ is peeled, $r$
	is still in the graph, so if $v_i$ belonged to $r$ its degree would be at
	least two. If $v$ is lost, it becomes lost after the removal of $r$, and thus
	after the first removal. In both cases, the value of $v$ depends only on
	vertices lost after the first removal.
\end{proof}

We can thus fix arbitrarily the values of the vertices lost before the first
removal (we keep their initial values, which are zero, or random in the case of
filters), and solve the removed equations using the remaining lost vertices as
unknowns. We first assign values to the pivots using the initial values of all
lost vertices, which gives the constant terms of the removed equations. Then, we
associate a unit bit vector with each unknown, and we propagate the vectors in
reverse peeling order along the edges peeled after the first removal, computing
the vector of each pivot as the XOR of the vectors of the other two vertices of
its edge; the XOR of the vectors of the vertices of a removed edge is the row of
the corresponding equation. Since peeling advances through a fuse graph as a
wave, the other two vertices of an edge are usually pivots of edges peeled
shortly afterwards: by storing the vectors of the pivots in reverse peeling
order, rather than by vertex, we make the propagation cache-friendly. We solve
the resulting dense $m'\times u$ system by
Gaussian elimination, XOR the solution into the values of the unknowns, and
perform a final assignment.

Without a bound on the number $u$ of unknowns, this procedure fails only if the
system is unsolvable, but we do not need all of them. We take as unknowns the
most recently lost vertices, in blocks of $256$ (the propagation is
bit-parallel): we start with $m'+32$ unknowns, rounded up to a multiple of
$256$, and double their number while the restricted system is unsolvable, up to
$4m'+2048$, so that attempts on unsolvable systems fail fast. In practice, the
restricted system is almost always solvable with just a few more than $m'$
unknowns, but the number of unknowns needed has a heavy tail, so a fixed number
of unknowns would make a significant fraction of solvable systems fail. The
propagation and the elimination cost $O(mu/w)$ and $O(m'^2u/w)$ operations on
$w$-bit words, where $m'$ is at most $6$\% of the edges (less than $0.5$\% for long
graphs with large segments).

\paragraph{The solvability threshold.} To size the graphs we need to know down
to which density augmented peeling succeeds, that is, the threshold of
solvability. When peeling gets stuck, the remaining edges form the
\emph{$2$-core} of the graph, its largest subgraph in which every vertex has
degree at least two: everything outside the core can be peeled, and nothing
inside it can, so the system of the graph is solvable if and only if the system
of its core is. Peeling thus succeeds only if the core is empty, whereas a core
with more edges than vertices, that is, with more equations than unknowns,
makes the system unsolvable for some values; for random
$3$-XORSAT~\cite{DuM3XT,PiSSTkX} the converse holds too, and we assume that it
holds for fuse graphs (our simulations confirm the assumption). Both thresholds
depend thus only on the size of the core.

When segments are large, the neighborhood of a vertex looks like a tree, and
the size of the core can be computed recursively, a technique known as
\emph{density evolution}~\cite{MolCRYBF}, which is also behind the analysis of
the threshold of peelability of fuse graphs~\cite{WalPCOT}: an edge belongs to
the core if each of its vertices belongs to some other edge of the core, and a
vertex belongs to the core if it belongs to at least two edges of the core. In
a fuse graph these probabilities depend only on the segment.

Consider a fuse
graph with $m$ edges whose first vertices lie in $\ell$ segments of size $s$,
and let $x = \ell s/m$ be the density of these segments. As $s\to\infty$, a
vertex of segment $i\in[0\..\ell+2)$ is the first vertex of a Poisson number of
edges with mean $1/x$ if $i<\ell$, and the same holds for the second vertex if
$1\leq i\leq \ell$, and for the third if $i\geq 2$. Let $y_i$ be the probability
that a vertex of segment $i$, reached through one of its edges, belongs to some
other edge of the core: the expected number $\Lambda_i$ of other edges of the
core containing the vertex accounts for the three positions it can occupy, and
$y$ is the largest fixed point of
\[
	y_i = 1 - e^{-\Lambda_i},\qquad \Lambda_i = \frac1x\bigl([i<\ell]\,y_{i+1}y_{i+2} +
	[1\leq i\leq \ell]\,y_{i-1}y_{i+1} + [i\geq 2]\,y_{i-2}y_{i-1}\bigr),
\]
where $y_i=0$ outside $[0\..\ell+2)$. The core has \[s\sum_i
	\Prob[\mathrm{Poisson}(\Lambda_i)\geq 2]\] vertices and \[(m/\ell)\sum_{j<\ell}
	y_jy_{j+1}y_{j+2}\] edges: the threshold of peelability is the largest $x$ for
which the core is nonempty, and the threshold of solvability the largest $x$
for which it has more edges than vertices.

We write the threshold of solvability as $c_3\ell/(\ell+\delta(\ell))$, where
$c_3 = 1/0.917935\ldots \approx 1.08940$ is the inverse of the threshold of
random $3$-XORSAT: for $\ell=1$ the graph is a random $3$-partite hypergraph,
whence $\delta(1)=2$. Solving the equations above numerically, we find that
$\delta(\ell)$ decreases to ${\approx}\,0.805$ as $\ell$ grows: the two
partially loaded final segments are worth about $0.8$ segments, so they cost
about $1.2$ segments of space. The threshold of peelability, instead, is within
$10^{-5}$ of $c_3$ for $\ell\geq 24$, as predicted by threshold
saturation~\cite{KRUTSSCWCLEPWB}, which Walzer brought to hashing-based data
structures~\cite{WalPCOT}; for large segments, thus, the advantage of linear
solving is only the boundary term $\delta(\ell)/\ell$. For finite $s$, however,
peeling is much more sensitive to the fluctuations of the load of segments: for
example, with $10^6$ keys and $s=2^{11}$ ($\ell\approx 535$) peeling failed in
all our trials at density $1.106$, whereas augmented peeling succeeded in almost
all trials down to density $1.094$.

The thresholds above are limits for $s\to\infty$. For a given $s$, we model the
probability that the system of a graph without duplicate edges is solvable as a
normal distribution function of the density $x$, with constants fitted by
simulation. Its mean is the threshold of solvability plus a correction of order
$1/s$, which accounts for the fluctuations of the load of the segments and grows
slowly with their number. Its variance is the sum of the finite-size window of
a system with $\ell s$ variables and of a term that decreases quickly with $s$.

\paragraph{Sizing and sharding.} Duplicate edges make a system unsolvable, and
by~\eqref{eq:approx} a graph has no duplicate edges with probability
approximately $\exp(-m(m-1)/2\ell s^3)$. For a shard of $m$ keys we thus choose
$s$ and $\ell$ so as to minimize the number of vertices $(\ell+2)s$, under the constraint that
the probability of success, duplicate edges included, is at least $p^{1/S}$,
where $S$ is the number of shards and $p=0.7$ is the target probability that a
construction attempt succeeds.

Duplicate edges also limit sharding, as in the case of peelable shards. Since
$\ell s\approx xm$, the expected number of duplicate edges over all shards is
$S\cdot m^2/2\ell s^3\approx n/2xs^2$, which does not depend on $S$: the segment
size must thus grow as $\sqrt n$, whatever the number of shards, and smaller
shards contain fewer segments, increasing the fraction
$(2-\delta(\ell))/(\ell+\delta(\ell))$ of space lost to the final segments. On
the other hand, the cost of the dense system grows quickly with the size of the
shard. As a compromise, up to $2\cdot 10^7$ keys we use $\min\bigl(8,
	2^{\lfloor\lg(n/200\,000)\rfloor}\bigr)$ shards (at least one), or $16$
shards if they contain at least $400\,000$ keys (above a few million keys the
segment size is at least $2^{12}$, so smaller shards would contain too few
segments).

Above $2\cdot 10^7$ keys peelable shards of at least $10^7$ keys are available,
and augmented peeling brings little: its shards would be either large, and slow
to solve, or short, and with a larger overhead.

\subsection{Subsigning}
\label{sec:subsign}

We propose a technique that we call \emph{subsigning} to speed up
construction, and, marginally, query time. The idea is to use just a part of
the bits of the signature, the \emph{local signature}, to generate edges (and
hashes for filters) for a shard, as the reduced number of keys in a shard makes
the probability of a collision much smaller. Most of the operations discussed in
Section~\ref{sec:peelsig} use signatures, so being able to reduce their size
means reducing memory usage and operations.

One has, however, to be careful that subsigning does not lead to unwanted
collisions of local signatures. Within a shard of $m$ keys, $t$-bit local
signatures are all distinct with probability at least $e^{-m^2/2^{t+1}}$ by the
argument following~\eqref{eq:approxlow}, so with $S$ shards of about $n/S$ keys
the probability of a duplicate local signature is at most about
$1-e^{-n^2/(S\,2^{t+1})}$, which decreases as the number of shards grows. For
example, if we use $128$-bit signatures and we subsign with $64$ bits, the
probability stays below $5$\% up to about two hundred billion keys with MWHC,
but only up to about ten billion keys with fuse graphs, which shard less. In
practice, one treats duplicate local signatures similarly to duplicate
signatures, with the difference that the user can circumvent the problem by
using more bits for subsigning. In the case of filters, however, as we will now
see, this is not necessary.

\subsection{Duplicate Signatures and Filters}

A further useful observation is that in the case of filters, if the hashes used by
the filter to check for false positives are derived from the sharding signature,
and not directly computed from the keys, as pioneered by the Sux4J
implementations~\cite{Sux4J31}, duplicate signatures can simply be removed. Keys
with the same signature are mapped to the same hash, so there will be no
difference in the behavior of the filter. In fact, we can even deduplicate local
signatures if they are used for generating both edges and hashes.

Sorting the signatures to find duplicates slightly slows down the construction,
but we can use one of the many recent excellent parallel radix sort
implementations~(see, e.g.,~\cite{OKFTEPPIPRS,AWFEIPSMSA}). If we also take care
to use the first bits after the sharding bits (if any) to compute the segment
for fuse graphs using a monotone function, 
then sorting for duplicates can play the same role as the count sort
by segment. In fact, if the following further bits are used to select the first
vertex, the radix sort provides even more locality to the generation phase,
recovering part of the overhead of radix sorting with respect to count
sorting. The strategy used by Graf and Lemire~\cite{GrLBFF} to generate an edge
indeed uses fixed-point arithmetic, thus satisfying this property,
although they do not exploit it in their implementation.

\subsection{Subsigning and Duplicate Edges}
\label{sec:subdup}

The analysis of duplicate edges we developed in the previous sections
makes the implicit assumption that edges are drawn uniformly and
independently at random among the possible edges. This assumption,
which is customary in the literature, is however false when
subsigning: in that case, the edge associated with a key is a deterministic
function of its $t$-bit local signature, and when the number of
possible edges is larger than $2^t$ the function simply cannot reach
all of them. As we will now see, this apparent defect is in fact an
advantage, as it makes duplicate edges \emph{less} likely than
predicted by the analysis, and, at large scale, impossible.

Let us describe the function in detail in the case $k=3$. Let $s$ be
the segment size, which we assume to be a power of $2$, and $\ell$
the number of possible first segments. Given a $t$-bit local
signature $x$, we compute
\[
	v_0 = \biggl\lfloor \frac{x\ell s}{2^t} \biggr\rfloor \qquad
	o_1 = x \bmod s \qquad
	o_2 = \lfloor x / s\rfloor \bmod s,
\]
where the first vertex $v_0$ is computed in fixed-point
arithmetic, and $o_1$, $o_2$ are offsets given by the two lowest blocks
of $\lg s$ bits of $x$; the remaining two vertices are obtained by
XORing the offset of $v_0+s$ with $o_1$, and the offset of $v_1+s$
with $o_2$.\footnote{This is the strategy used in our implementation;
	the strategy of Graf and Lemire~\cite{GrLBFF} is analogous, and the
	discussion applies with minimal changes.} Note that the edge and the
triple $\langle v_0, o_1, o_2\rangle$ determine each other, as the
offsets can be recovered by XOR.

Suppose now that two distinct local signatures $x$ and $x'$ generate
the same edge. Then, they must agree on the triple above: agreement
on $o_1$ and $o_2$ implies that $x$ and $x'$ agree on their $2\lg s$
lowest bits, so, as they are distinct, $|x - x'|\geq s^2$; and agreement on $v_0$
implies $|x - x'|\cdot\ell s< 2^t$. We conclude that if
\begin{equation}
	\label{eq:inj}
	\ell s^3 \geq 2^t
\end{equation}
the function mapping local signatures to edges is \emph{injective}:
in other words, once~\eqref{eq:inj} holds a duplicate edge can only
be caused by a duplicate local signature.

When instead $\ell s^3 < 2^t$, collisions are possible, but they are as rare as
they can be: no map from local signatures to edges can do better than the one
above, which is balanced up to rounding.

\begin{proposition}
	\label{prop:balanced}
	Let $x$, $x'$ be a uniformly distributed pair of \emph{distinct} $t$-bit
	local signatures, and let $N=2^t$. Then, for every function $f$ from
	$N$ to a set $E$ of edges,
	\[
		\Prob\bigl[\,f(x)=f(x')\,\bigr]\geq \frac1{|E|} - \frac1N.
	\]
	Moreover, the map $x\mapsto\langle v_0,o_1,o_2\rangle$ defined above has
	$|E| = \ell s^3$, and every edge has $q$ or $q+1$ preimages, where
	$q=\lfloor N/\ell s^3\rfloor$; hence, its collision probability is
	\[
		\frac{q\bigl(2N-(q+1)\ell s^3\bigr)}{N(N-1)},
	\]
	the minimum over all functions from $N$ to a set of $\ell s^3$ edges.
\end{proposition}

\begin{proof}
	Let $n_e$ be the size of the preimage of the
	edge $e$, so that $\sum_e n_e = N$. The number of ordered pairs of
	distinct signatures that collide is $\sum_e n_e(n_e-1)$, whence
	\[
		p \coloneqq \Prob\bigl[\,f(x)=f(x')\,\bigr] =
		\frac{\sum_e n_e(n_e-1)}{N(N-1)} = \frac{\sum_e n_e^2 - N}{N(N-1)}.
	\]
	If $|E|>N$ the bound is negative and there is nothing to prove, so we can
	assume $|E|\leq N$. By convexity (equivalently, by Cauchy--Schwarz)
	$\sum_e n_e^2\geq N^2/|E|$, with equality if and only if all preimages have
	the same size. Therefore, as $N/|E|-1\geq 0$,
	\[
		p \geq \frac{N^2/|E| - N}{N(N-1)} = \frac{N/|E|-1}{N-1}\geq
		\frac{N/|E|-1}N = \frac1{|E|}-\frac1N,
	\]
	which proves the first claim.

	For the second claim, note first that $v_0$ ranges in $\ell s$ and $o_1$,
	$o_2$ range in $s$, so $|E| = \ell s^3$. Fix an edge $\langle v_0, o_1,
		o_2\rangle$. The condition $v_0 = \lfloor x\ell s/2^t\rfloor$ confines
	$x$ to the interval $\bigl[\lceil v_0N/\ell s\rceil \.. \lceil (v_0+1)N/\ell
		s\rceil\bigr)$, whose length $L$ is $\lfloor N/\ell s\rfloor$ or $\lceil
		N/\ell s\rceil$. The conditions $o_1 = x\bmod s$ and $o_2=\lfloor
		x/s\rfloor\bmod s$ fix the residue of $x$ modulo $s^2$, and an interval
	of length $L$ contains $\lfloor L/s^2\rfloor$ or $\lceil L/s^2\rceil$
	integers of each residue class. Since $q=\lfloor N/\ell
		s^3\rfloor=\lfloor\lfloor N/\ell s\rfloor/s^2\rfloor$, both possible
	values of $L$ lie in $[qs^2\..(q+1)s^2]$, so every preimage has $q$ or $q+1$
	elements. Now, for every integer $n$ we have $(n-q)(n-q-1)\geq0$, that is,
	$n(n-1)\geq 2qn-q(q+1)$, with equality if and only if $n$ is $q$ or $q+1$.
	Summing over the edges, as $\sum_e n_e=N$, we obtain that for every function
	from $N$ to a set of $\ell s^3$ edges
	\[
		\sum_e n_e(n_e-1)\geq 2qN-q(q+1)\ell s^3 = q\bigl(2N-(q+1)\ell s^3\bigr),
	\]
	with equality for our map.
\end{proof}

The first claim is essentially the classical bound of Carter and
Wegman~\cite{CaWUCHF} for universal classes of hash functions, whose proof
contains the same argument; the intermediate bound $(N/|E|-1)/(N-1)$ is its
refinement by Sarwate~\cite{SarNUCHF}, who observes that it can be attained
only if $|E|$ divides $N$; the second claim gives the minimum in the general
case.

The collision probability of our map is zero when $\ell s^3\geq 2^t$, as it must
be by~\eqref{eq:inj}, and it exceeds the lower bound $1/\ell s^3-2^{-t}$
significantly only close to this boundary (if $\ell s^3\leq 2^{t-1}$, by less
than $8$\%). The net result is that in the computation of the probability of a
duplicate edge the pairwise collision probability $1/(c m\cdot s_c(m)^2)$ must
be replaced with the exact value of Proposition~\ref{prop:balanced}, and the
bound~\eqref{eq:dupfuse} on the number of shards must be enforced only as long
as $\ell s^3 < 2^t$, where $\ell$ and $s$ are the number of segments and the
segment size of a shard of $n/S$ keys; beyond that point, duplicate edges from
distinct local signatures are impossible, and sharding is limited only
by~\eqref{eq:logballsbins} and by the minimum shard size.

The practical consequences are pleasant. With $t=64$, $c = 1.105$ and the
estimate~\eqref{eq:nln} for the segment size, condition~\eqref{eq:inj} holds
exactly when the segment size is at least $2^{19}$, that is, when shards contain
at least ${\approx}10^8$ keys. In the peeling-based regime we never create
shards smaller than $10^7$ keys, so the only window in which duplicate edges can
actually appear is that of shard sizes between $10^7$ and ${\approx}10^8$ (the
augmented-peeling regime of Section~\ref{sec:small} uses much smaller shards,
and duplicate edges are accounted for when sizing the graphs). In that window
the failure probability must be computed using the collision probability of
Proposition~\ref{prop:balanced} (close to $1/\ell s^3 - 2^{-t}$ away from the
boundary): our implementation evaluates the resulting probability directly for
each candidate amount of sharding, and chooses the finest sharding for which it
does not exceed a fixed budget $\eta$ (we use $\eta=5\%$). Thus, at
$10^{11}$--$10^{12}$ keys, where the uniform analysis predicts a probability of
a duplicate edge per construction ranging from a few percent to more than
$80$\%, shards contain more than $10^8$ keys, and no construction can fail
because of a duplicate edge.

We remark that with full $128$-bit signatures the number of possible edges is
always much smaller than $2^t$, so the uniform analysis of the previous sections
remains valid. Full signatures are however necessary for functions based on fuse
graphs beyond about ten billion keys (to keep the probability of a duplicate
local signature below $5$\%), as we discussed in Section~\ref{sec:subsign}, because for
functions a duplicate \emph{local signature} is fatal (two keys would be forced
to return the same output); in that case there is no injective regime, and our
implementation applies the same budget $\eta$ to the retry probability given by
the uniform analysis, which limits sharding more than in the subsigned case.
In our experiments (not shown in the figures), full signatures make the
construction of \VFilter and \VFunc about $20$\% and $10$\% slower,
respectively, whereas query times are essentially
unchanged.

\paragraph{A cautionary tale.} Since in a filter we are free to choose the
hash associated with a key, one might be tempted to derive it from the edge
instead of the local signature: keys with the same edge would then generate the
same equation, and duplicate edges could be removed like duplicate local
signatures, so that no duplicate edge could ever make a construction fail. 
Unfortunately, a key outside the key set whose edge coincides with the edge of
a key is a false positive with certainty, and since there are at most $m$ key
edges out of $\ell s^3$ possible ones, the false-positive rate becomes
\[
	2^{-b} + \frac m{\ell s^3} = 2^{-b} + \frac{2\lambda}m,
\]
where $\lambda = m^2/2\ell s^3$ is the expected number of pairs of keys of a
shard with the same edge. The second term cannot be reduced by increasing $b$,
and it is large exactly when duplicate edges are frequent, that is, when
removing them would be useful: in the augmented-peeling regime it is about
$2^{-14}$ at $10^4$ keys and $2^{-22}$ at $10^6$ keys, so with $b=16$ the
false-positive rate at $10^4$ keys would be about five times the nominal one. When,
instead, the second term is negligible (between $2^{-30}$ and $2^{-36}$ in the
peeling regime), duplicate edges are rare, and the probability of a retry is
already bounded by $\eta$.

\section{Experiments}

We ran experiments on integer key sets ranging from medium-sized to very large
(a trillion keys); in this section and in the next one keys are always $64$-bit
integers, as they have the shortest hash time. The experiments on monotone
minimal perfect hash functions and on compressed static functions, which use
different key sets, are described in Sections~\ref{sec:mmphf}
and~\ref{sec:csf}.
We have in mind use cases with billions of keys and a sizable output, such as
index functions,\footnote{An \emph{index function} (also known as an
	\emph{order-preserving minimal perfect hash function}) maps elements of a
	list to their position in the list.}
but, as in~\cite{DDHSWRFSSRAM}, we mostly perform experiments on filters, as
their queries perform essentially the same operations as those of static
functions with the same output width, and benchmarking software is available
from the authors. For BuRR, however, filter queries perform the same operations
as function queries only on keys (see Section~\ref{sec:protocol}).

The choice of competitors is based on the extensive comparison
of~\cite{DDHSWRFSSRAM}, which evaluates on $10^8$ keys a dozen filter
constructions, from Bloom, cuckoo, Morton, and quotient filters to ribbon- and
peeling-based ones. In that comparison, BuRR dominates all other constructions
for space overheads below $2$\%, whereas constructions based on hypergraph
peeling (Xor filters, i.e., MWHC, and coupled filters, i.e., fuse graphs) have
the fastest queries for overheads between $4$ and $30$\%. The other competitive
constructions are either filters only, and thus cannot represent static
functions, or are competitive only if their additional operations (e.g.,
insertions and deletions) are needed. The only Pareto-optimal constructions of
that comparison that can represent static functions are thus BuRR and those
based on hypergraph peeling, which are our main terms of comparison. We also
include \PHunc, a static function based on a minimal perfect hash function, a
kind of construction that the comparison of~\cite{DDHSWRFSSRAM} does not
consider.

We used an \texttt{m8a.metal-24xl} Amazon AWS instance with $96$
AMD\textregistered{} EPYC\texttrademark{} 9R45 cores and $384$\,GiB of RAM,
endowed with a $24\,000$\,GiB Amazon EBS \texttt{gp3} SSD volume with a
provisioned throughput of $2\,000$\,MiB/s and $16\,000$ IOPS. The code was compiled using
\texttt{g++} 16.2.0 and \texttt{rustc} 1.100.0, optimizing for the native
processor (see Section~\ref{sec:protocol} for the effect of vectorization).

\begin{figure}[tp]
	\centering
	\includegraphics[width=0.49\textwidth]{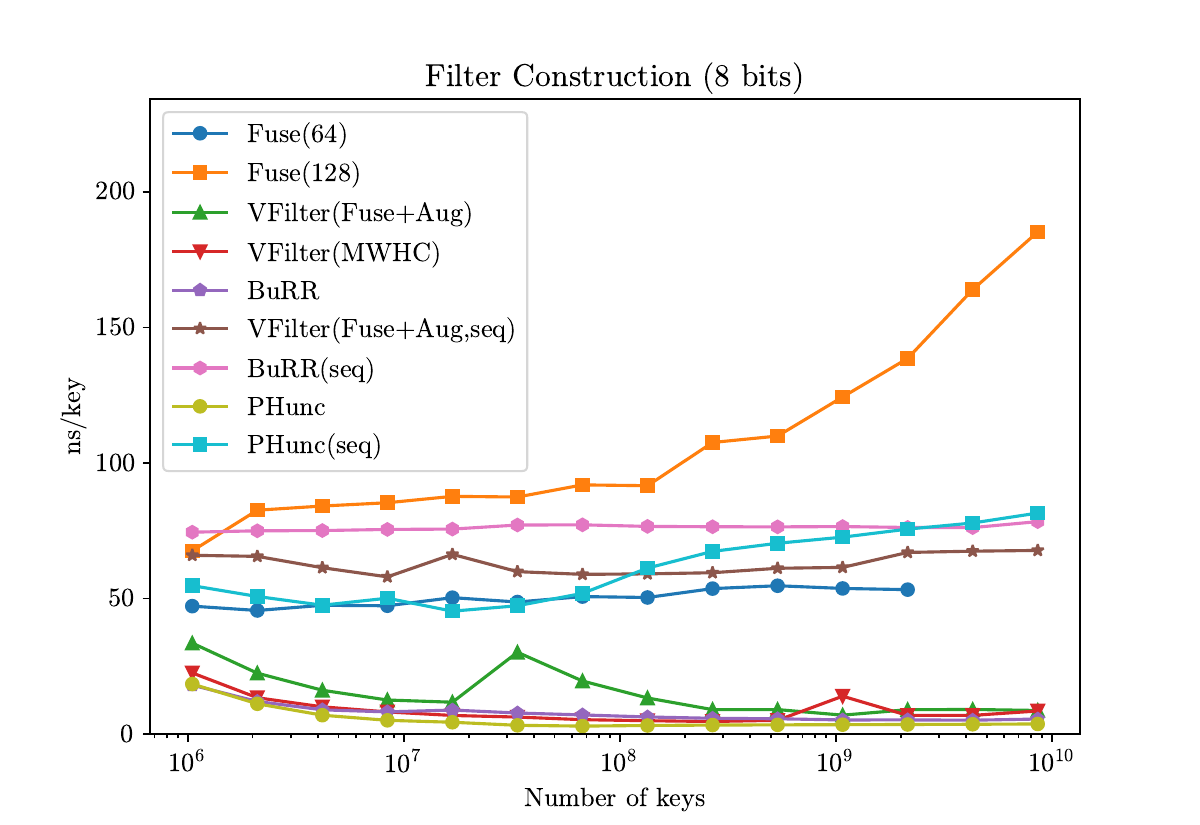} \includegraphics[width=0.49\textwidth]{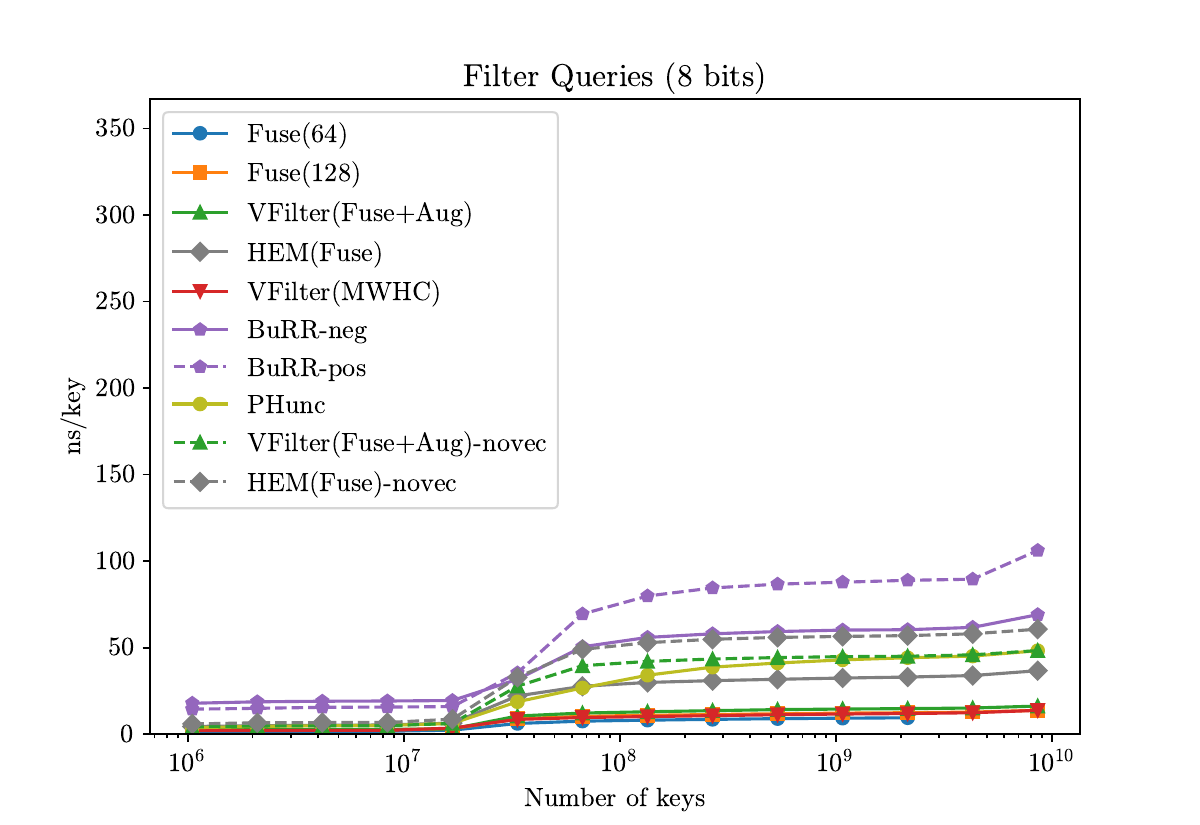}\\
	\includegraphics[width=0.49\textwidth]{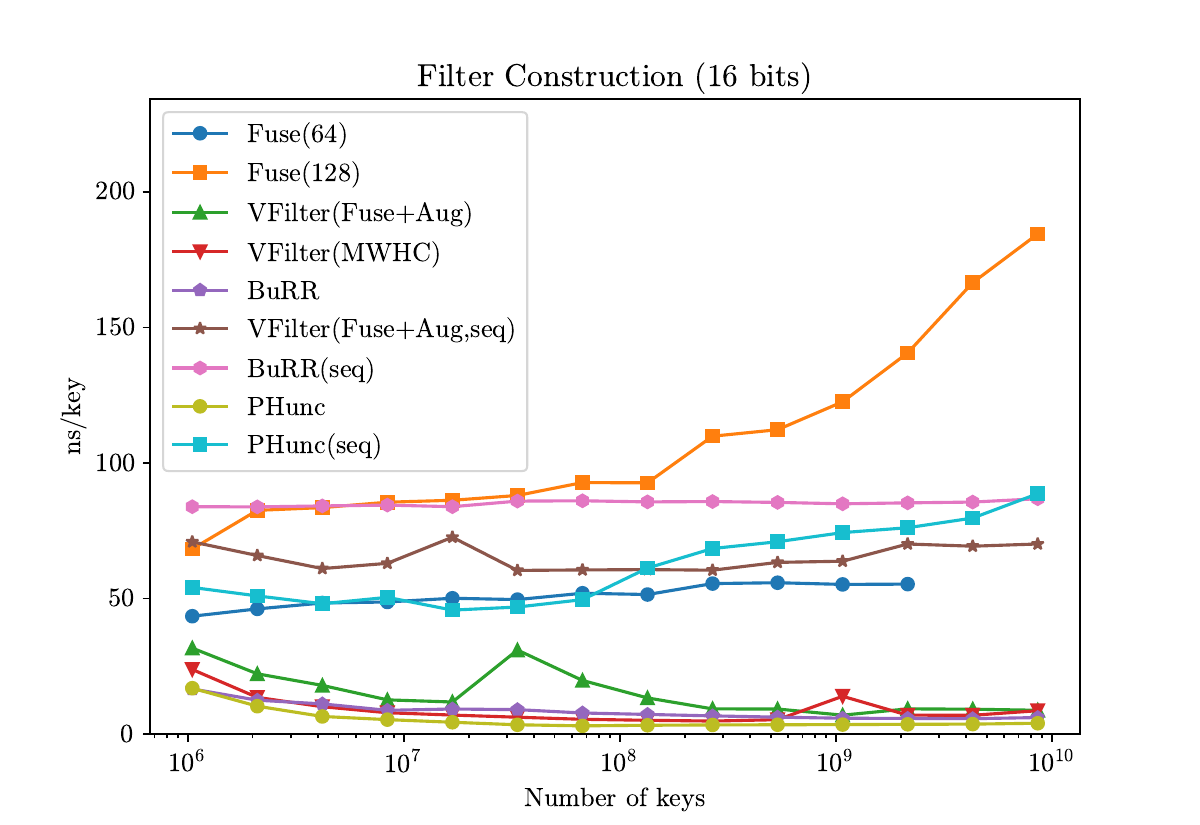} \includegraphics[width=0.49\textwidth]{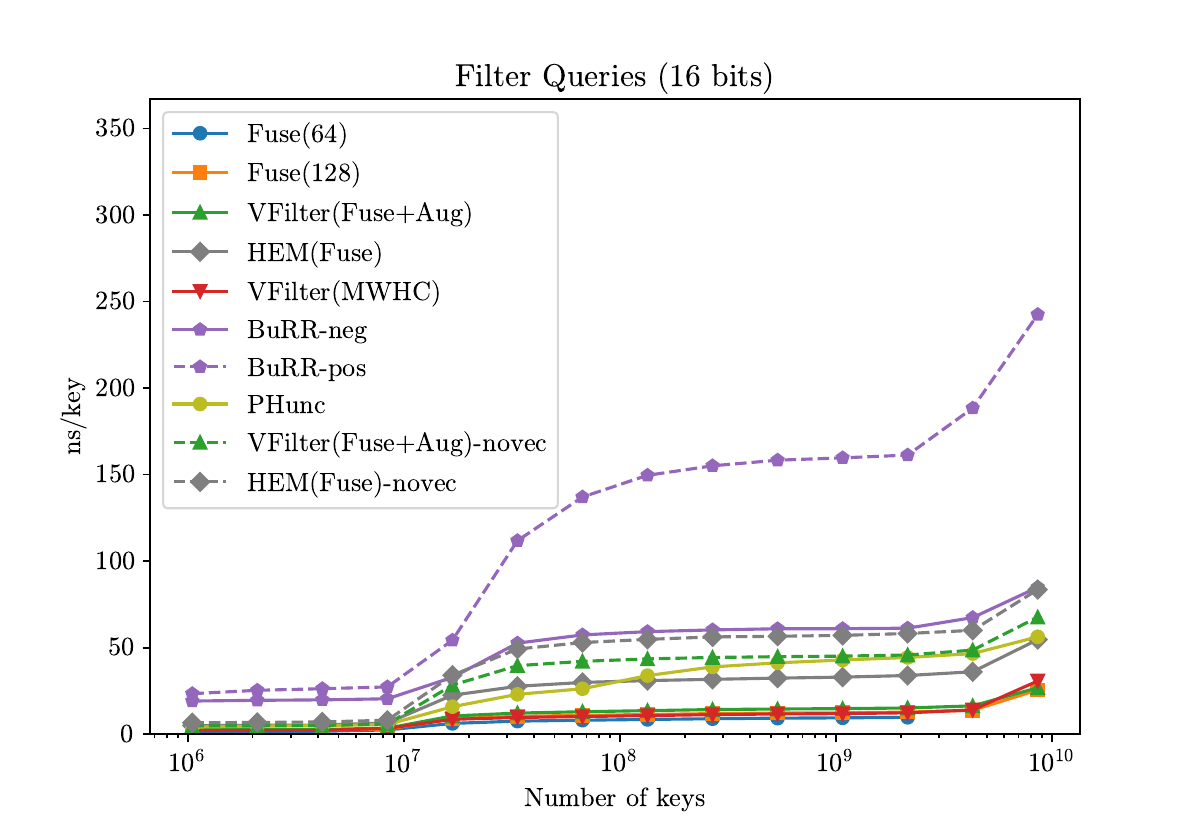}\\
	\includegraphics[width=0.49\textwidth]{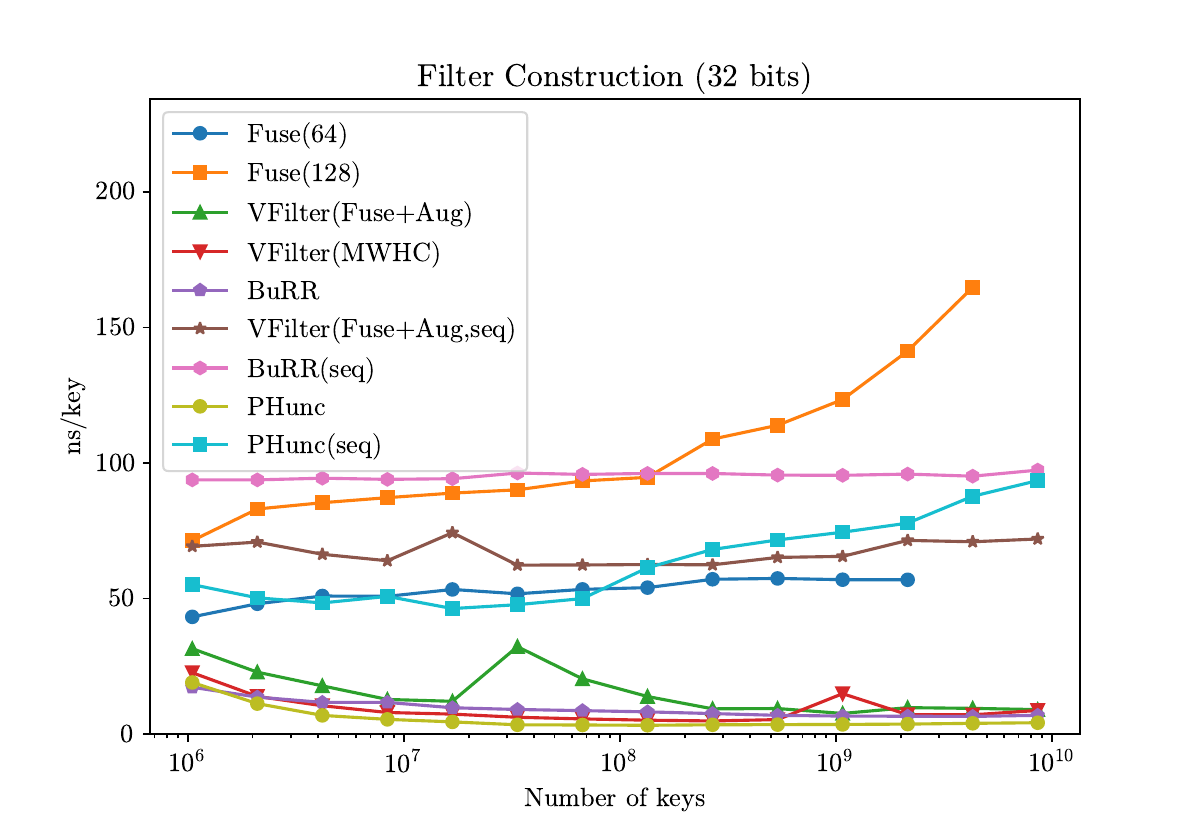} \includegraphics[width=0.49\textwidth]{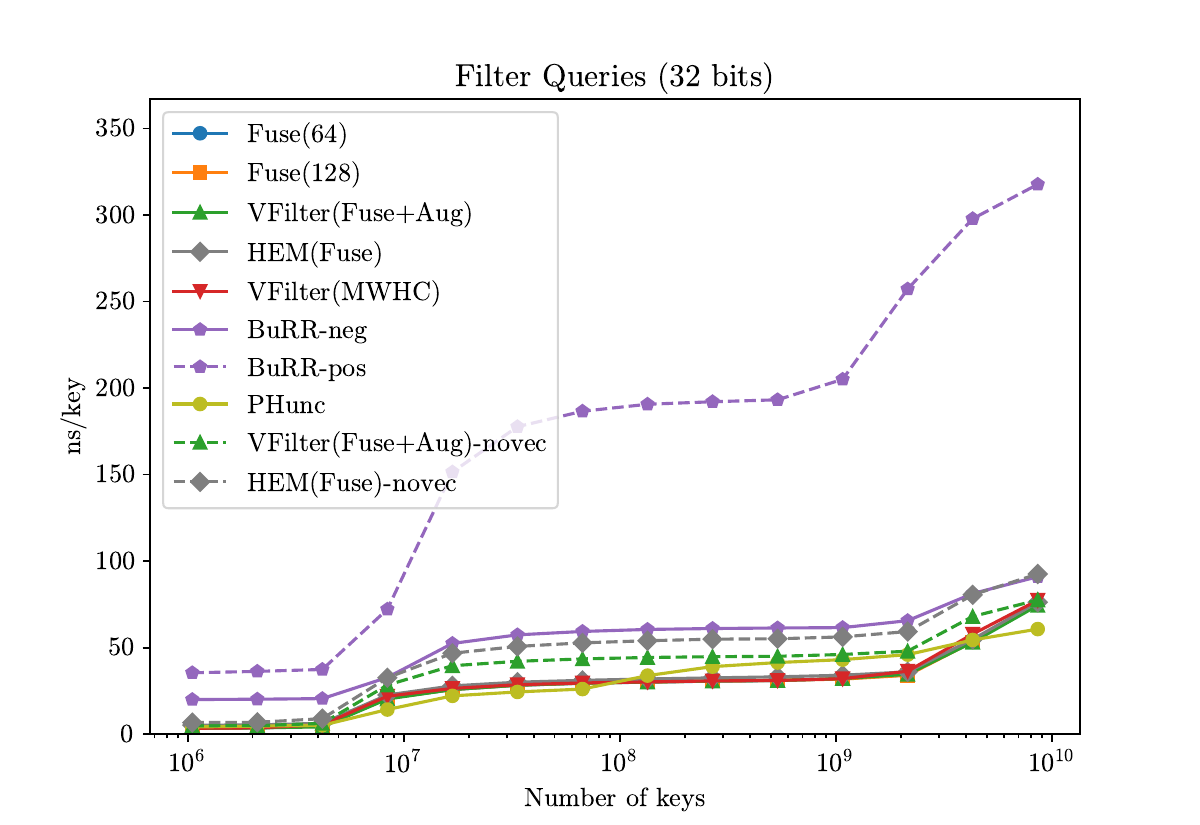}
	\caption{Construction and query times for power-of-two
		output widths. The lines with suffix -novec are the query times of
		\VFilter and HEM(Fuse) compiled with the loop vectorizer disabled.}
	\label{fig:poweroftwo}
\end{figure}

\begin{figure}[tp]
	\centering
	\includegraphics[width=0.49\textwidth]{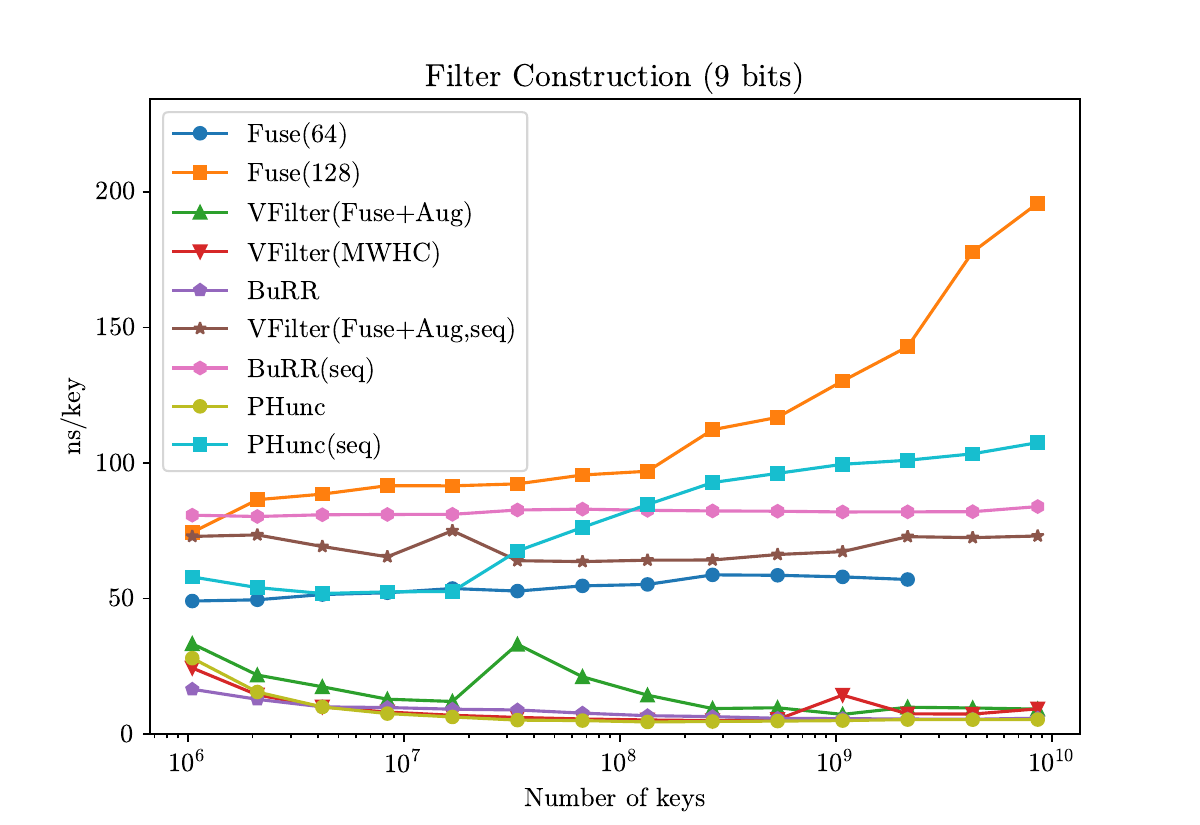} \includegraphics[width=0.49\textwidth]{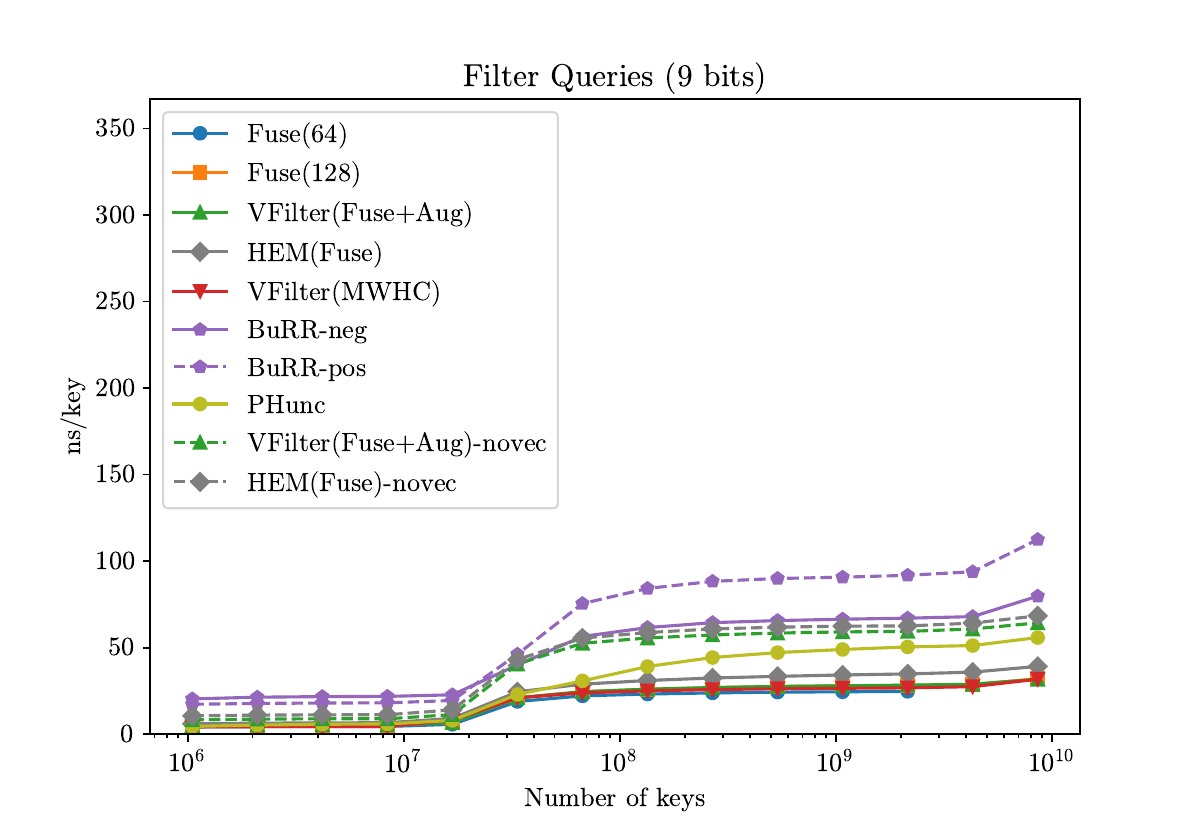}\\
	\includegraphics[width=0.49\textwidth]{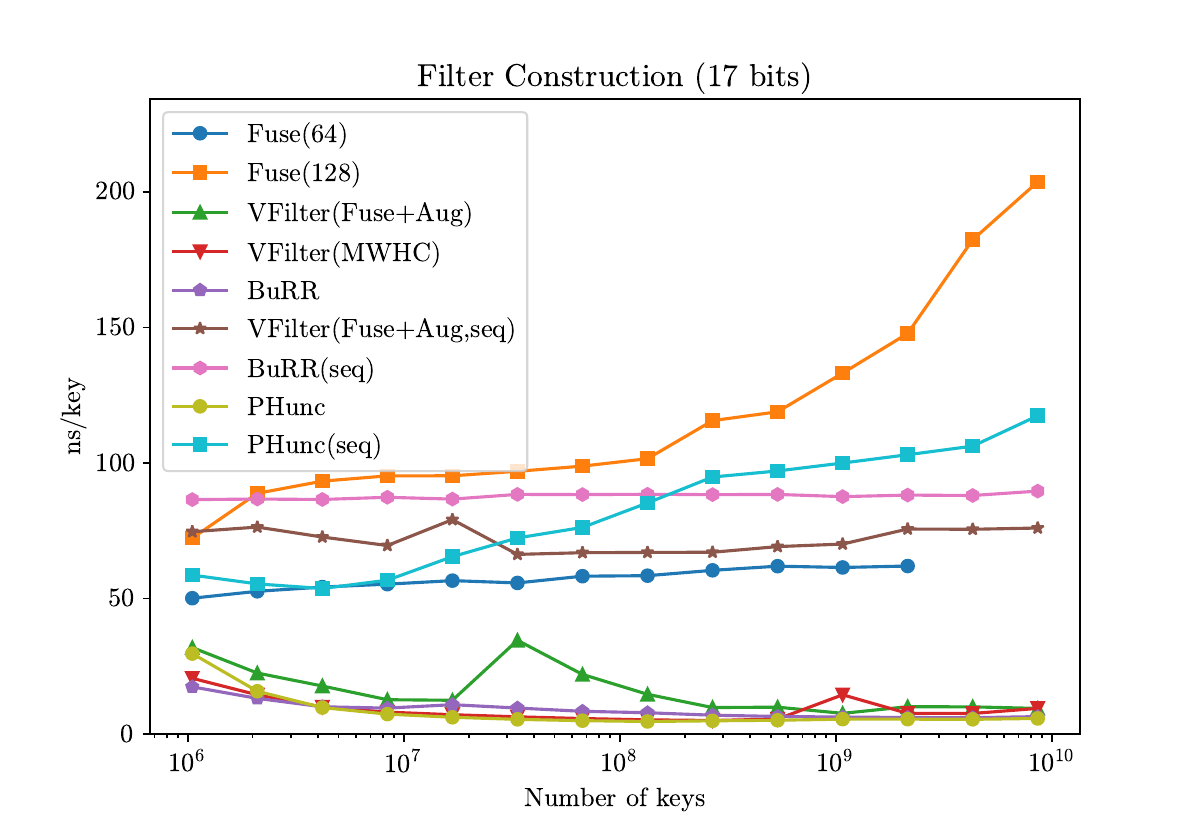} \includegraphics[width=0.49\textwidth]{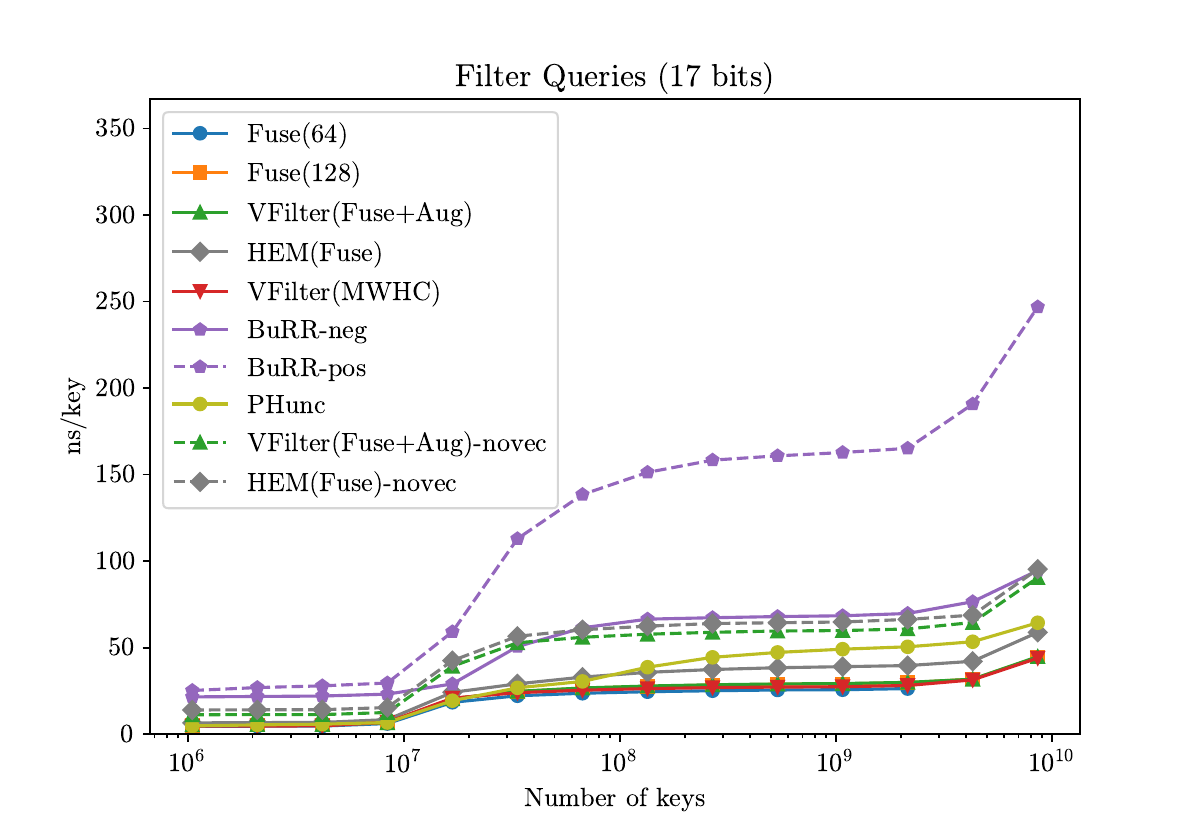}\\
	\includegraphics[width=0.49\textwidth]{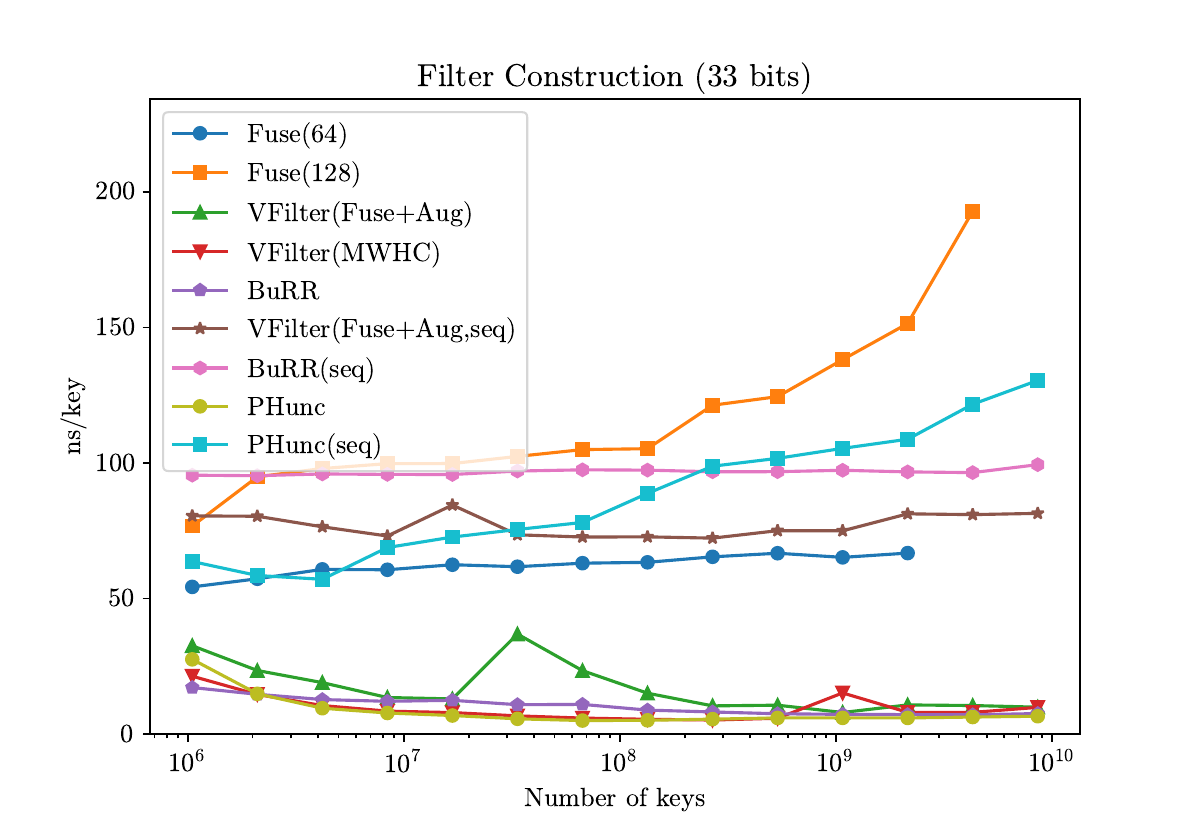} \includegraphics[width=0.49\textwidth]{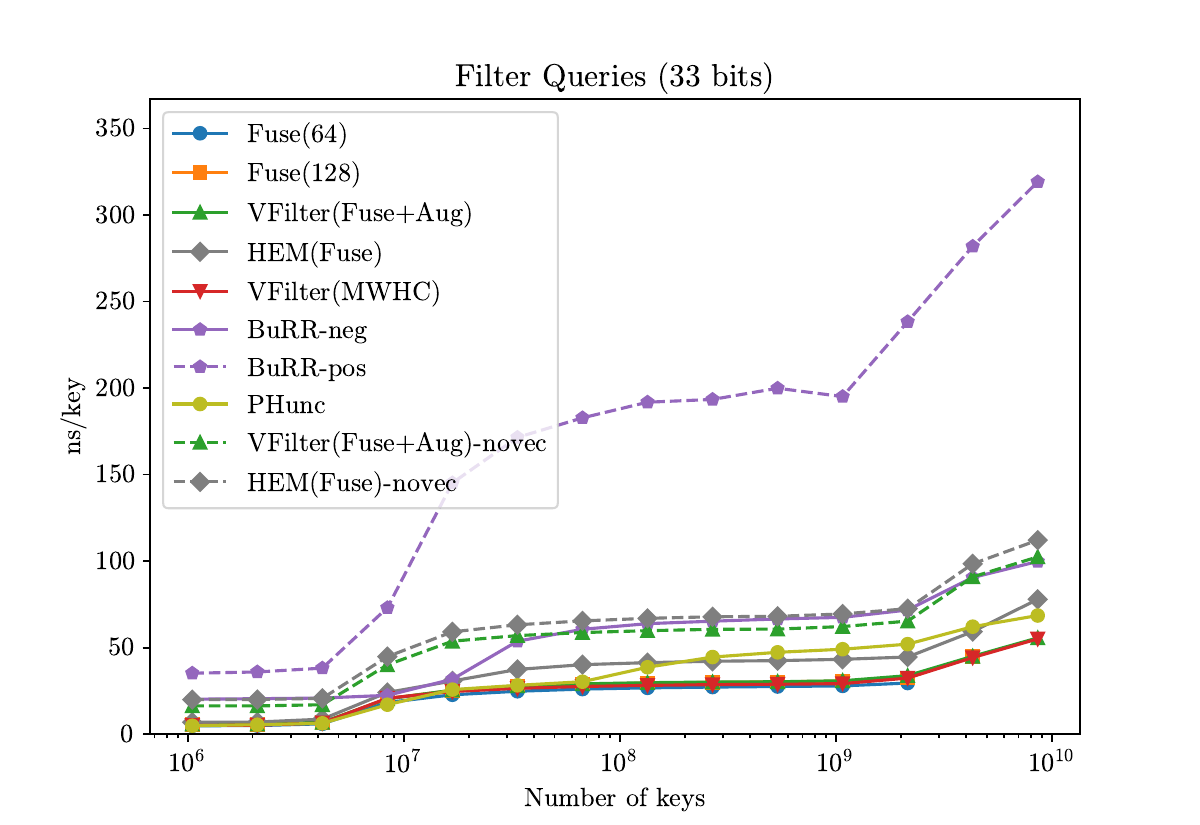}
	\caption{Construction and query times for power-of-two-plus-one
		output widths. The lines with suffix -novec are the query times of
		\VFilter and HEM(Fuse) compiled with the loop vectorizer disabled.}
	\label{fig:poweroftwoplusone}
\end{figure}

\begin{figure}[tp]
	\centering
	\includegraphics[width=0.49\textwidth]{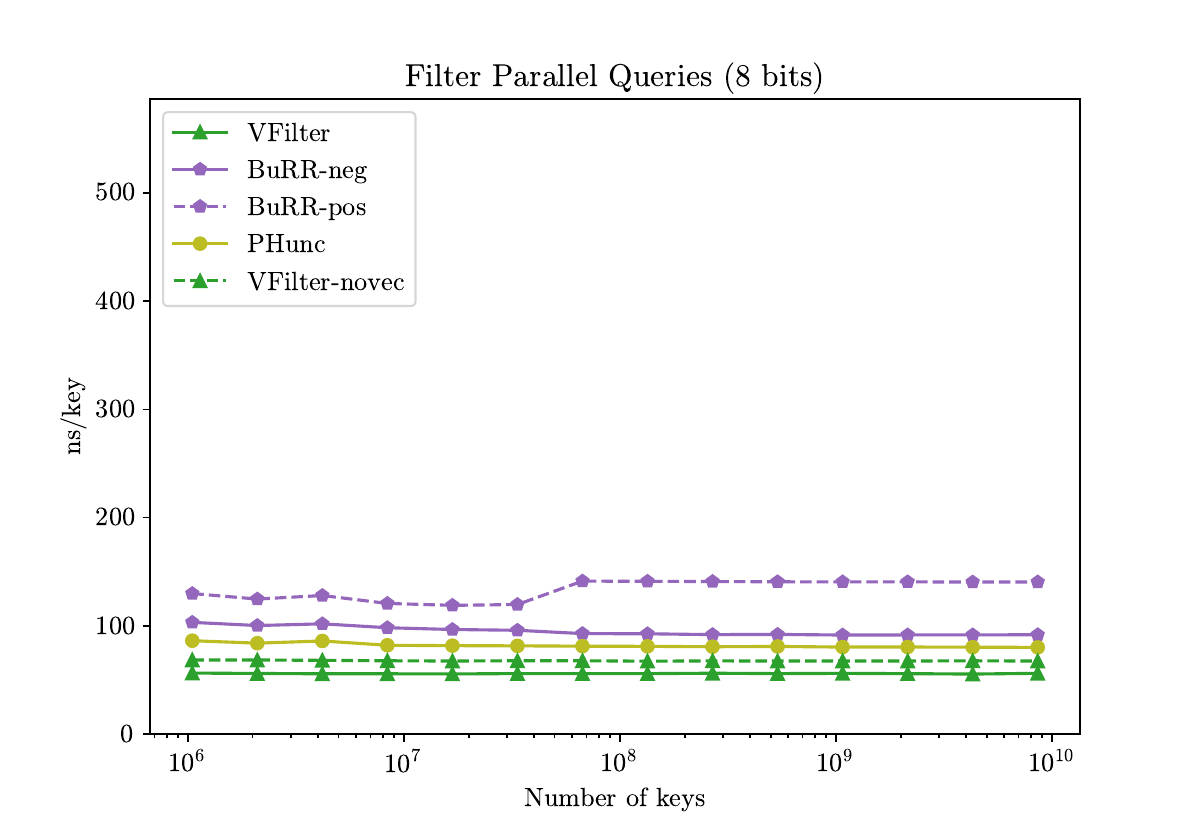} \includegraphics[width=0.49\textwidth]{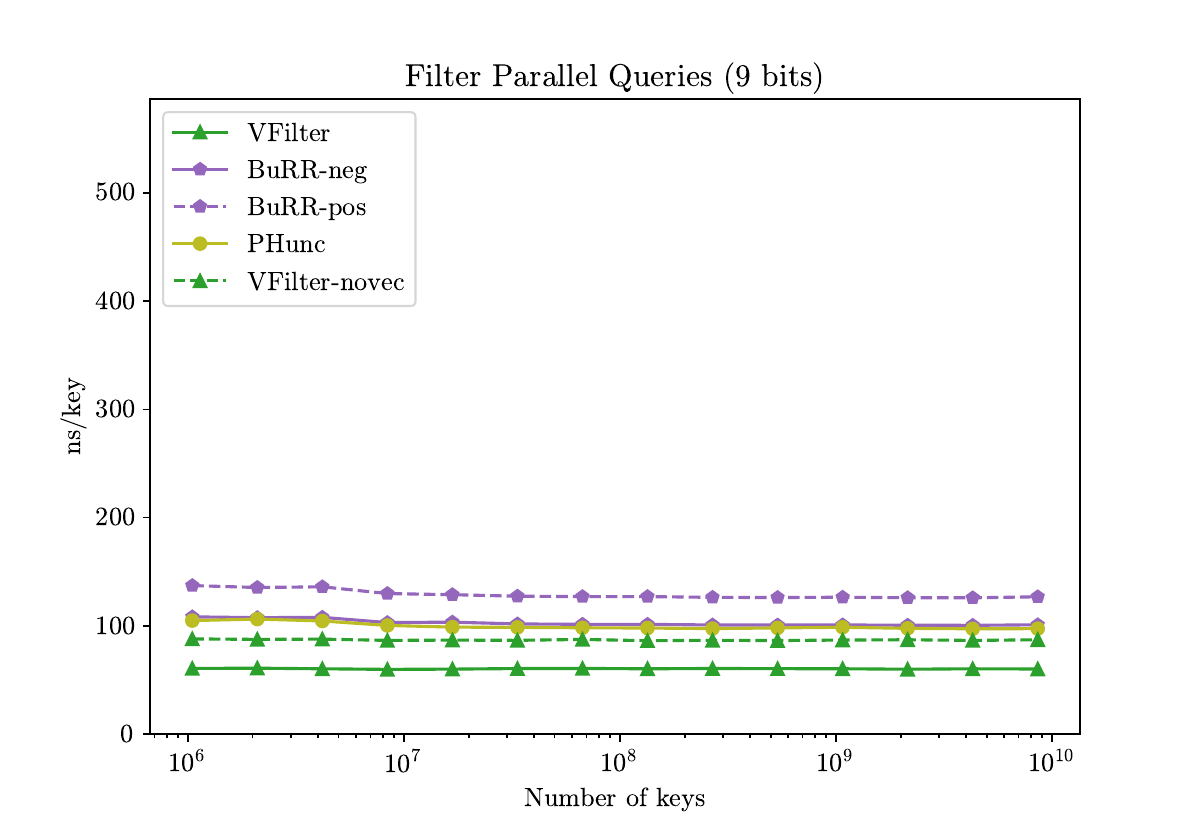}\\
	\includegraphics[width=0.49\textwidth]{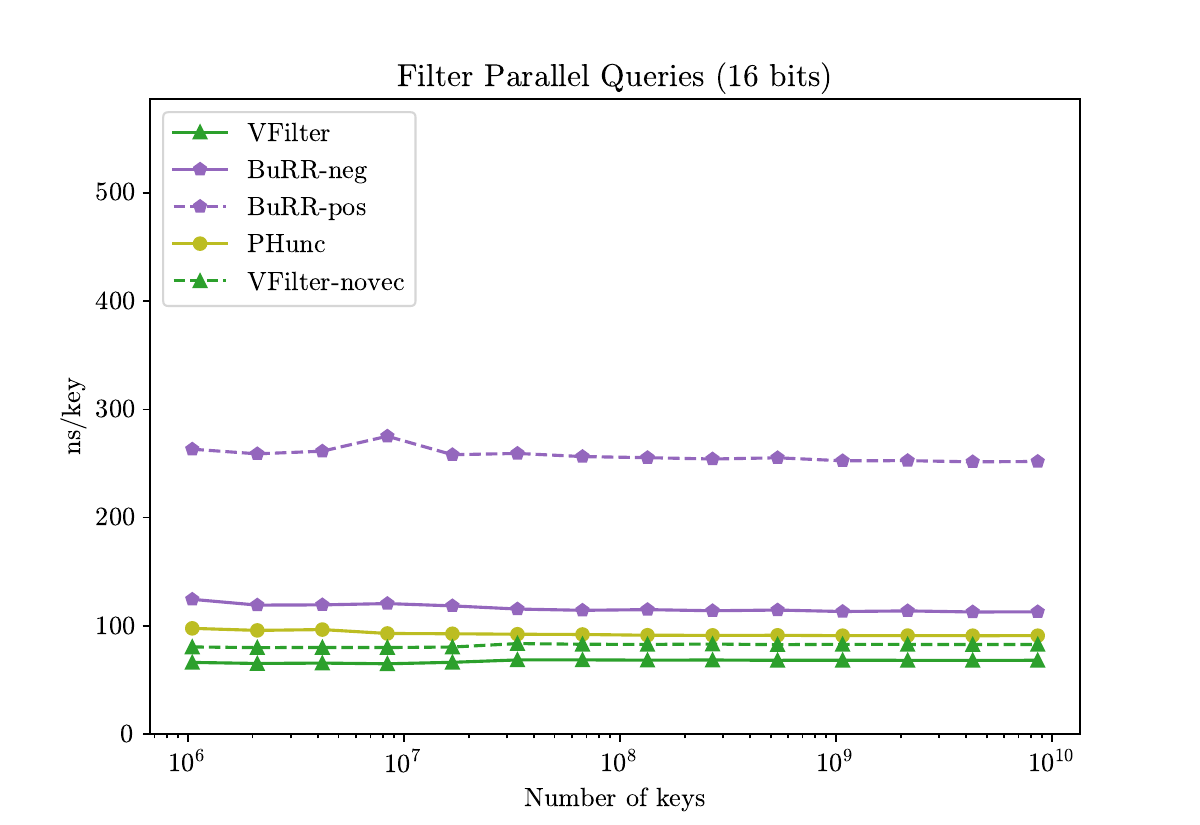} \includegraphics[width=0.49\textwidth]{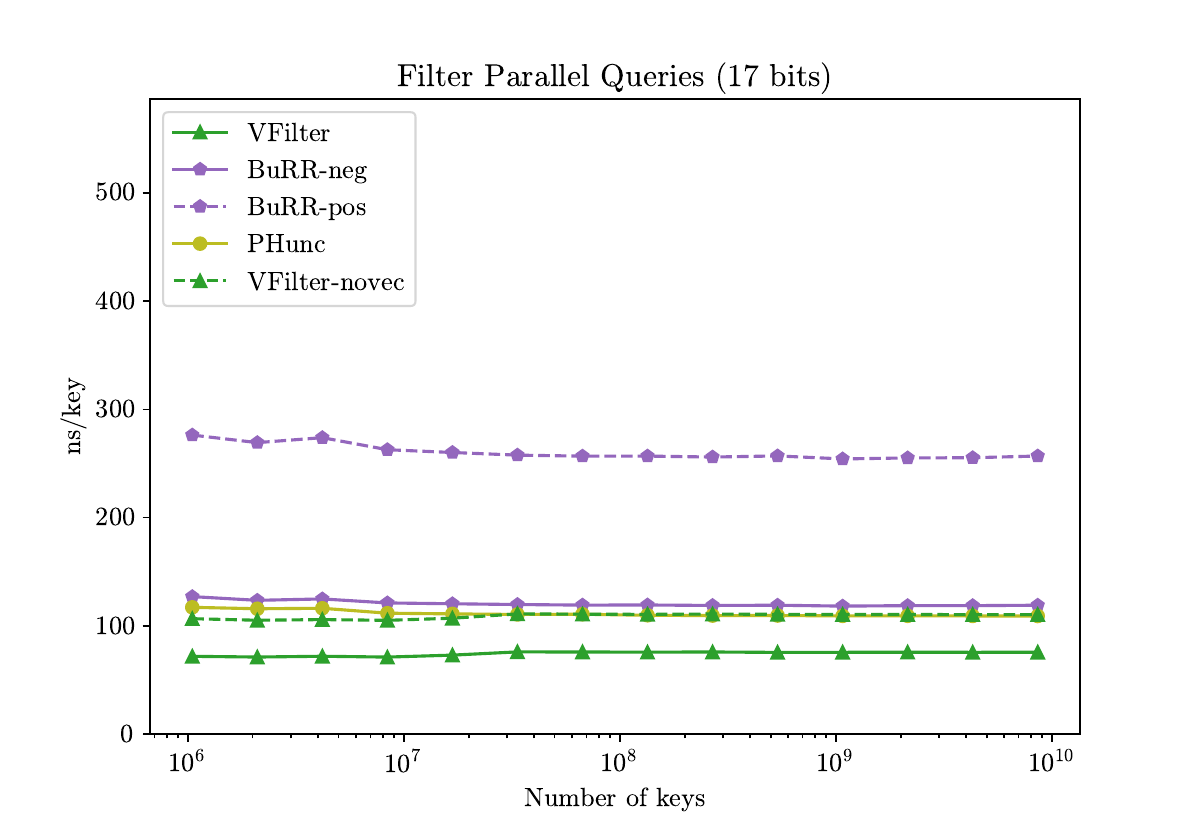}\\
	\includegraphics[width=0.49\textwidth]{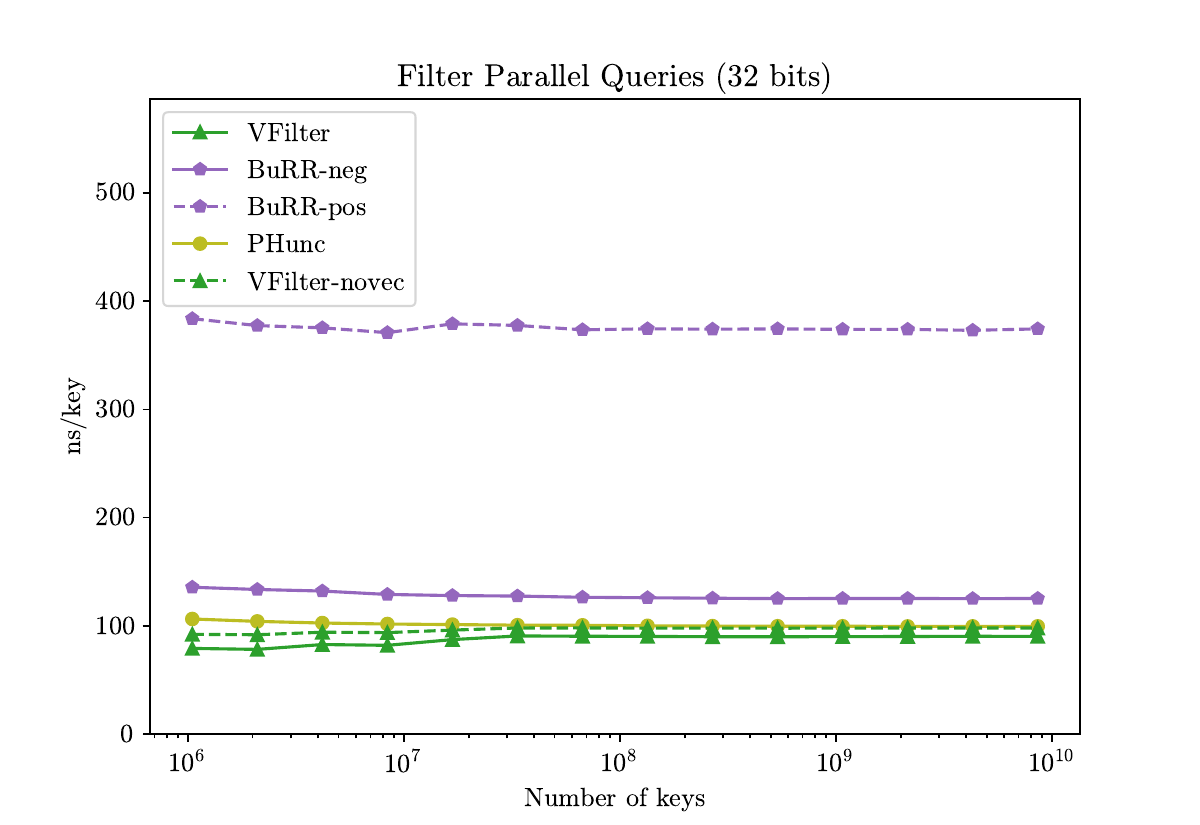} \includegraphics[width=0.49\textwidth]{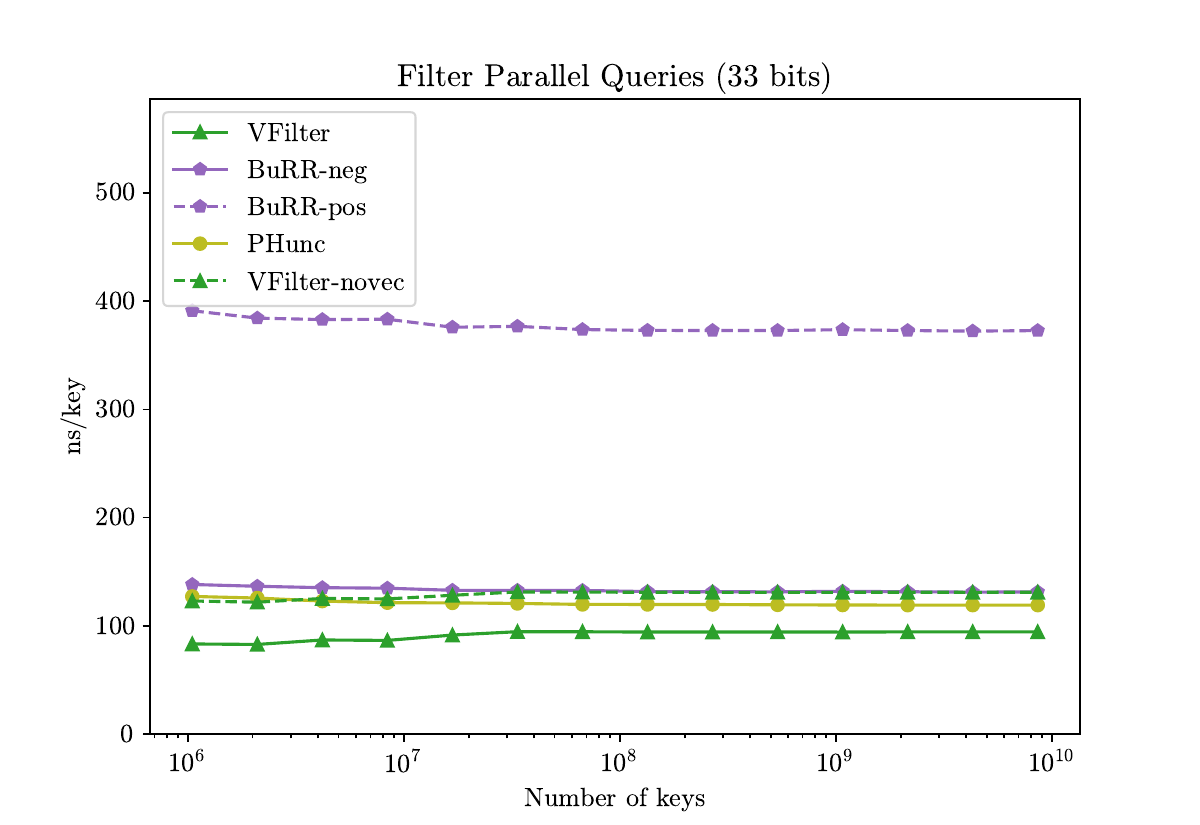}
	\caption{Query times under the highly parallel workload of the BuRR paper:
		$2^{33}/n$ structures of $n$ keys queried at random by $96$ threads.
		Since positive and negative queries to \VFilter and \PHunc take the
		same time, we show only the latter. The lines with suffix -novec are the
		query times of \VFilter compiled with the loop vectorizer disabled.}
	\label{fig:stress}
\end{figure}

\begin{figure}[t]
	\centering
	\includegraphics[width=0.49\textwidth]{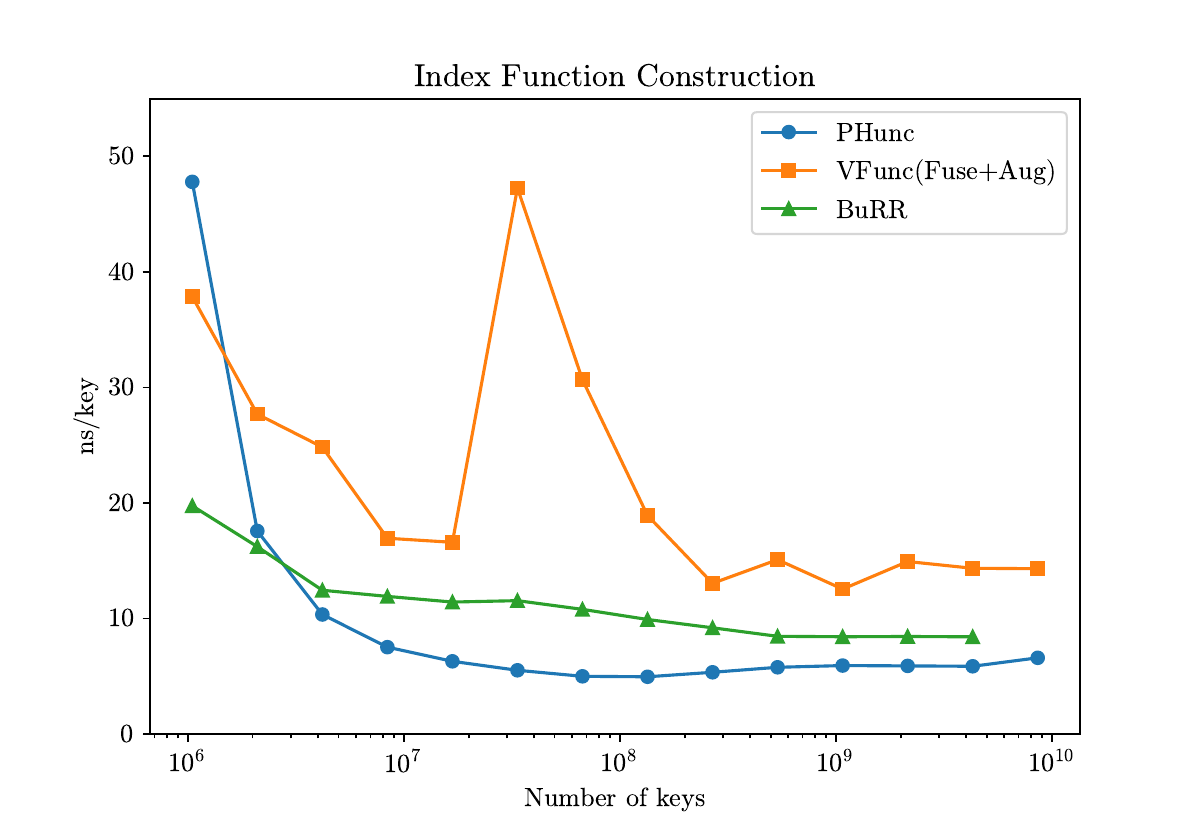} \includegraphics[width=0.49\textwidth]{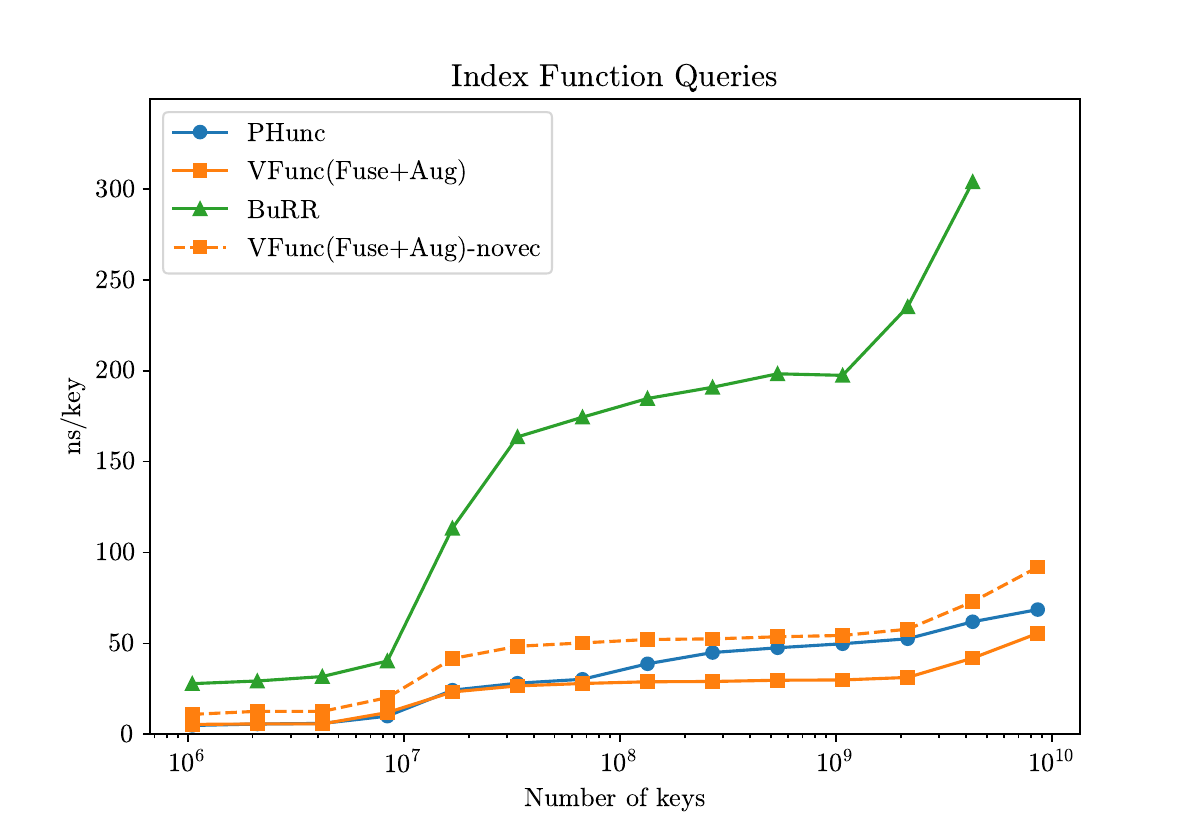}
	\includegraphics[width=0.49\textwidth]{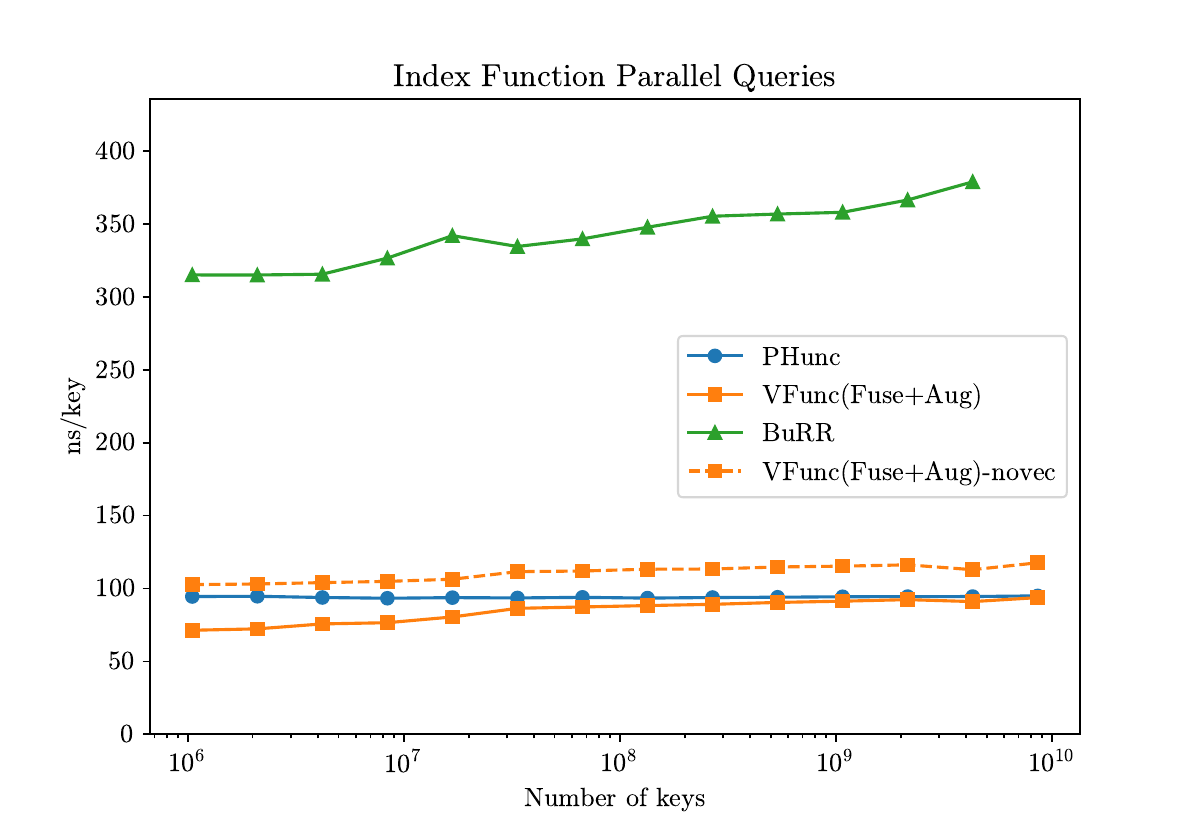}
	\caption{Construction and query times for index functions (top), and
		query times under the highly parallel workload of
		Figure~\ref{fig:stress} (bottom). BuRR is used in retrieval mode, and
		all constructions use $32$ threads. The lines with suffix -novec are
		the query times of \VFunc compiled with the loop vectorizer disabled.
		The construction of BuRR with $2^{33}$ keys failed for insufficient
		memory, both in the single-structure and in the parallel test.}
	\label{fig:func}
\end{figure}

\begin{figure}[t]
	\centering
	\includegraphics[width=0.49\textwidth]{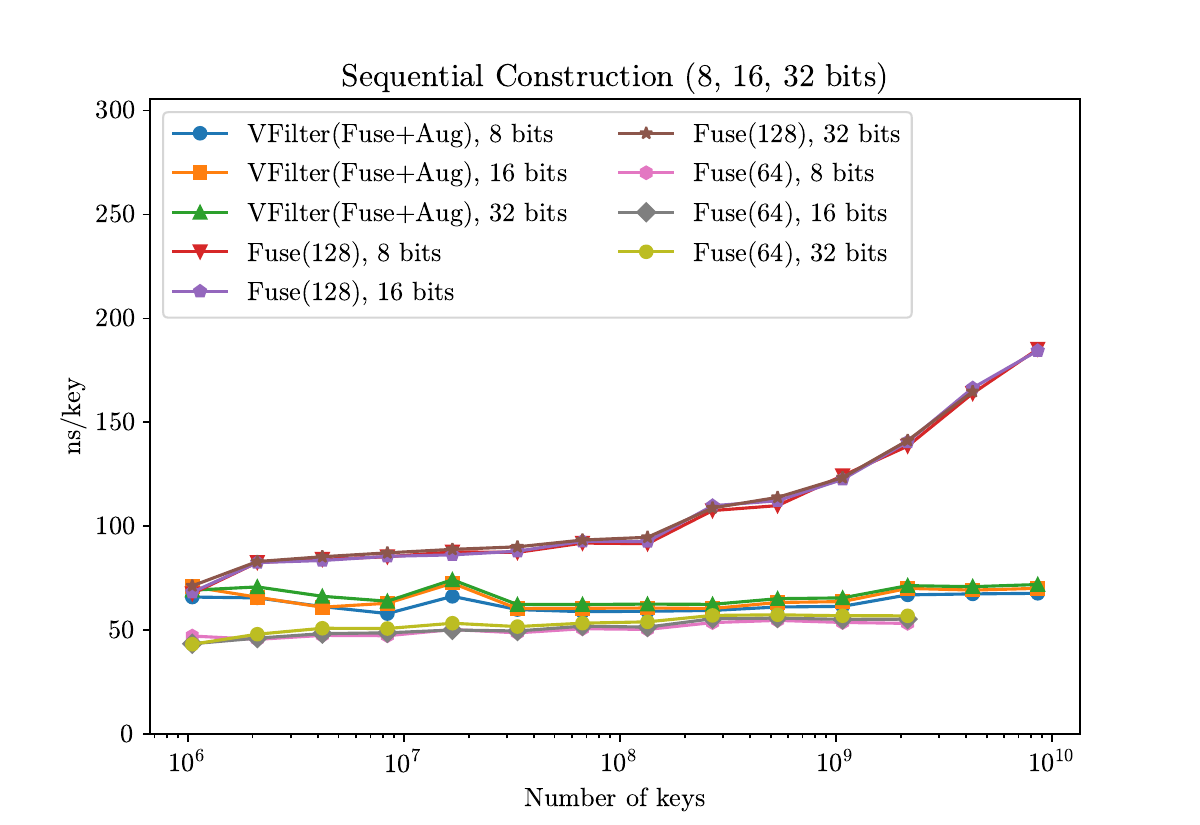} \includegraphics[width=0.49\textwidth]{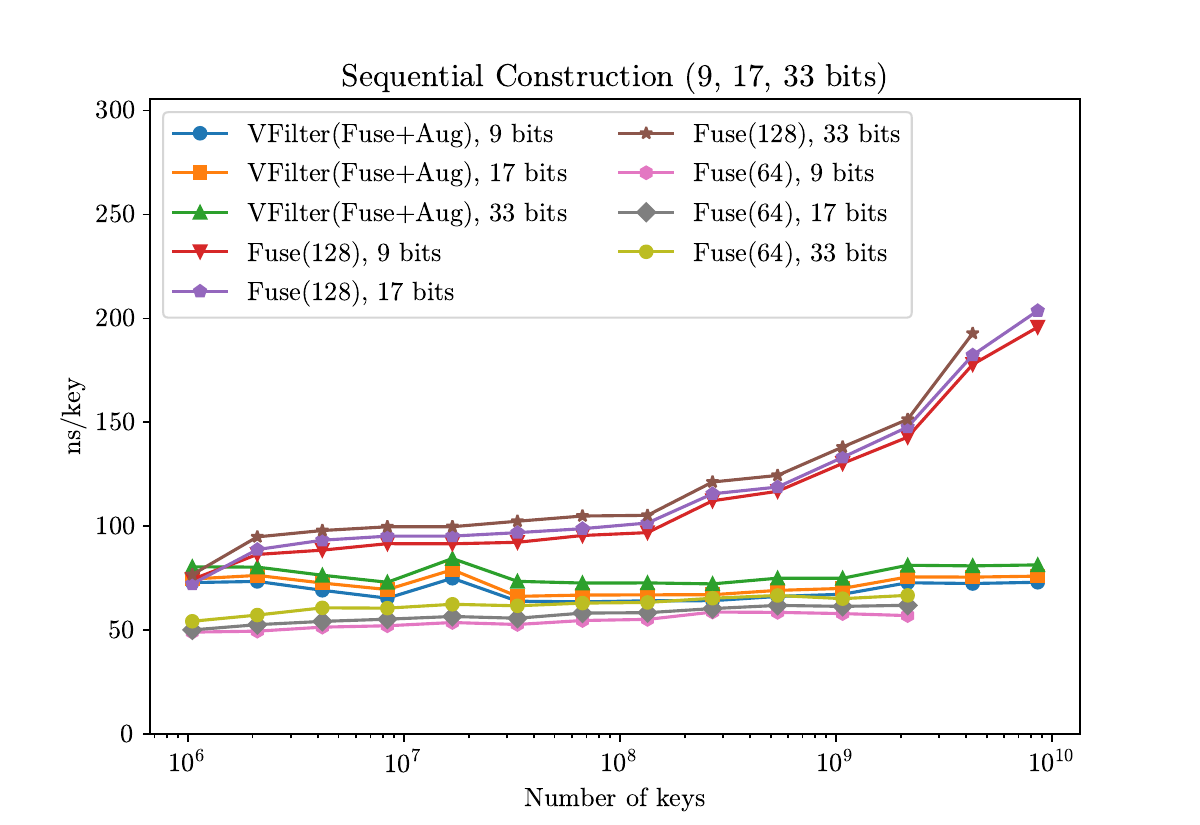}
	\caption{Construction time in the sequential case for sharded (\VFilter) and
		unsharded fuse graphs with $64$- and $128$-bit signatures (computed in
		parallel); unsharded fuse graphs with $64$-bit signatures cannot be
		built beyond $2^{31}$ keys. The steeper increase at $2^{32}$ keys is
		due to the low-memory visits we must force
		in the unsharded case to build structures with $2^{32}$ or more keys
		within the available memory; with $32$- and $33$-bit values, building a
		structure with $2^{33}$ keys failed for insufficient memory.}
	\label{fig:comp}
\end{figure}

We have the following setups:
\begin{itemize}
	\item For power-of-two output widths $b$ ($8$, $16$, and $32$ bits), we
	      compare BuRR with unsharded fuse graphs with $64$- and $128$-bit
	      signatures, labeled Fuse(64) and Fuse(128); with \VFilter, both in
	      the version based on the MWHC construction, labeled VFilter(MWHC), and
	      in the version based on fuse graphs, labeled VFilter(Fuse+Aug)
	      (augmented peeling is used up to $2^{24}$ keys); and with \PHunc,
	      which in this case maps each key to a pseudorandom $b$-bit value.
	      Unsharded fuse graphs with $128$-bit signatures are not useful in
	      practice, as they cannot be solved in parallel, but their benchmarks
	      highlight the cost of different components of the construction. All
	      constructions but BuRR store their values in a slice of the unsigned
	      integer type of the output width.
	\item Unsharded fuse graphs can only be solved sequentially (signatures are
	      computed in parallel); for the other constructions we report the
	      parallel construction time and, for BuRR, \VFilter, and \PHunc, also
	      the sequential one (labeled seq). For BuRR, we report query times both
	      on negative and on positive queries (labeled BuRR-neg and BuRR-pos;
	      see Section~\ref{sec:protocol}). The results are displayed in
	      Figure~\ref{fig:poweroftwo}.
	\item We perform the same tests for power-of-two-plus-one output widths ($9$,
	      $17$, and $33$ bits). In this case, all constructions but BuRR store
	      their values in a bit array. The results are displayed in
	      Figure~\ref{fig:poweroftwoplusone}.
	\item We repeat the query tests under the highly parallel workload of the
	      BuRR paper~\cite{DDHSWRFSSRAM}, which reports that when all hardware
	      threads query at random many structures, BuRR becomes faster than
	      hypergraph-based constructions. To make the workload comparable
	      across key sets, we keep the total number of keys fixed: for each
	      size $n$ we build $2^{33}/n$ structures of $n$ keys, which are then
	      queried at random by $96$ threads, both on negative and on positive
	      queries (see Section~\ref{sec:protocol}). The results are displayed
	      in Figure~\ref{fig:stress}.
	\item We also compare static functions mapping integers to themselves,
	      built using $32$ threads: \PHunc, that is, a very fast minimal
	      perfect hash function, PHastR~\cite{VigP}, followed by a bit array;
	      \VFunc (labeled VFunc(Fuse+Aug)); and BuRR used as a static function,
	      that is, in its retrieval mode, in which the solution stores the
	      values of the keys in place of their hashes. The interest of this
	      experiment stems from a consideration of Demaine, Meyer auf der Heide,
	      Pagh, and P{\u{a}}tra{\c{s}}cu~\cite{DHPDDDPSU} that an MPHF is
	      stronger than a static function, as one can obtain a static function
	      from an MPHF by indexing an array with the result of the MPHF, but not
	      the other way around; this consideration is in contrast with an
	      observation by Dietzfelbinger and Pagh~\cite{DiPSDSRAM} that many
	      kinds of retrieval data structures issue independent memory accesses,
	      which are parallelized efficiently by modern processors. The results
	      are displayed in Figure~\ref{fig:func}, which also shows query times
	      under the highly parallel workload described above. Note that when
	      several values are associated with each key, as with the columns of a
	      table, a single MPHF can index several arrays, amortizing its cost,
	      whereas a static function pays its space overhead on every output
	      bit.
	\item In Figure~\ref{fig:comp} we compare the sequential construction time
	      for fuse graphs in the sharded version (\VFilter) and in the unsharded
	      versions with $64$- and $128$-bit signatures (in the unsharded case
	      only solving is sequential, as signatures are computed in parallel).
	\item We also measure directly the cost of the sharding scheme we replaced.
	      The row HEM(Fuse) of Table~\ref{tab:detail}, and the corresponding
	      curves in the query plots, report a structure that differs from
	      \VFilter{} only in its layout: it uses the same signatures, shard
	      selection, construction of fuse-graph edges, expansion factor, segment
	      size, number of shards, and representation of values. Shards,
	      however, keep their natural, unequal sizes, so their starting
	      positions must be stored in an array and loaded at query time, and
	      each shard carries a local seed. A shard that peels at the first
	      attempt has local seed zero; its stored hashes are then those of
	      \VFilter{}, and its edges too if it has as many vertices as a shard of
	      \VFilter{}. The difference between the two rows thus measures the
	      cost of the HEM layout: loading the offset/seed word of the shard, and
	      computing the edge from the values it contains rather than from
	      constants. The only exceptions are key sets of up to $2^{24}$ keys,
	      which \VFilter{} solves by augmented peeling: HEM(Fuse) solves its
	      shards by peeling, so it must use the larger expansion factor and
	      segment size of peeling, but a query performs the same operations on
	      either graph.
	\item The row HEM(GOV3) reports instead HEM as it is actually deployed,
	      namely the GOV3 construction of Sux4J~\cite{GOVFSCCSMPHF,Sux4J541},
	      queried by the C implementation distributed with Sux4J so that the
	      comparison is not affected by the language. GOV3 solves its shards by
	      lazy Gaussian elimination, which is cubic and thus forces shards of
	      about $1\,500$ keys: its offset/seed array is therefore several
	      megabytes at a billion keys, rather than the few hundred bytes of the
	      previous row. The first row thus measures the cost of the HEM layout
	      alone; the second adds the cache misses of the much finer sharding of
	      GOV3 on large key sets and, since a local seed must change the edge, a
	      second hash of the signature after the offset/seed word has been
	      loaded (in Sux4J, a second round of SpookyHash).
\end{itemize}

Detailed numbers for $n=2^{20}$ and $n=2^{30}$ are reported in
Table~\ref{tab:detail}.

\begin{table}[tp]
	\centering
	\caption{Construction and query time (ns/key) at $n=2^{20}$ and
		$n=2^{30}$ for power-of-two and power-of-two-plus-one output widths. The
		Parallel columns report construction with $32$ threads and the query
		times of Figure~\ref{fig:stress} ($2^{33}/n$ structures of $n$ keys
		queried by $96$ threads), which we measured only for \VFilter, BuRR, and
		\PHunc. The rows VFilter and MWHC correspond to VFilter(Fuse+Aug) and
		VFilter(MWHC) in the figures; for HEM(Fuse) and HEM(GOV3) we report only
		query times. The rows with suffix -novec report the query times of
		\VFilter and HEM(Fuse) compiled with the loop vectorizer disabled.}
	\label{tab:detail}
	\small
	\setlength{\tabcolsep}{3.5pt}
\begin{tabular}{lrrr@{\hspace{10pt}}rrr@{\hspace{18pt}}rrr@{\hspace{10pt}}rrr}
\toprule
& \multicolumn{6}{c}{Sequential} & \multicolumn{6}{c}{Parallel} \\
\cmidrule(lr){2-7}\cmidrule(lr){8-13}
 & 8 & 16 & 32 & 9 & 17 & 33 & 8 & 16 & 32 & 9 & 17 & 33 \\
\midrule
\multicolumn{13}{l}{\emph{Construction, $n=2^{20}$}} \\
BuRR & 74 & 84 & 94 & 81 & 87 & 96 & 18 & 17 & 17 & 17 & 17 & 17 \\
VFilter & 66 & 71 & 69 & 73 & 75 & 81 & 34 & 32 & 32 & 33 & 32 & 33 \\
PHunc & 55 & 54 & 55 & 58 & 59 & 64 & 19 & 17 & 19 & 28 & 30 & 28 \\
MWHC & --- & --- & --- & --- & --- & --- & 23 & 24 & 23 & 24 & 21 & 21 \\
Fuse(64) & 47 & 44 & 43 & 49 & 50 & 54 & --- & --- & --- & --- & --- & --- \\
Fuse(128) & 68 & 68 & 71 & 74 & 72 & 77 & --- & --- & --- & --- & --- & --- \\
\midrule
\multicolumn{13}{l}{\emph{Queries, $n=2^{20}$}} \\
BuRR-neg & 17.9 & 19.2 & 20.0 & 20.4 & 21.6 & 20.3 & 103.3 & 124.6 & 135.7 & 108.4 & 127.1 & 138.3 \\
BuRR-pos & 14.5 & 23.4 & 35.5 & 17.3 & 25.3 & 35.3 & 129.9 & 263.2 & 383.7 & 137.3 & 276.2 & 391.0 \\
VFilter & 2.7 & 2.7 & 3.5 & 5.0 & 5.4 & 5.8 & 56.5 & 66.3 & 79.3 & 60.8 & 71.9 & 83.4 \\
VFilter-novec & 4.3 & 4.8 & 4.9 & 8.4 & 11.3 & 16.4 & 68.6 & 80.6 & 92.2 & 88.0 & 106.7 & 123.0 \\
PHunc & 3.8 & 4.1 & 4.2 & 4.5 & 4.8 & 4.9 & 86.4 & 97.8 & 106.4 & 104.9 & 117.2 & 127.3 \\
MWHC & 2.2 & 2.2 & 3.4 & 4.3 & 4.7 & 5.1 & --- & --- & --- & --- & --- & --- \\
HEM(Fuse) & 4.5 & 5.0 & 5.2 & 6.1 & 6.6 & 6.9 & --- & --- & --- & --- & --- & --- \\
HEM(Fuse)-novec & 6.0 & 6.7 & 6.8 & 10.7 & 14.0 & 20.0 & --- & --- & --- & --- & --- & --- \\
HEM(GOV3) & 15.7 & 17.5 & 17.6 & 23.7 & 27.5 & 33.0 & --- & --- & --- & --- & --- & --- \\
Fuse(64) & 1.6 & 1.7 & 3.4 & 4.2 & 4.5 & 4.9 & --- & --- & --- & --- & --- & --- \\
Fuse(128) & 2.3 & 2.4 & 3.5 & 4.8 & 5.2 & 5.6 & --- & --- & --- & --- & --- & --- \\
\midrule
\multicolumn{13}{l}{\emph{Construction, $n=2^{30}$}} \\
BuRR & 77 & 85 & 95 & 82 & 88 & 97 & 5 & 6 & 7 & 6 & 6 & 7 \\
VFilter & 62 & 64 & 66 & 67 & 70 & 75 & 7 & 7 & 8 & 7 & 8 & 8 \\
PHunc & 73 & 74 & 74 & 100 & 100 & 105 & 4 & 4 & 4 & 5 & 6 & 6 \\
MWHC & --- & --- & --- & --- & --- & --- & 14 & 14 & 15 & 14 & 15 & 15 \\
Fuse(64) & 54 & 55 & 57 & 58 & 61 & 65 & --- & --- & --- & --- & --- & --- \\
Fuse(128) & 124 & 123 & 123 & 130 & 133 & 138 & --- & --- & --- & --- & --- & --- \\
\midrule
\multicolumn{13}{l}{\emph{Queries, $n=2^{30}$}} \\
BuRR-neg & 60.2 & 61.0 & 61.6 & 66.5 & 68.4 & 67.5 & 91.6 & 113.3 & 125.3 & 100.9 & 118.3 & 132.0 \\
BuRR-pos & 87.8 & 159.7 & 205.1 & 90.7 & 162.8 & 195.1 & 140.7 & 252.5 & 373.8 & 126.5 & 254.2 & 373.6 \\
VFilter & 14.6 & 14.8 & 32.2 & 28.0 & 29.3 & 30.9 & 56.2 & 68.4 & 90.1 & 60.5 & 75.6 & 94.3 \\
VFilter-novec & 44.8 & 45.1 & 46.1 & 59.1 & 59.9 & 62.2 & 67.6 & 83.0 & 98.0 & 87.0 & 110.6 & 131.0 \\
PHunc & 43.0 & 42.9 & 43.1 & 49.0 & 49.1 & 49.1 & 80.5 & 91.0 & 99.7 & 98.8 & 109.3 & 119.3 \\
MWHC & 11.7 & 11.9 & 32.1 & 26.6 & 27.5 & 29.2 & --- & --- & --- & --- & --- & --- \\
HEM(Fuse) & 32.5 & 33.0 & 34.1 & 34.2 & 39.0 & 43.3 & --- & --- & --- & --- & --- & --- \\
HEM(Fuse)-novec & 56.5 & 57.1 & 56.3 & 62.5 & 64.9 & 69.5 & --- & --- & --- & --- & --- & --- \\
HEM(GOV3) & 99.0 & 101.1 & 102.3 & 103.9 & 108.1 & 114.2 & --- & --- & --- & --- & --- & --- \\
Fuse(64) & 9.3 & 9.5 & 32.1 & 24.5 & 25.7 & 27.9 & --- & --- & --- & --- & --- & --- \\
Fuse(128) & 12.2 & 12.4 & 31.9 & 27.9 & 29.1 & 30.7 & --- & --- & --- & --- & --- & --- \\
\bottomrule
\end{tabular}
\end{table}


In all these tests BuRR has an almost negligible space overhead
(between $0.1$ and $0.2$\% on large key sets) whereas \VFunc/\VFilter have a
space overhead of $10.5$\% from about two billion keys ($\epsilon=0.001$; the
overhead is slightly larger for smaller key sets), and the MWHC construction has
a space overhead between $23.3$ and $24.2$\% ($\epsilon=0.01$). For \PHunc the overhead is the
space used by PHastR, about two bits per key, that is, between $6$ and $25$\%
of the output width, decreasing as the width grows (in
Figure~\ref{fig:func}, where the width is $\lceil\lg n\rceil$, between $6$
and $10$\%).

We test key sets of size $2^{20}$, $2^{21}$, \ldots, $2^{33}$. Not all
constructions reach the largest size. Unsharded fuse graphs with $64$-bit
signatures stop at $2^{31}$, as with $2^{32}$ keys they would need more than
$2^{32}$ vertices, whereas their implementation uses $32$-bit vertices;
moreover, by~\eqref{eq:approx}, at $2^{32}$ keys the probability of a duplicate
signature, which makes a construction attempt fail, would exceed $39$\%.
Unsharded fuse graphs with $128$-bit signatures and $32$- or $33$-bit values
stop at $2^{32}$, as with $2^{33}$ keys their construction runs out of memory.

We used the parallel BuRR implementation made available by the
authors~\cite{BLSPCBRR}, suitably modified to run the same tests as the Rust
code.\footnote{\url{https://github.com/vigna/BuRR/commit/678e43f3dd40f2fcdb50d9c65625ae68f8f5c2f3}}
Since BuRR has several parameters whose influence on the behavior of the filter
is not always predictable, in particular in combination, we used the settings
that the authors call a “fast filter configuration”, which should provide a
“good all-around preset for filters” (it uses, in particular, interleaved
solution storage, which we found to be the fastest choice in almost all
configurations); BuRR static functions use the same configuration in retrieval
mode. The code for BuRR provided by the authors has the output width fixed at
compile time, which might be acceptable for filters, but is entirely
unrealistic for static functions (e.g., index functions have an output width
that depends on the size of the key set): in Figure~\ref{fig:func} we use a
different BuRR binary for each width. The comparison is thus slightly unfair to
\VFilter and \VFunc, whose output width is chosen at run time.

For all other constructions except HEM(GOV3), we use our Rust implementation,
version 0.15.5 of \texttt{sux}. HEM(Fuse) and the Rust port of the parallel benchmark of the BuRR paper
are built on \texttt{sux} within the benchmark framework described next.
The implementation uses Rust traits to abstract over the size of the signatures
($64$ or $128$ bits) and the sharding/edge generation logic (MWHC, fuse, etc.),
making it possible to obtain monomorphized, optimized versions of all
constructions (including bare fuse graphs) using the same code base.

\subsection{Measurement Protocol}
\label{sec:protocol}

The whole benchmark campaign is driven by a single Python framework, which
builds every binary on demand, runs every data point, and generates the figures
and tables of this paper (except for Table~\ref{tab:offline}, whose runs are
described in the documentation of the framework); it is distributed as free software, together
with pinned submodules of every competing
implementation.\footnote{\url{https://github.com/vigna/ecs-bench}}
We describe here the protocol it implements.

\emph{Keys.} Keys are generated deterministically and identically for all
implementations: a structure on $n$ keys is built on the integers $[0\..n)$.
Queries use a different sequence, namely the Weyl sequence $x_{i+1} = x_i +
	\texttt{0x9E3779B97F4A7C15} \bmod 2^{64}$, so that essentially all queries are
\emph{negative}, which is the realistic workload for a filter. The exceptions
are MMPHFs (Section~\ref{sec:mmphf}): their keys are sorted random $64$-bit
integers (with the same distribution, but from different generators in the Rust
and C++ code) or strings, and queries are keys sampled from the key set; and
BuRR static functions (Figure~\ref{fig:func}), which are queried on keys, as
their benchmark checks the retrieved values (the cost of a query to a static
function does not depend on whether the queried element is a key).

\emph{Positive queries.} For filters based on hypergraphs, the cost of a query
does not depend on whether it is a key: a query is straight-line code that
retrieves and combines the same values in both cases, and then compares the
result with the fingerprint of the key, so it performs exactly the operations
of a query to a static function with the same output width, plus a comparison.

This is not true of BuRR: a BuRR filter query computes the
bits of the result one at a time, each from a different pair of words of the
solution, and returns as soon as a bit differs from the corresponding bit of
the fingerprint. A negative query thus stops on average after two bits, whereas
a positive query computes all bits, performing exactly the operations of a
query to a BuRR static function with the same output width, plus a comparison
per bit. We thus measure BuRR also on positive queries, obtained by mapping
each element $x$ of the Weyl sequence to $\lfloor xn/2^{64}\rfloor$, which is a
key, as keys are the integers $[0\..n)$, so that the query loop differs from the
negative one by a single multiplication; we label the two cases BuRR-neg and BuRR-pos, and the latter is the relevant one when
BuRR is used as a static function. In our tests BuRR-pos is indeed representative
of the query time of a BuRR static function with the same output width: computing
the value of a key without comparing it with the fingerprint changes the query
time by less than $1$\% in median, and by at most $13$\% at all sizes and
widths; and at $2^{32}$ keys a BuRR static function with $32$-bit values
(Figure~\ref{fig:func}) answers queries in $304$\,ns, against $298$\,ns for
BuRR-pos.

\emph{Queries.} Each query run performs $10^8$ queries ($2\cdot 10^7$ for
BuRR-CSF and for LeMonHash-VL), repeated $5$ times in the same process; the
reported figure is the minimum of the five repetitions.
Queries are single-threaded, except in the parallel test described below. A
single-threaded query to \PHunc retrieves the value of the key, as for the other
static functions (used as a filter, it would just compare the value with the
hash of the key).

\emph{Vectorization.} Since the queries of our benchmarks are independent, the
compiler can vectorize the query loops, computing several queries at a time
with vector instructions. To show the effect of this optimization, we also
measured \VFilter, \VFunc, HEM(Fuse), and the LCP-based MMPHFs of
Section~\ref{sec:mmphf} compiled with the loop vectorizer of LLVM disabled
(\texttt{-C llvm-args=-vectorize-loops=false}); in the figures, their query
times appear as dashed lines labeled with the suffix -novec, and in the text we
say that queries are \emph{not vectorized}. BuRR is not affected by the same
change, as its query times change by at most $4$\% when compiled with the loop
vectorizer of GCC disabled (\texttt{-fno-tree-loop-vectorize}), and neither is
\PHunc, whose query times change by at most $2.2$\%.

\emph{Parallel queries.} For the test of Figure~\ref{fig:stress} we use, for
BuRR, the benchmark used in the BuRR paper for the same purpose, and, for
\VFilter, a Rust port of the same benchmark.\footnote{The original benchmark
	derives the structure to query from the same random word as the key, so
	that the structure is a function of the key, and only $n$ distinct pairs
	are queried; we fixed this in both versions.} The port also runs \PHunc,
queried as a filter by comparing the retrieved value with the hash of the key.
For the parallel test of Figure~\ref{fig:func}, it runs \VFunc and \PHunc on
positive queries only; for the same test, we extended the BuRR benchmark so
that it builds BuRR static functions and checks the retrieved values. For each
size $n$, we build
$2^{33}/n$ structures on the same keys $[0\..n)$ with different seeds, using
$96$ threads (if there are fewer structures than threads, each structure is
built using several threads). Then, $96$ threads perform $10^8$ queries each on
random pairs formed by a key and a structure; pairs are generated in blocks
outside of the timed loop, and the reported time, from a single run, is the sum
of the query times of all threads divided by the number of queries. Positive
queries use random keys in $[0\..n)$, and negative queries random integers in
$[n\..2n)$.

\emph{Construction.} Each data point is a single build. All builds use the same
seed, so for each key
set all output widths of a filter share the same hypergraphs, and the same
retries, if any. In the augmented-peeling regime, where by design an attempt
succeeds with probability at least $70$\% (Section~\ref{sec:small}), the
expected construction time is thus less than $1.5$ times that of a successful
attempt; in our runs, no VFilter(Fuse+Aug) construction needed a retry. The
reported time is the total elapsed time of the builder, which for our
structures runs from the start of the computation of the
signatures to the completed structure, and thus includes hashing, sharding,
sorting, peeling, and any retry. For BuRR we report the corresponding total,
which likewise includes hashing, insertion, and back-substitution. Only the
generation of the input is excluded, as different implementations have
different ways to feed the keys (e.g., the BuRR code assumes all keys are in
RAM, whereas our implementation can also read them sequentially from a stream,
as in our sequential constructions; our parallel constructions read them from
RAM, as BuRR does); hashing, which every construction performs, is included. Parallel construction was given $32$ threads for every
implementation that supports it (the signatures of unsharded fuse graphs were
computed using $16$ threads).

\emph{Space.} Every builder reports the in-memory occupancy of the structure it
has just built; when the builder serializes the structure, the framework
cross-checks this figure against the size of the file on disk, and flags a
disagreement larger than $1$\%.

\emph{Machine state.} All cores were pinned at $4.50$\,GHz using the
\texttt{performance} governor, with idle states disabled, so that measurements
are not affected by frequency scaling. Transparent huge pages were in
\texttt{madvise} mode, the default of the operating system (Amazon Linux 2023):
since none of the programs we test requests huge pages, all of them use
$4$\,KiB pages.

\subsection{Results}
\label{sec:results}

The main results are the following:
\begin{itemize}
	\item When the structure fits in the cache ($2^{20}$ keys),
	      hypergraph-based constructions are much faster than BuRR: a negative
	      query to BuRR takes $18$--$22$\,ns, and a positive one
	      $15$--$36$\,ns, against $2.7$--$5.8$\,ns for \VFilter ($4.3$--$16.4$\,ns
	      when queries are not vectorized) and $1.6$--$5.6$\,ns for unsharded
	      fuse graphs.
	\item On large key sets, where memory accesses dominate, positive queries,
	      which are the relevant case for static functions, are much slower
	      with BuRR: at $2^{30}$ keys BuRR-pos is $3.2$--$11$ times slower than
	      \VFilter, and $1.5$--$4.4$ times slower when queries are not
	      vectorized, as a positive query computes all the bits of the result,
	      so its cost grows with the output width. On negative queries BuRR is
	      $1.9$--$4.1$ times slower at $2^{30}$ keys, but only $1.1$--$1.4$
	      times slower when queries are not vectorized, and with $9$-, $17$-,
	      and $33$-bit values it is even faster at a few sizes between
	      $2^{23}$ and $2^{25}$ keys (with $33$-bit values, also at $2^{32}$
	      and $2^{33}$ keys).
	\item \VFilter (i.e., $\epsilon$-cost sharding plus fuse graphs) is
	      slightly slower than unsharded fuse graphs with $64$-bit signatures,
	      but more than half of the difference is due to hashing keys to
	      $128$-bit signatures, which \VFilter needs for large key sets (a
	      shard then uses only the $64$-bit local signatures): \VFilter is
	      within ${\approx}20$\% of unsharded fuse graphs with $128$-bit
	      signatures.
	\item Under the highly parallel workload of Figure~\ref{fig:stress}, query
	      times depend little on the size of the structures, as the total
	      number of keys is fixed, and memory accesses are slower, so the
	      differences we observe with a single query thread shrink. BuRR is
	      $1.4$--$1.9$ times slower than \VFilter on negative queries and
	      $2.1$--$4.8$ times slower on positive queries; when queries are not
	      vectorized, the factors become $1.0$--$1.5$ and $1.4$--$4.2$, and with
	      $33$-bit values BuRR is at most $12$\% slower on negative queries.
	      On our platform we thus do not observe the behavior
	      reported in~\cite{DDHSWRFSSRAM}; we believe that the main reason is a
	      more modern memory architecture (DDR5 vs.~DDR4).
	\item The cost of HEM is significant. At $2^{30}$ keys, the HEM layout
	      alone (HEM(Fuse)) makes queries $2.2$ times slower than those
	      of \VFilter at $8$ and $16$ bits, $22$--$40$\% slower at
	      power-of-two-plus-one widths, and $6$\% slower at $32$ bits; when
	      neither is vectorized, it makes them $6$--$27$\% slower. HEM as it is
	      actually deployed (HEM(GOV3)) is $3.2$--$6.8$ times slower than
	      \VFilter, and $1.8$--$2.2$ times slower when \VFilter is not
	      vectorized.
	\item Sequential construction of BuRR is slower than that of \VFilter, but
	      BuRR construction parallelizes extremely well, providing almost
	      constant per-key construction time: with $32$ threads, BuRR builds
	      faster than \VFilter at every size, and faster than the MWHC
	      construction up to $2^{21}$ keys and from $2^{30}$ keys on.
	\item Hypergraph-based constructions become increasingly slower as the
	      key set becomes larger due to bad locality, but as soon as sharding
	      starts, construction time per key drops very quickly. Augmented
	      peeling makes \VFilter{} sharded from $2^{20}$ keys on, but at
	      $2^{25}$ keys, where shards are peeled and must contain at least
	      $10^7$ keys, the number of shards drops from $16$ to $2$, and
	      construction becomes slower until sharding catches up.
	\item MWHC shards more finely than the fuse graphs used by \VFilter, and
	      its parallel construction is faster at all sizes but $2^{30}$ keys,
	      where \VFilter builds twice as fast (see below); only at $2^{33}$
	      keys the better locality of fuse graphs makes up for the coarser
	      sharding, and the two constructions take the same time.
	\item The construction time of MWHC has ups and downs: as the key set
	      grows, bad cache behavior makes the construction slower, but when the
	      sharding bounds allow one more sharding bit, shards become half as
	      large, which improves locality (and, up to $2^{22}$ keys, makes it
	      possible to use twice as many threads). Which bound binds depends on
	      the size: evaluating \eqref{eq:logballsbins} and~\eqref{eq:dupmwhc}
	      with $\epsilon=0.01$, $\theta=10\%$, $c=1.23$, and $\eta=5\%$, the
	      balls-and-bins bound~\eqref{eq:logballsbins} is the binding
	      constraint up to $2^{26}$ keys, the two bounds agree at $2^{27}$ and
	      $2^{28}$ keys, and the duplicate-edge bound~\eqref{eq:dupmwhc} binds
	      beyond that point. The spike at $2^{30}$ keys is not structural: two
	      consecutive construction attempts failed because of a duplicate edge,
	      an event of probability below $0.1$\%.
	\item Sharding not only enables parallelism, but also speeds up sequential
	      construction (Figure~\ref{fig:comp}): from about a billion keys on,
	      \VFilter{} builds $1.8$--$2.7$ times faster than unsharded fuse graphs
	      with $128$-bit signatures, because its shards are solved using
	      $64$-bit local signatures and $32$-bit vertices; indeed, after
	      hashing, its solving time is close to that of unsharded fuse graphs
	      with $64$-bit signatures, which, however, cannot be built beyond
	      $2^{31}$ keys.
\end{itemize}

Using a state-of-the-art MPHF followed by an array of values (\PHunc) is
competitive with \VFunc as long as the structures are small, but from $2^{27}$
keys on its single-threaded queries are $24$--$68$\% slower
(Figure~\ref{fig:func}). The advantage of \VFunc, however, comes from
vectorization: when queries are not vectorized, \VFunc is $1.7$--$2.2$ times
slower, and \PHunc is faster at all sizes. Under the highly parallel workload
the gap shrinks as the structures grow: \PHunc is $32$\% slower than \VFunc at
$2^{20}$ keys, but only $1$--$6$\% slower from $2^{27}$ keys on; when queries
are not vectorized, \PHunc is $9$--$24$\% faster at all sizes. Moreover,
\PHunc builds faster from $2^{21}$ keys on, by a factor of $2.1$--$2.6$ from
$2^{22}$ keys on (up to $8.5$ between $2^{25}$ and $2^{27}$ keys, where \VFunc
has fewer shards). In the end, which of \VFunc and \PHunc is preferable depends
on the application: on the width of the values, on whether queries can be
vectorized, on the parallel load, and on whether the index computed by \PHunc
can be reused.

BuRR, used as a static function, builds $1.4$--$4.1$ times faster than \VFunc,
and faster than \PHunc up to $2^{21}$ keys, but its queries are $3.4$--$7.5$
times slower than those of \VFunc ($2.0$--$4.2$ times when queries are not
vectorized) and $4.0$--$5.8$ times slower than those of \PHunc; under the
highly parallel workload, it is $3.3$--$4.4$ times slower than both.
The steep increase of the query time of BuRR
between $2^{23}$ and $2^{25}$ keys happens, as for the other structures, when
the structure no longer fits in the L3 cache; the increase is larger because
BuRR computes each bit of the value separately, using about $15$ instructions
per bit, from $2b$ consecutive words, where $b$ is the output width. Building a
BuRR static function takes up to $38$\% more time than building a BuRR filter
with $17$- to $33$-bit values (it is faster only with $33$-bit values, by at
most $9$\%, at a few sizes between $2^{22}$ and $2^{26}$ keys), and building
\VFunc takes $16$--$63$\% more time than building \VFilter, as the construction
of a static function must carry the value of each key along with the key.

Used as a filter, \PHunc builds in parallel $1.1$--$2.7$ times faster than
\VFilter (up to $9.3$ times from $2^{25}$ to $2^{27}$ keys, where \VFilter has
fewer shards). Its single-threaded queries are slower than those of \VFilter
on large key sets ($1.3$--$3.0$ times at $2^{30}$ keys; the only exception is
$2^{33}$ keys with $32$-bit values, where \PHunc is $18$\% faster), but when
queries are not vectorized \PHunc is $4$--$27$\% faster at $2^{30}$ keys.
Under the highly parallel workload \PHunc is $10$--$74$\% slower than \VFilter
at all sizes and widths; when queries are not vectorized, it is faster only
with $17$-bit values from $2^{25}$ keys on (by about $1$\%) and with $33$-bit
values, for which it also uses less space, from $2^{22}$ keys on (by up to
$10$\%).

We conclude that the ${\approx}10$\% space gain of BuRR comes at a significant
performance cost on positive queries and on sequential construction, in
particular for large key sets (and, for construction, for large output sizes);
the cost is even larger when BuRR is used as a static function, as all queries
are then positive, and their cost grows with the output width. Moreover,
$\epsilon$-cost sharding has a small impact on query time (\VFilter is within
${\approx}20$\% of unsharded fuse graphs with the same signatures), while
keeping space occupancy under control. The situation is different for filters
whose workload consists mostly of negative queries, as a negative query to BuRR
usually stops after testing a couple of bits: on large key sets, under a highly
parallel workload or when queries are not vectorized, BuRR comes reasonably
close to the query time of \VFilter, and its space savings might then be
preferable.

\section{Large--Scale Offline Construction}

\VFunc and \VFilter can be built using either in-memory or on-disk sharding. In
all previous tests all structures were built in memory, but we report in
Table~\ref{tab:offline} the offline construction and query times of very large
$2$-bit \VFilter instances, showing that our approach scales to trillions of
keys. Note that since shards are solved independently, the data structure could
be written to disk one shard at a time, without ever being materialized in
memory, and then memory-mapped; our implementation, however, solves shards in
parallel into an array in memory. For this reason we use $2$-bit filters: wider
outputs need proportionally more memory (for example, about $1$\,TiB with
$8$-bit values at $10^{12}$ keys).
Construction time per key grows sharply from $10^{10}$ to $10^{11}$ keys
because at $10^{10}$ keys the signatures ($160$\,GB) stay in the page cache,
whereas from $10^{11}$ keys on the shards must be read from disk, and solving
becomes I/O-bound (on average, fewer than five of the $32$ threads are busy).

\begin{table}[tp]
	\centering
	\caption{Offline construction and query time for very large $2$-bit \VFilter
		instances. Signatures are computed and stored on disk sequentially, and
		shards are solved using $32$ threads; construction time does not
		include hashing. All timings are in ns/key; query times are the minimum
		of five repetitions. At $10^{12}$ keys the peak memory usage was
		$264$\,GiB, of which $258$\,GiB are the filter itself.}
	\label{tab:offline}
	\begin{tabular}{lrr rr}
		\toprule
		Keys      & Shards   & Hashing & Construction & Query   \\
		\midrule
		$10^{10}$ & $2^7$    & $6.81$  & $7.17$       & $30.83$ \\
		$10^{11}$ & $2^9$    & $7.56$  & $49.67$      & $66.80$ \\
		$10^{12}$ & $2^{13}$ & $17.51$ & $60.25$      & $94.53$ \\
		\bottomrule
	\end{tabular}
\end{table}

\section{Monotone Minimal Perfect Hashing}
\label{sec:mmphf}

\begin{problem}[Monotone minimal perfect hashing]
\label{prob:mmphf}
We are given a set of keys $K$, $|K|=n$, out of a totally ordered universe $U$,
$|U|=u$. Our problem is to store information so that given $x\in U$ we can
compute in constant time a value that will be the rank of $x$ in $K$ (i.e., the
number of keys of $K$ smaller than $x$) if $x\in K$.
\end{problem}

A structure solving this problem is called a \emph{monotone minimal perfect hash
function} (MMPHF). A static function storing the ranks would solve this problem using
$n\lceil\lg n\rceil$ bits, but Belazzougui, Boldi, Pagh, and
Vigna~\cite{BBPMMPH,BBPTPMMPH2} showed that the order of the keys makes it
possible to use $O(n\log\log u)$ bits with constant-time queries. They also
provide a structure with query time $O(\log\log u)$ and space $O(n\log\log\log
u)$, which was proved to be space-optimal by Assadi, Farach--Colton, and
Kuszmaul~\cite{AFKTBMMPH}.

The simplest constant-time construction of~\cite{BBPMMPH}, \emph{LCP
bucketing}, divides the sorted keys into buckets of $B$ consecutive keys, and
computes the longest common prefix (LCP) of the keys of each bucket, seen as bit
strings. Since the keys are sorted, the LCPs of different buckets are distinct,
and the structure can be built using two static functions: the first maps each
key to the length of the LCP of its bucket and to its offset in the bucket; the
second maps the LCP of each bucket to the index of the bucket. To answer a
query, we evaluate the first function, obtaining a length $p$ and an offset $o$;
the prefix of length $p$ of the key is the LCP of its bucket, so the second
function maps it to the index $j$ of the bucket, and the rank is $jB+o$. The
first function stores at most $\lceil\lg(\lg u+1)\rceil+\lg B$ bits per key, and the
second $\lceil\lg(n/B)\rceil$ bits per bucket, so choosing $B$ of the order of
$\ln n$ minimizes the space.\footnote{Our implementation uses the smallest power
of two not smaller than $\lceil 1+c\ln 2+\ln n-\ln(1+\ln n)\rceil$, where
$c\approx1.11$ is the expansion factor of the underlying static functions,
so that the length and the offset can be packed in a single value $pB+o$
using a shift.} A \emph{two-step} variant~\cite{BBPTPMMPH2} exploits the skewed
distribution of LCP lengths: the first function stores the index of the length
in a small table of frequent lengths, or an escape value, and a second, much
smaller function stores the length for the keys whose bucket has an infrequent
LCP length.

Both constructions reduce MMPHFs to static functions. In the Java
implementations of Sux4J~\cite{Sux4J541} the underlying functions are GOV
functions, and thus use HEM sharding. In \texttt{sux}, \LcpMmphf and
\LcpTwoMmphf use instead \VFunc, so they inherit $\epsilon$-cost sharding: there are no offsets or local seeds, construction can be
parallel or offline, and it can be performed on a rewindable stream of keys
(e.g., a file) loading in memory only the first key of each bucket, which is
needed to build the second function. A query requires two or three \VFunc
queries and hashing a prefix of the key.

Recently, Ferragina, Lehmann, Sanders, and Vinciguerra introduced
LeMonHash~\cite{FLSVLMMPH}, the first new design of MMPHFs
since~\cite{BBPMMPH,BBPTPMMPH2}: the only intermediate development we are aware
of, the path-decomposed tries of Grossi and Ottaviano~\cite{GrOFCTPD}, makes the
trie-based constructions of~\cite{BBPTPMMPH2} faster. LeMonHash is a
\emph{learned} MMPHF: it keeps the bucketing scheme of~\cite{BBPTPMMPH2}, but a
piecewise linear approximation of the distribution of the keys, as in learned
indices, maps keys monotonically to buckets, and a retrieval data structure
(BuRR) stores the rank of each key within its bucket; a variant, LeMonHash-VL, handles
variable-length strings. When the model fits the distribution of the keys well,
LeMonHash uses much less space than LCP-based structures, and in the experiments
of~\cite{FLSVLMMPH} it is also faster. Those experiments, however,
compared C++ implementations hardwired for integers or byte sequences with the
Java implementations of~\cite{BBPTPMMPH2} from Sux4J, which turn any object
into a bit vector through a transformation strategy chosen at run time, and pay
a significant price for this generality. Moreover, the transformation strategy
used for strings converted them to UTF-16, doubling the length of ASCII keys,
and thus the amount of data processed by the Java structures (but not by the C++
ones).\footnote{A detailed analysis, with corrected benchmarks, is available at
\url{https://github.com/vigna/MMPHF-Experiments}.} Our Rust implementation of
LCP-based MMPHFs makes it possible to compare the two approaches under the same
conditions.

\begin{figure}[t]
	\centering
	\includegraphics[width=0.49\textwidth]{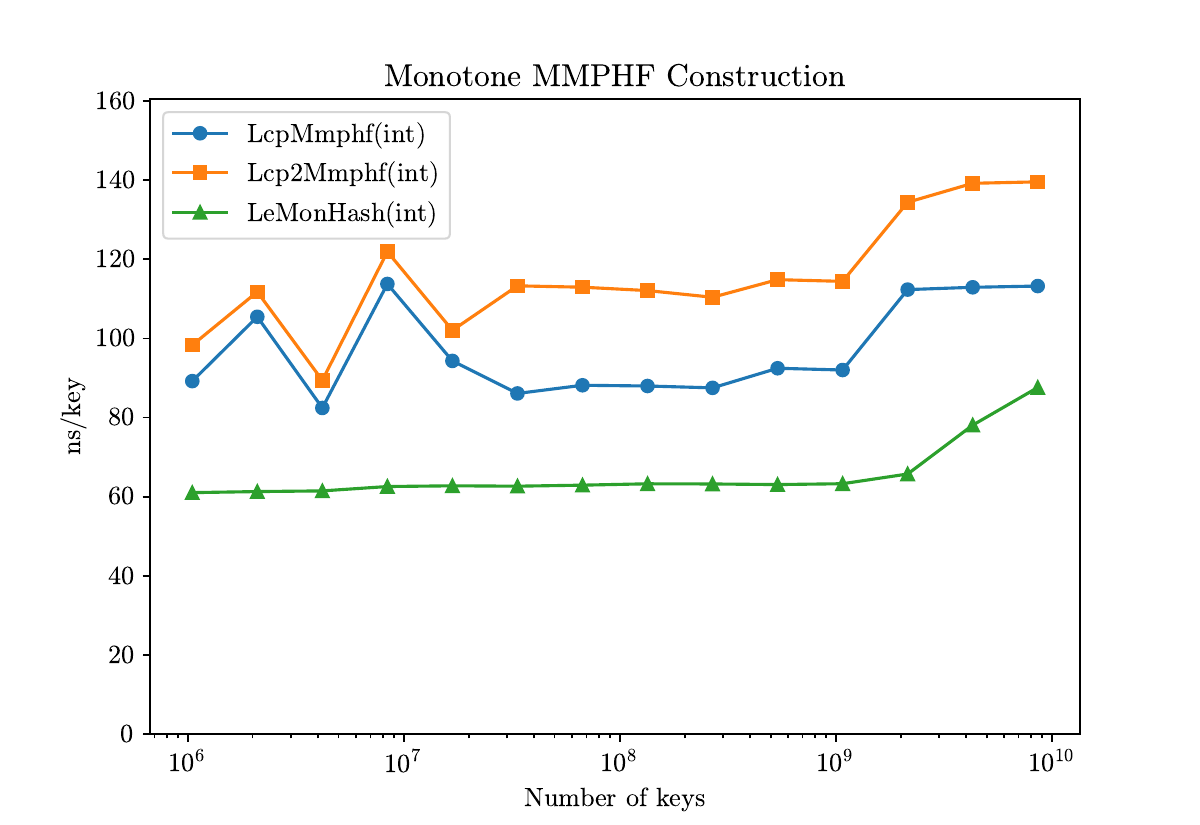} \includegraphics[width=0.49\textwidth]{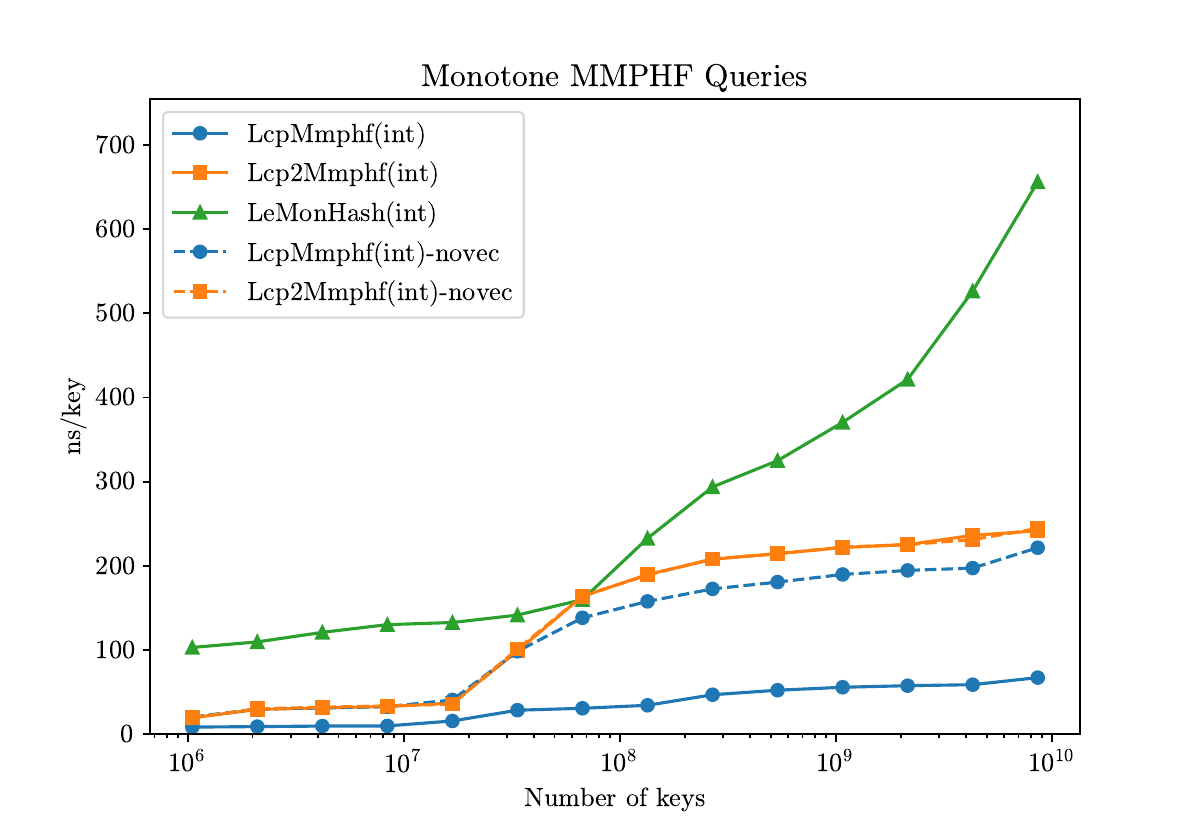}
	\caption{Construction and query times of monotone minimal
		perfect hash functions on sorted random $64$-bit integer keys. All
		constructions are sequential. The lines with suffix -novec are the query
		times of \LcpMmphf and \LcpTwoMmphf compiled with the loop vectorizer
		disabled.}
	\label{fig:mmphf}
\end{figure}

\begin{table}[tp]
	\centering
	\caption{Construction time (ns/key), space (bits/key), and query time
		(ns/key) of \LcpMmphf, \LcpTwoMmphf, and LeMonHash on sorted random
		$64$-bit integer keys (top), and of the same structures (LeMonHash-VL for
		LeMonHash) on $34\,866\,765$ terms, $105\,896\,555$ URLs, and
		$1\,070\,557\,254$ URLs (bottom). All constructions are sequential.}
	\label{tab:mmphf}
	\small
	\setlength{\tabcolsep}{4pt}
\begin{tabular}{lrrr@{\hspace{18pt}}rrr}
\toprule
 & \multicolumn{3}{c}{$n = 2^{20}$} & \multicolumn{3}{c}{$n = 2^{32}$} \\
\cmidrule(lr){2-4}\cmidrule(lr){5-7}
 & Constr. & Bits/key & Query & Constr. & Bits/key & Query \\
\midrule
LcpMmphf & 89.2 & 11.10 & 8.5 & 112.9 & 11.99 & 58.8 \\
Lcp2Mmphf & 98.3 & 8.98 & 19.6 & 139.2 & 9.83 & 236.2 \\
LeMonHash & 61.0 & 3.00 & 103.1 & 78.1 & 2.98 & 526.4 \\
\bottomrule
\end{tabular}

	\bigskip

\begin{tabular}{lrrr@{\hspace{18pt}}rrr@{\hspace{18pt}}rrr}
\toprule
 & \multicolumn{3}{c}{\texttt{trec-text.terms}} & \multicolumn{3}{c}{\texttt{uk-2007-05.urls}} & \multicolumn{3}{c}{\texttt{eu-2015.urls}} \\
\cmidrule(lr){2-4}\cmidrule(lr){5-7}\cmidrule(lr){8-10}
 & Constr. & Bits/key & Query & Constr. & Bits/key & Query & Constr. & Bits/key & Query \\
\midrule
LcpMmphf & 111 & 21.87 & 194 & 139 & 21.88 & 222 & 132 & 21.94 & 270 \\
Lcp2Mmphf & 206 & 13.19 & 174 & 109 & 17.92 & 229 & 113 & 16.95 & 286 \\
LeMonHash & 169 & 6.08 & 394 & 183 & 6.46 & 1665 & 190 & 6.92 & 2852 \\
\bottomrule
\end{tabular}
\end{table}

We used the machine and the methodology of Section~\ref{sec:protocol}, with
queries on keys of the set, and we built all structures using a single thread.
On sorted random $64$-bit integer keys (Figure~\ref{fig:mmphf} and
Table~\ref{tab:mmphf}) LeMonHash uses just $3$ bits per key, as its model fits
uniformly distributed keys perfectly, against $11.1$--$12$ bits per key for
\LcpMmphf and $9$--$9.9$ bits per key for \LcpTwoMmphf, and it is also faster to
build ($61$--$88$\,ns/key against $82$--$140$\,ns/key). Queries, however, are much
faster with LCP-based bucketing. At $2^{20}$ keys \LcpMmphf queries are twelve
times faster ($8.5$ vs.~$103$\,ns/key), and five times faster when queries are
not vectorized ($20.4$\,ns/key), whereas \LcpTwoMmphf queries, which are not
affected by vectorization, are five times faster ($19.6$\,ns/key). More
importantly, the query time of \LcpMmphf grows only slightly once the structure
no longer fits in the cache, as it should for a constant number of memory
accesses, whereas the query time of LeMonHash keeps growing.

In other words, in
practice the query time of LeMonHash is not constant, and the absolute gap
widens with the size of the key set.\footnote{The analysis of~\cite{FLSVLMMPH} bounds the query time of
LeMonHash by a constant when the keys are uniformly distributed and buckets are
assigned by linear interpolation, and by $O(\log\log_w(u/m))$, where $w$ is the
size of a machine word and $m$ the number of segments of the piecewise linear
approximation, when the approximation is combined with an optimal predecessor
structure. The implementation evaluated in~\cite{FLSVLMMPH}, which we use in the
same configuration (we only modified LeMonHash-VL so that it accesses strings
through views, avoiding copies), uses instead a piecewise linear approximation
with maximum
error $31$ and a predecessor search based on the Elias--Fano representation, so
neither bound applies to it.} \LcpTwoMmphf trades query time for space: beyond
$2^{24}$ keys its queries are $3.5$--$5.5$ times slower than those of \LcpMmphf,
though never slower than those of LeMonHash, except at $2^{26}$ keys, where they
are $2.5$\% slower; when queries are not vectorized, however, the two
structures have similar query times (\LcpTwoMmphf is at most $21$\% slower), as
\LcpMmphf loses most of its advantage.

On strings (Table~\ref{tab:mmphf}), where we used $34.9$ million terms from the
benchmarks of LeMonHash~\cite{FLSVLMMPH} and two sets of $105.9$ million and
$1.07$ billion URLs from the \texttt{uk-2007-05} and
\texttt{eu-2015} web crawls, part of the LAW
datasets,\footnote{\url{http://law.di.unimi.it/}} the gap widens with the size
of the key set also in relative terms: \LcpMmphf is twice as fast as
LeMonHash-VL on terms, $7.5$ times faster on $105.9$ million URLs, and $10.6$
times faster on a billion URLs (query times on strings are not affected by
vectorization). LeMonHash-VL uses $6.1$--$6.9$ bits per
key, against $21.9$ bits per key for \LcpMmphf and $13.2$--$17.9$ bits per key
for \LcpTwoMmphf.

LeMonHash is thus the structure of choice when space is the main concern and key
sets are not too large. When query speed matters, LCP-based MMPHFs built on
$\epsilon$-cost sharding are faster by up to an order of magnitude, and the
absolute gap grows with the size of the key set.

\section{Compressed Static Functions}
\label{sec:csf}

When the values of a static function are not uniformly distributed, storing $b$
bits per key wastes space.

\begin{problem}[Compressed static functions]
\label{prob:csf}
We are given a set of keys $K$, $|K|=n$, out of a universe $U$, and a function
$f:K\to\Sigma$, where $\Sigma$ is a finite set of values. Our problem is to
store information, using space close to $nH_0$, where $H_0$ is the empirical
entropy of the list of values $f(x)$, $x\in K$, so that given $x \in U$ we can
compute in constant time a value that will be $f(x)$ if $x\in K$.
\end{problem}

A structure solving this problem is called a \emph{compressed static function}.
Theoretical constructions have been described by Hreinsson, Krøyer, and
Pagh~\cite{HKPSCFCTA} and by Belazzougui and Venturini~\cite{BeVCSFA}.
Genuzio, Ottaviano, and Vigna~\cite{GOVFSCCSMPHF} (see also~\cite{GeVECSF})
proposed a simple practical construction: fix a prefix-free code for the values,
and for each key $x$ whose value has codeword $z_0z_1\cdots z_{\lambda-1}$ impose the $\lambda$
equations
\[
	w_{h_0(x)+i}\oplus w_{h_1(x)+i}\oplus w_{h_2(x)+i}=z_i \qquad 0\leq i<\lambda
\]
on bit variables $w_i$, as in~\eqref{eq:MWHC}. The space is then about $c$ times
the sum of the lengths of the codewords, where $c$ is the expansion factor of
the underlying construction. To answer a query, one reads three windows of
$\Lambda$ bits starting at $h_0(x)$, $h_1(x)$, and $h_2(x)$, where $\Lambda$ is the
maximum codeword length, and computes their XOR: the result starts with the codeword of $f(x)$,
followed by meaningless bits, and since the code is prefix-free the codeword can
be decoded nonetheless. To keep $\Lambda$ small, rare values are \emph{escaped}: they
are represented by an escape codeword followed by the value written explicitly.
The GOV construction for compressed functions uses a length-limited canonical
Huffman code, HEM sharding, and lazy Gaussian elimination.

Recently, Hermann, Lehmann, Vinciguerra, and Walzer~\cite{HLVWLSFDS} introduced
\emph{learned static functions}, in which a model predicts from features
associated with each key a distribution of the values, from which a key-specific
code is derived. As a non-learned baseline, they describe a compressed static
function, which we will call BuRR-CSF, that stores the codewords of a Huffman
code computed on the global distribution of values using a variable-length
variant of BuRR, and which they report as outperforming GOV in construction time,
query time, and space on all their datasets. We will return to this claim in
Section~\ref{sec:lsf}.

\subsection{\CompVFunc}

Our implementation, \CompVFunc, follows the design of the GOV construction for
compressed functions, but it is built on
\VFunc: the shard of a key is chosen by $\epsilon$-cost sharding, the triple
$h_0(x)$, $h_1(x)$, $h_2(x)$ is the local edge of the key in a fuse graph, and
the $\lambda$ equations associated with a key become $\lambda$ edges that are
translates of the same edge, one for each bit of the codeword. A query performs
three unaligned reads of $\Lambda$ bits, an XOR, and a lookup in the decoding table of a
length-limited canonical Huffman code; an escaped value requires three further
reads of the explicit value, which lies next to the codeword.

$\epsilon$-cost sharding applies verbatim: the only difference is that the
number of equations of a shard is the sum of the lengths of the codewords of its
keys, rather than the number of its keys. The balls-and-bins
bound~\eqref{eq:logballsbins} is applied to the number of keys, and the
variance due to the different lengths of the codewords is covered by the check
of the realized size of the largest shard that our implementation performs after
sharding (Section~\ref{sec:eps}). Thus, as in the case of \VFunc, the
structure stores no offsets or local seeds, and a query does not have to wait
for a first memory access before computing the address of the windows to read.
At the time of writing, \CompVFunc uses only fuse graphs, without augmented
peeling.

\subsection{Correlated Edges}

The translated edges of a key are not independent, and long codewords give rise
to \emph{clumps} of edges: for example, a $64$-bit escaped value generates more
than $64$ translates of the same edge. Clumps make the load of the segments of
the fuse graph fluctuate more than in the case of independent edges, and thus
raise the peelability threshold. If the graphs are sized without taking
correlation into account, on values containing $1$\% of escaped $128$-bit
values, constructions start to fail catastrophically.

To quantify the effect, let $\lambda_i$ be the number of edges contributed by
the $i$-th key of a shard (i.e., the length of its codeword, including the
literal following an escape), so that the shard has $m=\sum_i\lambda_i$ edges
(i.e., equations), and let
\[
	\kappa = \frac{\sum_i \lambda_i^2}{\sum_i \lambda_i},
\]
which is the expected number of edges contributed by the key of a random edge
($\kappa=1$ if each key contributes a single edge). Clumps multiply by $\kappa$
the variance of the load of a segment, that is, of the number of edges whose
first vertex lies in the segment. The first vertices of the edges of a key lie
in the same segment, except for the few translates that cross its end; thus,
up to those translates, the load of segment $j$ is the sum $N_j$ of the
$\lambda_i$ over the keys that choose segment $j$ for the first vertex of their
edge. Each key chooses this segment uniformly and independently among the
first $\ell$ segments of the graph: thus, the expected value of $N_j$ is
$m/\ell$, as in the case of independent edges, but its variance is
\[
	\sum_i \lambda_i^2\,\frac1\ell\Bigl(1-\frac1\ell\Bigr) = \kappa\,\frac
	m\ell\Bigl(1-\frac1\ell\Bigr),
\]
that is, $\kappa$ times the variance in the case of independent edges. The
relative standard deviation of the load is thus approximately
$\sqrt{\kappa\ell/m}=\sqrt{x\kappa/s}$, where $s$ is the segment size and
$x=\ell s/m$ is the density of the first $\ell$ segments.

Simulations show that the increase of the peelability threshold is
proportional to $\sqrt{\kappa/s}$, times a slowly growing function of the
number of segments. Fitting the point at which $70$\% of the attempts succeed,
as for the law for independent edges of Section~\ref{sec:fuse}, we obtain the
law
\[
	\frac{\ell s}{m}\geq 1.0949 + 0.1844\sqrt{\frac{\kappa}{s}}\,\bigl(\ln
	\hat\ell\bigr)^{1.0573},
\]
where $\hat\ell$ is an estimate of the overall number of segments. \CompVFunc
sizes the fuse graph of each shard as if its edges were independent, computes
the maximum of $\kappa$ over all shards, and increases $\ell$, when necessary,
to satisfy the law; moreover, each failed attempt increases the density by
$0.01$, so that the construction eventually succeeds even on distributions
that make the law too optimistic. The correction increases space only on
distributions with clumps.

\subsection{Experimental Results}

\begin{figure}[t]
	\centering
	\includegraphics[width=0.49\textwidth]{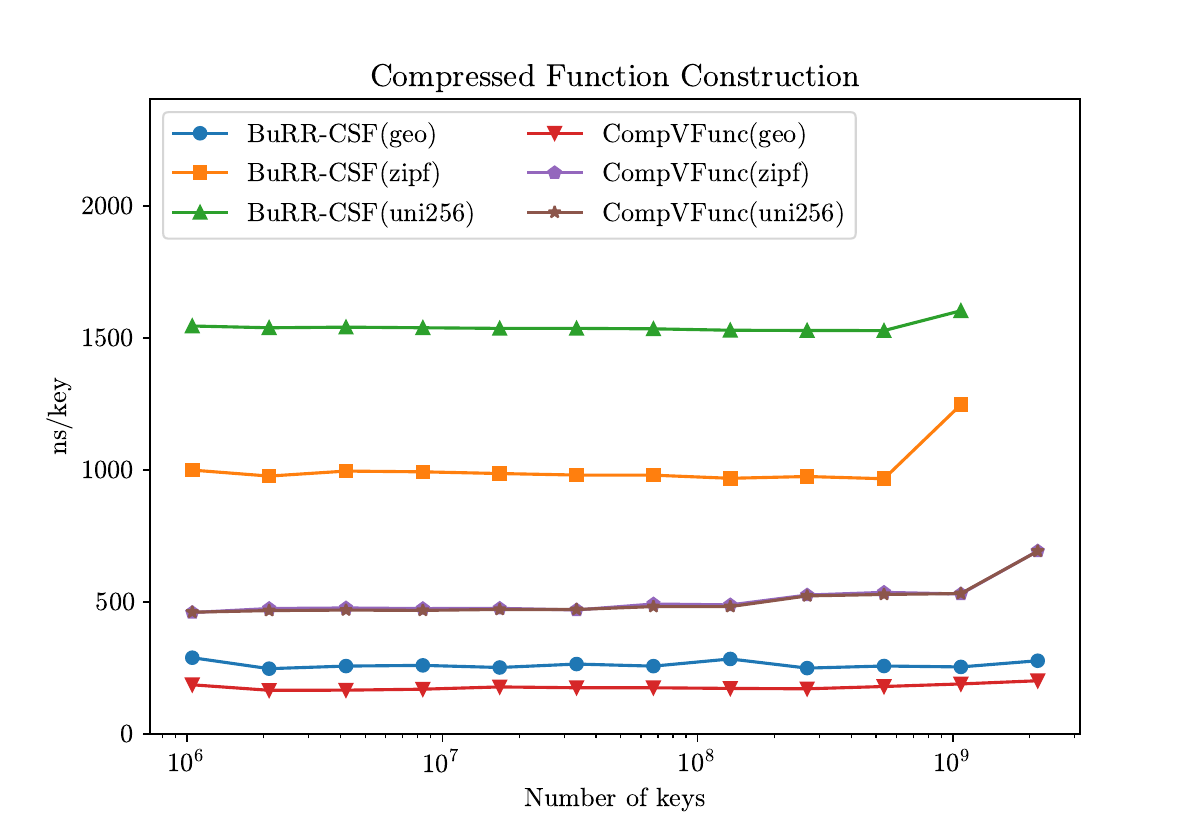} \includegraphics[width=0.49\textwidth]{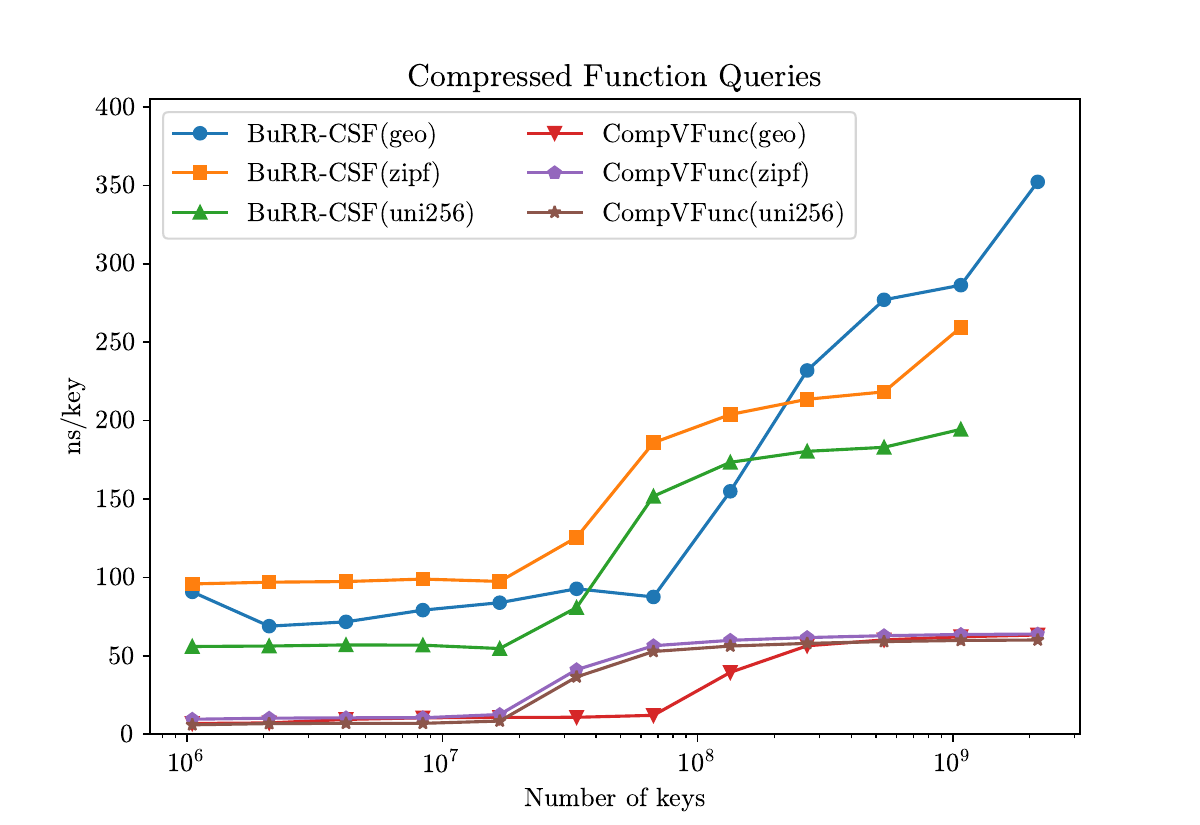}
	\caption{Construction and query times of \CompVFunc and BuRR-CSF
		on three distributions of values. All constructions are sequential.}
	\label{fig:csf}
\end{figure}

\begin{table}[tp]
	\centering
	\caption{Space (bits/key), overhead with respect to the entropy, and query
		time (ns/key) of \CompVFunc and of GOV as implemented in Sux4J
		(GOV3Comp) at $n=2^{20}$ (top) and $n=2^{28}$ (bottom).}
	\label{tab:gov}
	\small
	\setlength{\tabcolsep}{4pt}
\begin{tabular}{lrrr@{\hspace{18pt}}rrr@{\hspace{18pt}}rrr}
\toprule
 & \multicolumn{3}{c}{Geometric} & \multicolumn{3}{c}{Zipf} & \multicolumn{3}{c}{Uniform} \\
\cmidrule(lr){2-4}\cmidrule(lr){5-7}\cmidrule(lr){8-10}
 & Bits/key & Ovhd\% & Query & Bits/key & Ovhd\% & Query & Bits/key & Ovhd\% & Query \\
\midrule
CompVFunc & 2.27 & 13.5 & 6.9 & 9.28 & 14.4 & 9.6 & 8.99 & 12.3 & 6.0 \\
GOV3Comp & 2.28 & 13.8 & 35.9 & 9.13 & 12.5 & 39.6 & 8.87 & 10.9 & 23.4 \\
\bottomrule
\end{tabular}

\bigskip

\begin{tabular}{lrrr@{\hspace{18pt}}rrr@{\hspace{18pt}}rrr}
\toprule
 & \multicolumn{3}{c}{Geometric} & \multicolumn{3}{c}{Zipf} & \multicolumn{3}{c}{Uniform} \\
\cmidrule(lr){2-4}\cmidrule(lr){5-7}\cmidrule(lr){8-10}
 & Bits/key & Ovhd\% & Query & Bits/key & Ovhd\% & Query & Bits/key & Ovhd\% & Query \\
\midrule
CompVFunc & 2.23 & 11.3 & 56.4 & 9.09 & 12.1 & 61.7 & 8.94 & 11.7 & 57.9 \\
GOV3Comp & 2.28 & 14.1 & 110.5 & 9.01 & 11.1 & 161.9 & 8.85 & 10.7 & 104.1 \\
\bottomrule
\end{tabular}
\end{table}

\begin{table}[tp]
	\centering
	\caption{Construction time (ns/key), space (bits/key), and query time
		(ns/key) of \CompVFunc and BuRR-CSF at $n=2^{20}$ (top) and $n=2^{29}$
		(bottom). All constructions are sequential.}
	\label{tab:csf}
	\small
	\setlength{\tabcolsep}{4pt}
\begin{tabular}{lrrr@{\hspace{18pt}}rrr@{\hspace{18pt}}rrr}
\toprule
 & \multicolumn{3}{c}{Geometric} & \multicolumn{3}{c}{Zipf} & \multicolumn{3}{c}{Uniform} \\
\cmidrule(lr){2-4}\cmidrule(lr){5-7}\cmidrule(lr){8-10}
 & Constr. & Bits/key & Query & Constr. & Bits/key & Query & Constr. & Bits/key & Query \\
\midrule
CompVFunc & 187.4 & 2.27 & 6.9 & 461.0 & 9.28 & 9.6 & 462.3 & 8.99 & 6.0 \\
BuRR-CSF & 290.0 & 2.04 & 90.8 & 1000.7 & 8.20 & 95.9 & 1546.5 & 8.02 & 56.0 \\
\bottomrule
\end{tabular}

\bigskip

\begin{tabular}{lrrr@{\hspace{18pt}}rrr@{\hspace{18pt}}rrr}
\toprule
 & \multicolumn{3}{c}{Geometric} & \multicolumn{3}{c}{Zipf} & \multicolumn{3}{c}{Uniform} \\
\cmidrule(lr){2-4}\cmidrule(lr){5-7}\cmidrule(lr){8-10}
 & Constr. & Bits/key & Query & Constr. & Bits/key & Query & Constr. & Bits/key & Query \\
\midrule
CompVFunc & 180.9 & 2.23 & 60.2 & 537.5 & 9.09 & 62.8 & 529.6 & 8.94 & 59.1 \\
BuRR-CSF & 258.5 & 2.04 & 277.1 & 967.9 & 8.19 & 218.3 & 1529.1 & 8.02 & 183.0 \\
\bottomrule
\end{tabular}
\end{table}

We compared \CompVFunc with BuRR-CSF, using the authors' code, and with the GOV construction for compressed functions, as
implemented in Sux4J (\texttt{GV3CompressedFunction}, labeled GOV3Comp in
Table~\ref{tab:gov}; we call it simply GOV in this section), on three distributions
of values: a geometric distribution (the number of trailing zeros of a random
$64$-bit integer, with entropy $2$ bits), a Zipf distribution with exponent one
over $2\,000$ values (entropy $8.12$ bits), and a uniform distribution over $256$
values (entropy $8$ bits). \CompVFunc and GOV use the same length-limited
Huffman code, without the entropy-based truncation of the code that \CompVFunc
applies by default. As in the case of MMPHFs, all constructions are
sequential. The query times of \CompVFunc change by at most $5$\% when the
loop vectorizer is disabled. GOV is built by Java code, and queried by the C code distributed
with Sux4J, so for GOV we report only queries and space, at $2^{20}$ and $2^{28}$
keys.

The comparison with GOV (Table~\ref{tab:gov}) shows the effect of
$\epsilon$-cost sharding and fuse graphs, as the two structures use the same
coding and essentially the same space (the difference is less than $3$\%):
\CompVFunc answers queries $3.9$--$5.2$ times faster at $2^{20}$ keys, and
$1.8$--$2.6$ times faster at $2^{28}$ keys. The effect includes the simpler
hashing that $\epsilon$-cost sharding makes possible: having no local seeds,
\CompVFunc hashes each key once, whereas GOV must hash the signature a second
time with the local seed of its shard.

BuRR-CSF (Figure~\ref{fig:csf} and Table~\ref{tab:csf}) is close to the entropy,
with an overhead between $0.2$\% and $2$\% up to $2^{29}$ keys, whereas the
overhead of \CompVFunc is between $11.3$\% and $14.4$\%. In exchange, \CompVFunc
is $1.3$--$1.6$ (geometric), $1.8$--$2.3$ (Zipf), and $2.9$--$3.3$ (uniform)
times faster to build, and answers queries $2$--$13$ times faster: $6$--$10$\,ns/key against $56$--$96$\,ns/key when the structures fit in the cache, and
$60$--$64$\,ns/key against $194$--$352$\,ns/key for the largest sizes. At
$2^{30}$ keys the space of BuRR-CSF jumps to $12.0$ (Zipf) and $11.7$
(uniform) bits per key, that is, $46$--$48$\% above the entropy, and its
construction slows down; we did not investigate the cause. With these two distributions, building BuRR-CSF with
$2^{31}$ keys exhausts the $384$\,GiB of core memory, so the corresponding curves
in Figure~\ref{fig:csf} stop at $2^{30}$ keys.

\subsection{On the Evaluation of Learned Static Functions}
\label{sec:lsf}

The claim of~\cite{HLVWLSFDS} that BuRR-CSF outperforms GOV in all three
metrics (space, construction time, and query time) contrasts with our
measurements of query time, in which GOV answers queries $1.3$--$2.5$
times faster than BuRR-CSF. The reason is an asymmetry in the benchmarks
of~\cite{HLVWLSFDS}. In each dataset, the key of an example is its index. All
structures except GOV hashed the index as an integer: the line of code that
would add the features of the example to the hash is commented out. GOV, on the
other hand, was given as keys strings containing the raw bytes of the feature
vector followed by an encoding of the index, from $9$ to about $220$ bytes per
key, stored in heap-allocated strings: each GOV query thus fetched a string
from a random memory location and hashed all its bytes. 

When GOV is given the
same keys as the other structures, its queries are as fast as or faster than
those of BuRR-CSF on every dataset (up to $2.3$ times faster), and the published
query times of GOV turn out to be $2.4$--$5.4$ times too large. The fix we
submitted was merged by the authors, a corrected version of the paper is
available on arXiv,\footnote{\url{https://arxiv.org/abs/2510.27588v3}} and the
editors of PVLDB have requested an erratum. All details, including the relevant
commits and the corrected timings, are available at
\url{https://github.com/vigna/LearnedStaticFunction}.

\section{Conclusions}

We presented two new data structures, the static function \VFunc and the static
filter \VFilter, based on $\epsilon$-cost sharding applied to a combination of
fuse graphs and augmented peeling. Their space is $10.5$\% larger than the
optimum on large key sets (slightly more on smaller ones), whereas BuRR is
almost optimal (its overhead is $0.1$--$0.2$\% on large key sets), and its
construction parallelizes even better. In exchange for the additional space,
on large key sets queries to our structures are faster than those to BuRR, by a
large factor on positive queries, which are those of static functions (only
negative queries to BuRR filters are in some cases as fast or faster, when
queries are not vectorized), and their sequential construction is faster;
their queries are also faster than those to HEM-sharded constructions, and
only slightly slower than those to unsharded fuse graphs, which do not scale.
Whether our structures or BuRR are preferable thus depends on whether query
speed or space is the main concern. Moreover, due to $\epsilon$-cost sharding
the construction of our structures can be carried out in parallel, or
offline with signatures stored on disk (our trillion-key filter was built on a
single machine using little more memory than the filter itself), and since
shards are solved independently it could also be distributed.

Our experiments also show that a fast minimal perfect hash function followed by
an array of values, as in \PHunc, is a strong alternative: it uses about two
bits per key more than the values, which is less than the overhead of \VFunc
for outputs of about twenty bits or more, it builds faster in parallel, and its
queries are faster than those of \VFunc when they are not vectorized, and within
a few percent of them under highly parallel workloads on large key sets. In all our experiments, the
structure with the fastest queries among those that scale to large key sets in
comparable space is one of \VFunc, \VFilter, and \PHunc. Which one is preferable can be decided only in
the context of a specific application, as it depends on the width of the
values, on whether queries come in batches that the compiler can vectorize, on
the parallel load, and on whether the index returned by the minimal perfect hash
function can be reused, for example to access several arrays of values.

The same machinery carries over to every data structure that reduces to static
functions. LCP-based monotone minimal perfect hash functions built on \VFunc
have a query time up to an order of magnitude smaller than that of LeMonHash
(and a significantly larger space); \CompVFunc, a compressed static function
built on $\epsilon$-cost sharding and fuse graphs, answers queries up to five
times faster than the HEM-sharded GOV construction for compressed functions in
the same space, and up to thirteen times faster than BuRR-CSF, which, like
BuRR, is closer to the optimum (up to $2^{29}$ keys, at most $2$\% above the
entropy, against $11$--$14$\%).

There are several open problems and further research directions. It might be
interesting to experiment with $4$-uniform fuse graphs: on current hardware, the
additional memory access per key would reduce performance significantly in
exchange for a minor decrease of the space overhead (by about five percentage
points), but the tradeoff might be different on future hardware.

More interestingly, it would be useful to close the theoretical gaps in the
construction of fuse graphs, and thus in our data structures. Walzer's spatial-coupling analysis~\cite{WalPCOT} explains the
asymptotic behavior, but precise probability bounds on the peelability of fuse
graphs at the finite sizes we use
would make it possible to finely tune the tradeoff between the locality of the
construction and the probability of duplicate edges. For example, slightly increasing the
overhead leads experimentally to a wider range of peelable segment sizes, but,
without theoretical guarantees, exploring the space of parameters is a daunting
task. Indeed, the contrast between the estimates~\eqref{eq:nln}
and~\eqref{eq:ln} suggests that the range of values of $\ell$ for which fuse
graphs are peelable increases significantly with $n$.

Similarly, two of the laws we use to size our structures lack a theoretical
analysis. The probability that augmented peeling succeeds on small shards
(Section~\ref{sec:small}) combines a threshold of solvability that we computed
by density evolution, without a rigorous proof, with a normal model whose
constants are fitted by simulation; precise bounds on the rank of the incidence
matrix of a fuse graph with a finite number of segments of finite size would
make it possible to size small shards with guarantees, rather than with
estimates. The law for correlated edges of Section~\ref{sec:csf}, which
determines the expansion factor of \CompVFunc, is purely empirical: an
analysis of how the peelability threshold depends on the fluctuations of the
load of the segments would be welcome.

\section*{Acknowledgements}
We thank Dario Moschetti for the Rust port of the lazy Gaussian elimination Java
code from Sux4J, Hans-Peter Lehmann for his help in using the BuRR code, and
Stefan Walzer, Stefan Hermann, and Markus Wünstel for suggesting augmented
peeling for small shards.
\bibliography{biblio}
\end{document}